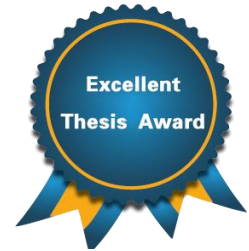

# Design and Implementation of Online Live Streaming System Using A 3D Engine

by

**Aizierjiang Aiersilan**

A thesis submitted in partial fulfillment of the requirements for the degree of

**Bachelor of Engineering**

Faculty of Software Engineering
**2021.5.13**

# Preface

With the development of the epidemic, people's demand for live video technology continues to emerge, and the requirements for low latency and high-quality transmission continue to increase. 5G can bring good news to users' high requirements for live video at the hardware level, but different design ideas still need to be tried at the software level in order to be able to live stream as much content as possible with the smallest possible delay. The WebcamTexture-related API in the Unity 3D (U3D) game engine used in the project provides developers with an efficient method to obtain external camera video data. Based on these basic interfaces, multi-peripheral and multi-content low-latency live video applications can be designed. Based on the control variable method, this study achieves the richness and diversity of video content while ensuring that the delay remains unchanged and the line is single. The interfaces related to world space, screen space and their interactive mapping in U3D are studied and analyzed. Multi-channel video frame data is obtained through the WebcamTexture interface and displayed on the canvas, so that the camera in the world space is aimed at the canvas, and the data of the camera in the world space is obtained as the basis for pushing the stream and sent, realizing a multi-peripheral and multi-content single-line low-latency live video streaming system. The affiliated repository is open source [here](#).

# Statement of Originality

I hereby declare that the thesis submitted is the result of my independent research conducted under the guidance of my supervisor. To the best of my knowledge, except where specific references and acknowledgments are made, the thesis does not contain any previously published or written research results by others, nor does it contain any material used for the award of any degree or diploma at any educational institution. Contributions from colleagues who worked with me on this research have been clearly acknowledged and expressed with gratitude in the thesis.

Abstract

DESIGN AND IMPLEMENTATION OF ONLINE LIVE
STREAMING SYSTEM USING A 3D-ENGINE

Aizierjiang Aiersilan

With the continuous development of communication network and related infrastructure, people's demand for live video streaming is increasing, and the requirements for low-latency and high-quality transmission are constantly improving. 5G network, which will be popularized in the nearest future, can bring good news for users' high requirements for live video streaming on the hardware level. At the software level, however, we still need a comprehensive method designed to stream as much content as possible with minimum latency. Existing methods for increasing streaming content is to directly use multi-channel, which does not fundamentally solve the problem of latency while merely adding a new channel into the network. I propose a method to obtain video data from multiple external cameras and stream them using only a single channel with lower latency based on a 3D Engine (eg. Unity 3D game engine). Based on the interfaces enabling the control of 3D virtual space in 3D Engines, a multi peripheral, multi content and low latency live streaming application is designed and achieved. The research is based on the Control Variable Method, which realizes the richness and diversity of streamed video content under the condition of ensuring the same delay in one single channel. This paper uses a 3D Engine to evaluate, illustrate and analyze the proposed method basically using World/Screen Space Camera, 3D Canvas and Webcam Texture APIs provided by the chosen 3D Engine to obtain multi-input video data.

For the implementation part, the affiliated project of this paper achieves the industry-level video streaming and broadcasting software application. The method is illustrated by the affiliated project via the implementation of a low-latency multi-channel live video streaming system.

Beginning with the underlying of RTSP protocol, analyzes the underlying protocol and video coding method applicable to the system, and then studied the client side technic, puts forward the requirements and objectives for the final software application, further describes the specific technical implementation principles of each module, and comprehensively describes the specific details of each module in the back-end, Finally, a live video streaming system based on Unity 3D (U3D) game engine is implemented, which supports data input of multiple external video capture devices. The system is composed of client application software based on U3D, long connection dedicated server based on WebSocket protocol, basic communication server based on HTTP protocol, persistent storage dedicated database based on MySQL, cache database based on Redis and load balancing server based on Nginx. The functions of each part are independent of each other, The function of other parts will not be affected if the specific parts are removed separately. This paper also expounds the solution ideas for the problems in the process of project implementation, discusses different solutions, and gives a mature solution for the affiliated project of this thesis.

One of the main features of the system is that the client of the live video streaming system captures multiple video data in the process, and transmits them through online live broadcast/streaming and one-to-one video call, so as to enrich the specific content of the live video with low latency, In the process of detailed design and implementation, the concept of "3D scene" built in U3D engine provides basic ideas. The video data obtained by multiple video capturing devices are mapped on the canvas in the virtual world space, and the content on the canvas is recorded by the built-in camera in world space for the use of encoded video transmission, which greatly reduces the generation of redundant data and forms a set of simple Director-Guided" live video network. At the end of this paper, the shortcomings of the affiliated project of this thesis are briefly described.

# TABLE OF CONTENTS

# 1. INTRODUCTION

## 1.1 BACKGROUND & SIGNIFICANCE

In recent years, with the continuous improvement of network infrastructure, many researchers have explored the potential impacts of 5G technology, especially as it approaches widespread deployment. The rapid development of live video streaming technologies is expected to benefit significantly from 5G advancements. I argue that the 5G era will greatly increase the demand for real-time information, with live video streaming being one of the most real-time-sensitive applications that directly impact users' daily lives. During the COVID-19 pandemic, which began in China in 2020 and quickly spread globally, many routine activities had to be conducted via live video streaming due to home isolation measures, further amplifying the demand for robust live streaming technologies. Among the critical challenges are the integration of live streaming technology with electronic games and the ability to handle complex live streaming tasks in daily life using multi-channel camera data. The live streaming system developed by me using the Unity 3D (U3D) engine aims to address these challenges and fill the existing gaps in the industry.

U3D game engine is a versatile, multi-platform game development tool developed by U3D Technologies. It enables developers to create interactive content such as 3D video games, architectural visualizations, and real-time 3D animations with relative ease. As a cross-platform game engine, U3D not only supports dual-platform compatibility for Windows and macOS but also allows games to be published across multiple platforms, including Windows, Mac, Wii, iPhone, WebGL (requiring HTML5), Windows Phone 8, and Android. In addition to its strong cross-platform capabilities, U3D offers an extensive array of third-party development kits, which facilitate rapid game development and iteration, while also enabling researchers to quickly implement theoretical concepts. This thesis leverages these platform advantages to explore and address current gaps in the industry.

The design and implementation of the online live streaming system based on U3D aims to enhance both the richness and innovation of live video streaming content, as well as its ease of use. By studying the implementation of contemporary live video streaming technologies within U3D, the affiliated project of this thesis introduces not only a basic live streaming module but also integrates multi-channel camera video streaming and the incorporation of 3D props with real-world content. These features address existing gaps in the industry and provide a feasible implementation path for embedded game live streaming technology, offering a more diverse range of live streaming options.

This thesis project has successfully developed a basic live video streaming system. I aim to use the affiliated project of this thesis as a foundation for proposing a viable implementation plan for low-latency multi-channel live video streaming technology, targeting the electric gaming, education, and media industries. Additionally, the project explores optimization strategies for reducing latency in multi-channel live video streaming, ensuring that live streaming, as it transitions into the 5G era, can be seamlessly integrated into users' daily lives. The innovations presented in the affiliated project of this thesis focus on external multi-camera video data streaming, the live streaming of content that merges 3D scenes with real-world imagery, and solutions for embedded live video streaming within the game development industry.

As the world evolves, so too do the technologies that shape our experiences. With the advent of the 5G era, not only will there be significant improvements in live video streaming latency, but a variety of new demands will also emerge, driven by the flourishing of live streaming technologies. The integration of multi-channel video data and the fusion of 3D scenes with real-world content will significantly enhance user engagement across a wide range of scenarios, fueling the growing demand for advanced live video experiences.

1.2 RESEARCH AT HOME & ABROAD

Since the advent of cable television, live TV has been a significant research focus in the field of satellite communications. If the combination of Alexander Graham Bell's telephone and Guglielmo Marconi's radio can be seen as the prototype of audio live streaming, then the launch of the "Telstar-1" satellite on July 10, 1962, marked a major milestone in live video technology. This development, which integrates audio and video information in real-time, gradually brought live video into the public's view. In the information age, the emergence of the Internet has made live video more flexible and accessible. In 1991, researchers from the Department of Computer Science at the University of Cambridge in the UK created the world's first live video streaming system with both server and client capabilities. The proposed system was designed to monitor a coffee pot remotely, refreshing the image every 20 seconds—a task with minimal latency requirements.

In recent years, the rapid development of computer network technology has led to new understandings of "live streaming latency." Advances in video compression and decoding technologies have established a reliable foundation for the real-time transmission of video over networks. More recently, with the introduction of 5G technology in China, the design and implementation of live video streaming systems have become a key area of interest for researchers in the field of communications. Furthermore, the 2020 COVID-19 pandemic has elevated live video streaming from a convenience to an essential ("rigid need") in both personal and professional contexts. Beyond "low latency," which remains a critical factor for enhancing user experience, there is now an increasing expectation for live video streaming to be convenient and diverse in its forms. As illustrated in Figure 1.2.1, platforms like Tencent Video Conference have become indispensable in modern life, with video calls transitioning from a luxury to a necessity.

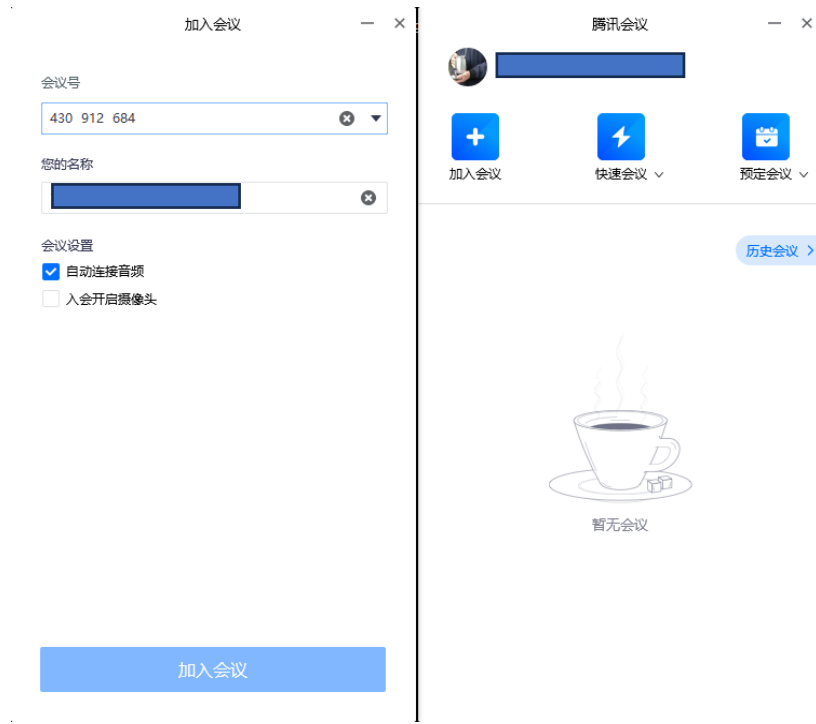

Figure 1.2.1 Showcase of the basic functions of Tencent Meeting

Research on live video technology is often conducted outside the confines of academia. While some renowned laboratories at institutions like Stanford University and Cornell University have been engaging in the study and research on live video, video compression, and encoding/decoding, the majority of advancements in live video technology are driven by enterprises, both domestically and internationally. These companies are focused on developing diversified functions, optimizing customer experience, and employing innovative technical approaches.

For instance, beyond network giants like Cisco and live video leaders such as Polycom, Avaya, Lifesize, and Sony, Zoom — a live video technology that has become deeply integrated into daily life — introduced several new features last year, including Deep Fake face-swapping, beauty filters, virtual backgrounds, simultaneous interpretation, and real-time subtitles. Zoom has not only achieved a leading technical position in "low latency," but has also closely aligned live video technology with the needs of everyday life, capturing a significant market share in the live streaming sector. By incorporating cutting-edge technologies like image processing, machine learning, and information security, Zoom has maximized the potential of live video. Moreover, Google has made significant contributions by open-sourcing WebRTC[1], providing a lightweight, open-source solution that has become highly popular among researchers in the live video streaming field.

---

[1] Web Instant Communications is an API that supports web browsers for real-time voice or video conversations. It was open sourced on June 1, 2011 and was included in the W3C recommended standard of the World Wide Web Consortium with the support of Google, Mozilla, and Opera. It transmits data based on the RTP protocol.

In the domestic market, companies like Huawei, ZTE, and Kedacom (Suzhou Keda) continue to lead in the development of new algorithms, formats, and technologies. Tencent, through its "Tencent Meeting" platform, is rapidly catching up to Zoom, leveraging existing technologies and algorithms to enhance the live video streaming experience. The recent merger of Douyu and Huya, two major live streaming platforms, has further solidified Tencent's dominant position in China's live video streaming industry.

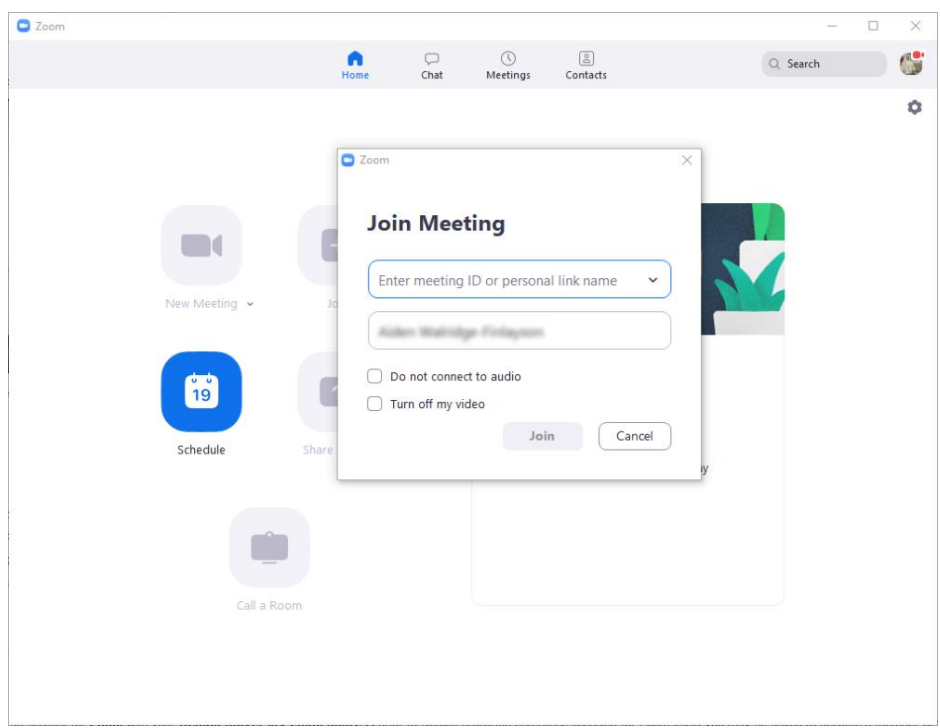

Figure 1.2.2 Showcase of the basic functions of Zoom Meeting

Live video technology is inherently dependent on advancements in network infrastructure. Faster network speeds enable a greater volume of information to be transmitted per unit of time, significantly enhancing the development of live video technology. This improvement is not only crucial from the fundamental aspect of "low latency," but also from the perspectives of increased information capacity and more sophisticated transmission methods.

In the current technological landscape of my country, 4G technology has been fully adopted, even in remote areas, where 4G Internet access is readily available. Meanwhile, 5G technology is at a critical stage of application and development. The widespread implementation of 5G will significantly boost network transmission speeds across the country and will directly drive the advancement of new media live streaming technologies.

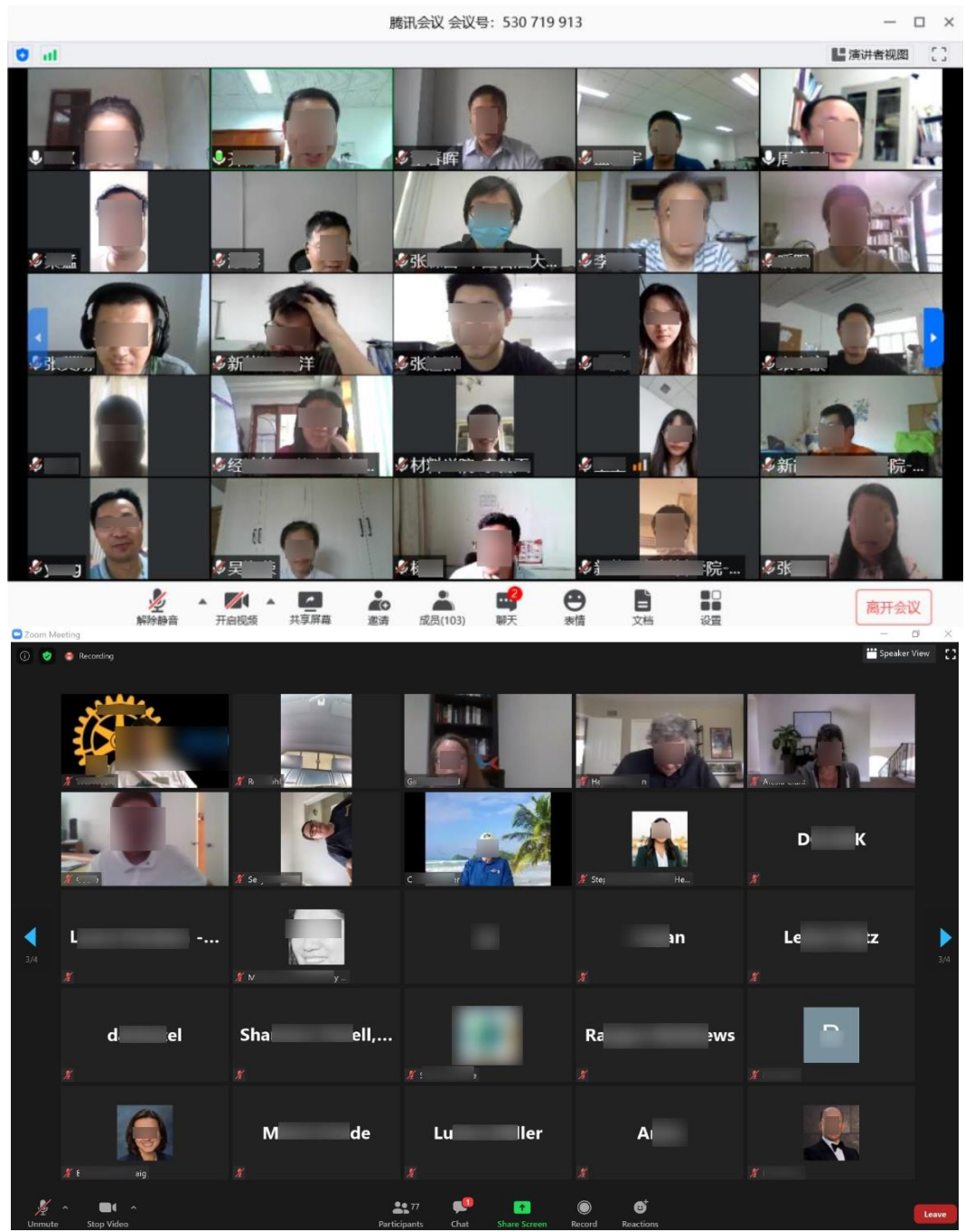

Figure 1.2.3 Comparison of Tencent Meeting (above) and Zoom Meeting (below)

However, it is important to recognize that the network environment is merely an external factor in the development of live video technology. Sustainable and stable progress in this field relies more fundamentally on internal advancements within the technology itself. According to materialist dialectics, internal factors are typically the primary source and driving force behind the movement and development of things, while external factors play a secondary role in influencing change. In the context of live video technology, internal factors such as encoding and decoding processes, compression techniques, multi-channel video data transmission with low latency, and the diversity of embedded systems (such as the AI system used by Zoom) are the true engines of its advancement. These internal elements form the foundation for the long-term success and innovation of live video technology. Conversely, external factors like network environment, bandwidth conditions, and hardware performance, while influential, serve as secondary conditions that can either enhance or constrain the development of these internal factors.

This thesis is based on the favorable external condition of robust network infrastructure available today. It focuses on the internal factors of live video technology, specifically examining issues related to the low-latency and stable transmission of multi-channel video frame data, and successfully implementing a corresponding live streaming system.

1.3 RESEARCH CONTENT

The affiliated project of this thesis is centered on the development of a live video streaming system for the U3D game engine platform, utilizing underlying algorithms for video encoding and decoding written in native C++. The system will focus on implementing the RTSP (Real-Time Streaming Protocol) streaming media protocol. The viewing end will employ the Easy Movie Texture commercial plug-in, developed specifically for the U3D engine, to design and implement a multi-channel live video streaming system with both receiving and streaming capabilities (including a streaming end, receiving end, and supporting server).

Just as the invention of the telephone turned ancient myths of "superb hearing" into reality, the advent of the Internet has made "clairvoyance" through live video technology a persistent area of fascination. In the current landscape of live video technology, numerous application-layer information transmission protocols are built upon the underlying TCP/IP protocol. The streaming media protocols most commonly accepted and used worldwide for live video include:

1. RTSP (Real-Time Streaming Protocol): A protocol with complex implementation that does not support web environments.
2. HLS (HTTP Live Streaming): A protocol with high cross-platform compatibility but significant latency.
3. RTMP (Real-Time Messaging Protocol): Known for its good real-time performance but limited cross-platform support.
4. RTP (Real-Time Transport Protocol): Typically used in conjunction with RTCP (RTP Control Protocol).

5. HTTP-FLV: A lightweight protocol that relies on Flash Video[2], with higher data consumption.
6. SRT (Secure Reliable Transport): An open-source Internet transmission protocol based on UDT (UDP-based Data Transfer Protocol), valued for its security, reliability, and low latency, but less suitable for large-scale content distribution.
7. PPP[3] (Point-to-Point Protocol): Suitable only for point-to-point information transmission.

Research indicates that major video platforms in China, such as YY, LeTV, iQiyi, Youku, Tudou, Sohu Video, and Huajiao Live, often use a combination of RTMP and HLS protocols for live streaming, with some platforms using RTMP in conjunction with HTTP-FLV. However, due to inherent limitations in RTMP, many mainstream platforms are also exploring or have begun technical trials with WebRTC (Web Real-Time Communication).

Given that the primary focus of the affiliated project of this thesis is the exploration and research of multi-channel live video streaming, with an emphasis on cross-platform compatibility and low latency, the RTSP protocol was selected as the streaming media protocol. RTSP is widely used in fields such as IPTV, surveillance, and mobile live video streaming. The underlying use of the UDP protocol meets the essential criteria for this research.

In the proposed live streaming system, when using the RTSP protocol for streaming, the streaming end acts as the "server," and the receiving end as the "client." In one-to-many scenarios, users can directly parse the RTSP push address on any platform to receive and play the video data streamed by the server. For integrated scenarios, where streaming and receiving occur simultaneously, the system supports one-to-one live video streaming.

Additionally, the streaming end, which is connected to multiple cameras, performs hardware-level detection of the number of connected cameras and their resolution before pushing video frame data. If multiple cameras are detected, the video frame data from external cameras will be received through multiple channels according to the user-specified camera numbers. The audio and video data are encoded using the AAC format and H.264 codec and then distributed via the RTSP protocol. To provide users with real-time information on each other's online status, identity, and live streaming status, a server dedicated to handling short connections is integrated into the system. This server processes simple user requests based on the HTTP protocol and records relevant data in a relational database.

---

[2] Flash Video: A video format launched by Adobe, referred to as FLV format. The entire FLV consists of The FLV Header, The FLV Body and other tags, so the loading speed is extremely fast. The suffix of the file packaged in FLV format is ".flv".

[3] PPP: This protocol is a link layer protocol designed for simple links such as transmitting data packets between equal units. This link provides full-duplex operation and delivers data packets in sequence. In live video streaming technology, such underlying protocols are not used directly but are encapsulated to implement specific application layer transmission protocols.

| Layers in OSI | Function | TCP/IP protocol suite |
|---|---|---|
| Application Layer | File transfer, email, file services, virtual terminal | TFTP, FTP, HTTP, WebSocket, SNMP, SMTP, DNS, RIP, Telnet, RTMP, RTP, RTSP, HLS |
| Presentation Layer | Data formatting, code conversion, data encryption | N/A |
| Session Layer | Disconnect or establish a connection with another node | N/A |
| Transport Layer | Provide end-to-end interface | TCP, UDP |
| Network Layer | Select a route for the data packet | IP, ICMP, OSPF, BGP, IGMP, ARP, RARP |
| Data Link Layer | Transmits addressed frames and error detection functions | SLIP, CSIP, PPP, MTU, ARP, RARP |
| Physical Layer | Transmit data in binary form on physical media | ISO2110, IEEE802, IEEE802.2 |

Table 1.3.1 Seven layers of OSI (Open Systems Interconnection) model.

The research process can be broadly divided into six stages. The first stage begins with "one-to-one" live video streaming, utilizing the OSI model (Table 1.3.1), to evaluate various protocols, including TCP, UDP, HTTP, and WebSocket, which encapsulate transport layer protocols at the application layer. By comparing these protocols' advantages and disadvantages for live streaming, an appropriate selection direction was determined. The second stage focuses on "one-to-many" video distribution, concentrating on the research and implementation of video frame data distribution mechanisms. Protocols such as RTMP, RTP, and RTSP were tested to determine their real-time efficiency, complexity, and applicability, leading to the selection of suitable methods, software, and plug-ins. In the third stage, the analysis and exploration of multi-channel video data acquisition methods were undertaken. Native JavaScript scripts were used to test video data acquisition from multiple external cameras across web platforms, Android devices, and Windows desktops. It was discovered that most browsers do not support multi-camera access, but U3D's 3D engine was effective for multi-camera video frame data acquisition. This led to the establishment of a development platform based on U3D for the project, enabling the integration of multiple external cameras within U3D. The fourth stage involved integrating the fragmented research outcomes from the previous stages to develop a comprehensive live video streaming system using U3D for Windows desktops. The proposed system incorporated "one-to-many" and "one-to-one" functionalities, multi-camera access, and real-time transmission capabilities. In the fifth stage, the Think PHP5 and Swoole frameworks were employed to build a general-purpose server for non-live streaming-related requests, and MySQL was used to design database tables for live streaming rooms and user management. The final stage focused on optimizing the client's peripheral functional modules to match the server interface, completing the integration and testing of client and server components. This stage also included the deployment of cross-platform applications for Android, web, and Linux operating systems, achieving the project's objectives. Throughout the project, the first three stages served as the experimental exploration phase, providing critical test results that informed the development of the subsequent stages. The final three stages were dedicated to applying theoretical research results in practice, using industry-standard frameworks such as Think PHP5, Swoole, and MySQL to build a robust backend system. Nginx was utilized for load balancing, enhancing the backend's scalability and applicability across various fields.

## 1.4 METHODS FOR IMPLEMENTION

This live video streaming system is developed using a front-end and back-end separation approach with a modular development structure on both ends. The modular design ensures that different modules are independent and do not interfere with one another, providing strong cohesion and flexibility for future maintenance and expansion.

On the back end, the system is divided into two major components: the server and the database. The server is further divided into an HTTP server and a WebSocket server, each catering to different conditions. The HTTP server provides methods for the front end to perform operations such as adding, deleting, modifying, and querying database information, serving as the logical service interface between the backend and the frontend. In contrast, the WebSocket server handles long-lived connections, offering real-time communication capabilities that eliminate the need for polling mechanisms required by the HTTP server. This is referred to as a state service. The backend database comprises both Redis, a non-relational database used for caching, and MySQL, a relational database. Redis is employed for storing simple key-value pairs, offering high efficiency in operations such as adding, deleting, modifying, and querying, which can be thousands of times faster than MySQL. Due to this significant efficiency gain, some engineers have even suggested replacing all MySQL databases with Redis. Additionally, the backend leverages a content distribution network (CDN) to manage user requests, providing load balancing and request forwarding to enhance overall performance and create a robust backend architecture suitable for commercial use.

The front end integrates several core technologies. The UDP-based RTSP video codec streaming technology is utilized for the live video module. For video playback, the U3D Easy Movie Texture plug-in is employed. The WebSocketSharp library is used for handling WebSocket requests, and U3D's built-in API supports the development of HTTP-related services and other technical components.

Overall, the project incorporates cutting-edge technologies. Both TypeScript and the Node.js operating MySQL third-party library represent advanced industry standards. Utilizing U3D for live video development is also an innovative approach, leveraging game development methodologies to enhance the system. Despite initial doubts about the feasibility of using a game engine for live streaming, theoretical research demonstrated that U3D could significantly improve development efficiency and practical value. Consequently, U3D was used to develop the client software for the proposed system.

In summary, the implementation technology selection process involved preliminary research and experimentation, leading to the adoption of various tools and frameworks that align with the system's requirements and the developer's practices. These include, but are not limited to, the WS-HTTP Server development framework for the backend, the U3D JSON parsing framework for the client, as well as U3D-specific plug-ins and libraries. These tools, derived from the experimental exploration stage, have provided substantial support for subsequent development and greatly enhanced the design and implementation of the entire system.

## 1.5 THESIS STRUCTURE

This thesis is structured into six chapters. The first three chapters focus on research and practice, conducting targeted and in-depth studies based on the project's requirements and objectives to derive feasible solutions and research results for practical applications. Chapters 4 and 5 address design and implementation, detailing the overall design and specific implementation of the live video streaming system.

**Chapter 1: Introduction**
This chapter presents the research background, value, and significance of the study. It reviews relevant research results and technologies from both domestic and international sources and outlines the main content and workflow of the graduation project.

**Chapter 2: Research on the RTSP Protocol**
Chapter 2 delves into the RTSP protocol, the live streaming protocol employed by the live video streaming client in the affiliated project of this thesis. It explores the protocol's specifics and details the research findings, enabling readers to gain a comprehensive understanding of the core elements of the live video client.

**Chapter 3: Research on Multi-Channel Video Capture Technology**
This chapter focuses on the technology for capturing video from multiple external camera devices. It explains the innovations introduced in the live video streaming client and discusses the advancements in multi-channel video capture technology.

**Chapter 4: Overall Design and Implementation**
Chapter 4 outlines the overall design plan and specific implementation strategies for the software. This chapter serves as a bridge between the research findings and the design and implementation phases, providing a comprehensive overview of the software's design.

**Chapter 5: Backend Architecture Design and Implementation**
This chapter describes the design and implementation of the backend architecture. It details the design schemes for each backend module and the methods used for their implementation. By the end of this chapter, the system will have undergone comprehensive development and testing, achieving a stable operational state.

**Chapter 6: Conclusion**
The final chapter summarizes the research results of the graduation project. It discusses the further research value and potential future directions for related technologies, providing a concluding overview of the project's contributions and implications.

# 2. RESEARCH & PRACITCE ON RTSP

## 2.1 LIVE VIDEO STREAMING PROTOCAL

In this chapter, I will introduce the live video protocol employed in the architecture of the affiliated project of this thesis: the Real-Time Streaming Protocol (RTSP). It will cover the protocol's composition and delve into the low-latency algorithms associated with RTSP in the context of live video. Additionally, we will explore how RTSP is implemented within U3D and provide an overview of the underlying architecture of the protocol itself.

### 2.1.1 PRELIMINARY STUDIES ON LIVE VIDEO STREAMING

The implementation and transmission of video streaming media over the Internet encompasses two primary concerns: encoding each video frame prior to network transmission and decoding and playback of the encoded video data by the receiving part. Generally, two methods are available for decoding: hardware decoding and software decoding. This project utilizes software decoding.

Despite the completion of encoding and decoding processes, the integration of data during streaming remains a complex and challenging task. During the experimental phase, various issues were identified on the client side when data integration between the transmitting and receiving ends was suboptimal. Given that the integrity of data received by the user's streaming device must be ensured, various verification mechanisms are necessary. In the affiliated project of this thesis, U3D's spatial capabilities are employed to streamline this verification process. During streaming, irrespective of whether video data is sourced externally, U3D's world space camera video capture data is used for real-time push, mitigating network jitter during the streaming process. However, during reception on the client side, network instability may lead to uneven increases in buffer queue sizes, resulting in buffer jitter, which manifests as playback jitter on the client end.

Since external network-induced jitter cannot be fully addressed at the software level, optimization is the achievable goal. This paper assumes a stable network environment for system use, although internet transmission is inevitably subject to variability due to factors such as ambient conditions and router performance. To minimize network jitter at the software level, the client-side design incorporates sub-packetization of each video frame data, utilizing a dedicated queue as a data stream storage library. During playback, video frame data is retrieved from this queue. The video frame presentation is managed by the playback module, which obtains data from an intermediate cache queue. Given that the U3D engine is employed for client development, the underlying mechanism renders video frame data on a clean canvas, achieving up to 90 frames per second, surpassing the minimum requirement of 24 frames per second. Consistent data retrieval from the cache queue ensures smooth video playback, enhancing user experience.

The following pseudo-code presents the server's stream-pushing and client's stream-pulling processes for video data packet retrieval, decoding, and playback. The algorithm implemented on both ends demonstrates efficiency in both time complexity and space complexity, addressing issues such as jitter, distortion, and sudden interruptions in video playback caused by network instability.

```
1.    void EnCodeFrame()
2.    { if (capture.grab()) // Using OpenCVForU3D captures device camera data
3.        {
4.            Decode the most recently acquired frame and return
5.            Convert an image from one space to another
6.            Convert to the corresponding color space
7.            Get pixels and store them into byte array;
8.            Use OpenCV to process array conversionTextension2D
9.            Use OpenCV to convert and compress TexturejpggridMode
10.           Record this frame as the latest frame to be sent
11.       }
12.   }
13.   void MsgSender(int contentLength, byte[] message)
14.   {
15.       for (int i = 0; i < clientSocketList.Count; i++)
16.       {
17.           // Send data to all online clients
18.           if (clientSocketList[i].isAvailable)) )
19.           {
20.               // The packet header indicates the size of the current packet
21.               btHolder = IntToByteArray(contentLength);
22.               // Packaging the data to be sent
```

```
23.            packet = addBytes(btHolder, message);
24.                // Send a packet to each client
25.            clientSocketList[i].Send(packet);
26.        }
27.        else
28.        {   // Record offline clients
29.            // Remove the client from the list
30.            clientSocketList.RemoveAt(i);
31.            i--;//Must!
32.        } } }
```

[The process of encoding, packaging and sending video data on the server]

Below is the presentation of unpacking and decoding process on the client side:

```
1.  // Decoded video frame packet queue
2.  Queue<byte[]> usableDataQueue = new Queue<byte[]>();
3.  // A lock used to prevent the decoding thread and the thread that displays the video
4.  object queueMutex = new object();
5.      // Unpacking
6.   void UnpackMultiMsg(byte[] rawData, int dataLen)
7.   {
8.      // Processing the acquired data
9.      newMessage = addBytes(buffer, rawData, dataLen);
10.     // Declare a pointer to the header only
11.     int headPointer = 0;
12.     Buffer.BlockCopy(newMessage, headPointer, bytePicker, 0, headLength);
13.     // Obtaining packet header information
14.     headCTL = ByteArrayToInt(bytePicker);
```

```
15.        // Convert the packet header information from big endian to local little endian to get the actual data
16.        IPAddress.NetworkToHostOrder(headCTL);
17.        while (headCTL <= newMessage.Length - headPointer - headLength)
18.        {
19.            headPointer += headLength; // Head pointer deviationshift
20.            // The decoded video frame data container is declared by the current packet size described in the packet header
21.            byte[] usableData = new byte[headCTL];
22.            Buffer.BlockCopy(newMessage, headPointer, usableData, 0, headCTL);
23.            lock (queueMutex)
24.            {
25.                // Enqueue available video frame data
26.                usableDataQueue.Enqueue(usableData);
27.                // Calculating the delay
28.                CalTimeSpan("usableDataQueue", $"sableDataQueue: {usableDataQueue.Count}");
29.            }
30.            isNewMat = true;
31.            startRecv = true;
32.            // Head pointer deviationshift
33.            headPointer += headCTL;
34.            if (newMessage.Length - headPointer < headLength)
35.                break;
36.            Buffer.BlockCopy(newMessage, headPointer, bytePicker, 0, headLength);
37.            headCTL = ByteArrayToInt(bytePicker);
38.            IPAddress.NetworkToHostOrder(headCTL);
39.        }
40.        buffer = new byte[newMessage.Length - headPointer];
```

```
41.        Buffer.BlockCopy(newMessage, headPointer, buffer, 0, newMessage.Length - headPointer);
42.      }
43.      // The amount of rendering to be performed per frame
44.      private void Update()
45.      {
46.          // Get Timestamp
47.          GetTimeStamp();
48.          // Project the video frame onto the canvas
49.          showFrameOnPlane();
50.          // Other logic detection and execution of user interaction
51.          Interaction();
52.      }
53.      // Flag used to record the status of the video frame data queue
54.      bool isQHasval = false;
55.      void showFrameOnPlane()
56.      {
57.          isQHasval = !newPackQueue.Count.Equals(0);
58.          if (isQHasval)
59.          {
60.              // Pop the video frame data that needs to be displayed on the canvas
61.              pixelData = newPackQueue.Dequeue();
62.          }
63.          if (isQHasval)
64.          {
65.              // Decode to get information from each image
66.              srcImg = Imgcodecs.imdecode(new MatOfByte(pixelData), 1);
67.
```

```
68.            if (textureHolder == null)
69.            {
70.                // Get each frame array according to image information
71.                textureHolder = new Texture2D(srcImg.width(), srcImg.height(), TextureFormat.BGRA32, false);
72.            }
73.            // Convert video frames from array to Texture2DTo be displayed on canvas
74.            OpenCVForU3D.U3DUtils.Utils.matToTexture2D(srcImg, textureHolder);
75.            // Apply and render to the texture
76.            textureHolder.Apply();
77.            liveHolder.GetComponent<Renderer>().material.mainTexture = textureHolder as Texture;
78.            pixelData = null;
79.        }
80.    }
```

[The process of client unpacking and decoding video data]

The core algorithms employed at both ends of the system reveal that, despite optimization at the algorithmic level for the unpacking and packing processes, there has been no significant improvement in CPU performance given the substantial volume of data involved. Clearly, the approach of packing data on the server side, unpacking it on the client side, and then dequeuing it for display on the canvas is not a viable method for real-time video data management. While utilizing a buffer queue as an intermediary layer to aggregate video frame data and ensure smooth playback without jitter on the client side is effective, it introduces another complex issue. The inclusion of such an intermediary layer imposes significant performance overhead on the playback module during the processes of "queuing each video frame data" and "retrieving each video frame data from the queue."

For instance, if the frame data is as large as 1920×1080 pixels and needs to be processed 24 times per second, the number of operations required during the queuing and dequeuing process increases exponentially. Additionally, the space overhead for storing one second of video data in the queue is 1920×1080×1024 bytes ÷ 8 × 24 times per second. The time overhead, considering twice the 24 operations, amounts to 48 such operations per second. Consequently, the CPU must exert significant effort to execute these instructions, resulting in the device heating up. This issue is particularly pronounced in mobile devices, where frequent queue operations can overwhelm the CPU, forcing it to continuously loop through tasks such as storage, decoding, and deletion of video frame data. Even when the client is run on a more robust desktop system rather than a mobile device, performance issues persist. As a software engineer, I argue that relying on hardware improvements to compensate for suboptimal software design and algorithmic inefficiencies is irresponsible during program development. While utilizing a higher-performance machine, such as a desktop, instead of a mobile device, may mitigate the issue of overheating, it does not address the underlying design flaws. Ultimately, resolving these issues requires developers to employ more efficient mechanisms at the algorithmic and design levels to achieve the desired functionality without compromising performance.

After the improvement, the queue buffering method was abandoned, and the client focused solely on receiving data. Upon receiving, the data was immediately decoded and displayed on the screen, addressing the video jitter issue during the decoding phase. Invalid video frames were replaced with subsequent frames to ensure continuity. According to the requirements of the RTSP protocol, the format of each frame was predefined. When the client acquired each frame, it was first decoded. Due to the control over network delay and data transmission volume, the server pre-encoded the transmitted data in H.264 format before streaming. Upon decoding, the client directly reads the header and body information of each packet, which is an RTP packet containing critical data displayed on the canvas. The RTSP protocol framework used in the affiliated project of this thesis encompasses three protocols: RTSP, RTP, and RTCP (Real-time Control Protocol). RTSP primarily manages connection establishment, termination, and other control functions. The RTCP protocol enables session participants to periodically send RTCP control packets to all other participants during an RTP session. The actual media data transmission in live video streaming is handled by the RTP protocol.

By using the RTSP protocol to encode and decode video frames in H.264 format and leveraging RTP packets for efficient video frame transmission, the affiliated project of this thesis's live streaming module effectively addresses the challenge of low latency. In the following section, we will delve into the strategies for controlling real-time streaming transmission delays, providing deeper insights into the workings of RTSP.

2.1.2 REAL-TIME STREAMING TRANSMISSION DELAY

As an application-layer protocol built on TCP, the RTSP protocol offers reliable stability in video data transmission. Unlike HTTP, where the client sends requests and the server responds, RTSP supports bidirectional communication, allowing both the client and server to initiate requests. This bidirectionality makes RTSP particularly effective for one-to-many applications that transmit multimedia data over IP networks. In RTSP, both the client and server manage multimedia streaming data—such as audio and video—by parsing packet header data. The protocol also supports simultaneous control of multiple streaming sessions. Although RTSP is not strictly bound to a specific network communication protocol for transmission, the server can opt to use either TCP or UDP for streaming content. While RTSP's syntax and operations are similar to those of HTTP, it does not prioritize time synchronization, which contributes to its lower latency.

In the RTSP protocol, the interaction between the client (stream-pulling end) and the server (stream-pushing end) involves six distinct stages. These stages are illustrated in the interaction diagram below. The six stages are: OPTIONS, DESCRIBE, SETUP, PLAY, Data Stream, and TEARDOWN. Notably, the "Data Stream" stage encompasses both the video stream data transmission and the transmission of control information for the video stream, which originally consisted of two separate data streams.

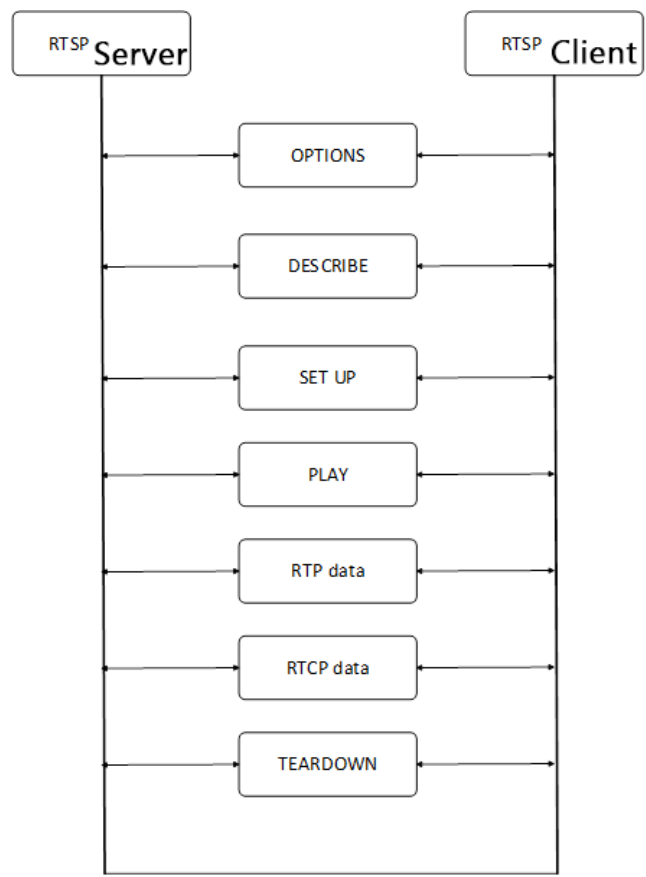

Figure 2.1.2-1 RTSP Streaming between the server and the client.

As is shown in Figure 2.1.2, the six stages in the RTSP protocol involve bidirectional communication, where each stage operates on a "client request + server response" mechanism.

First, during the OPTIONS phase, the client sends an OPTIONS request to the server to inquire about the available interfaces and methods. Upon receiving this request, the server responds with a message packet that lists the methods currently accessible to the client.

Next, in the DESCRIBE phase, the client requests the video streaming media description file from the server, typically through a URL beginning with "rtsp://". The format of this request message is based on the Session Description Protocol (SDP), as specified in RFC4566[4]. The SDP, also referenced in the SIP protocol[5], is not a transport protocol but rather a protocol for transmitting information. It plays a critical role during the session initialization process, transmitting the capability list of session participants to coordinate the parameters for both parties involved.

As a widely adopted commercial real-time streaming control protocol, RTSP operations are conducted via a message-response mechanism between the server and the client, where both parties can initiate requests and send responses. This symmetrical characteristic is fundamental to RTSP. The RTSP architecture relies on RTP for data transmission and RTCP for control transmission, utilizing either TCP or UDP for completing data transfers. Being a text-based protocol, RTSP uses UTF-8 encoding (RFC2279) and ISO10646 character sequences, following the general message format defined by RFC822, where each line ends with a CRLF. When the server receives the SDP-formatted message[6] from the client, it understands the client's request for the video streaming media description file and responds with a corresponding SDP message. This response contains essential details such as the video streaming media type, attributes, decoder information, and frame rate available on the server.

---

[4] Request For Comments (RFC) is a series of numbered documents. The documents collect information related to the Internet, as well as software files of the UNIX and Internet communities. The serial number after the RFC indicates the number of the protocol in the document.

[5] SIP (Session Initiation Protocol) is a multimedia communication protocol developed by IETF (Internet Engineering Task Force). It is a text-based application layer control protocol used to create, modify and release sessions of one or more participants. SIP is an IP voice session control protocol derived from the Internet. It is flexible, easy to implement and easy to expand. The protocol message format and other contents are detailed in RFC5621.

[6] See [RFC4566](RFC4566) for more details.

In the SET UP phase, the client initiates a connection request to the server to establish a session and prepare for receiving audio and video data. The client's request specifies whether the data packets should be transmitted via UDP or TCP, identifies the RTP and RTCP ports, and indicates whether the transmission should be unicast[7] or multicast[8]. Upon receiving the request, the server parses the information and assigns the appropriate ports for sending control data and audio/video data based on the client's specifications.

In the PLAY phase, with all preconditions met, the client is ready to begin playback, having completed all necessary preparations. However, the crucial video streaming data is yet to be obtained. The client then sends a request to the server to begin streaming the video. Similar to the HTTP response mechanism, the server responds with a "200 OK" status code, indicating that the video is ready and can be transmitted to the client. The server then begins streaming data to the client through the ports established during the SET UP phase, marking the commencement of the most critical stage in the entire process.

At this point, the server continuously sends two types of message packets to the client: RTP-based packets, which carry the video frame data for playback, and RTCP-based packets, which handle control information related to the video stream.

---

[7] Unicast: Data transmission to a specific host. Unicast, broadcast, and multicast are terms that describe different transmission forms.

[8] Multicast: It sends data to a specific group of hosts (multicast group). It is a special case of broadcast. Broadcast is when a host sends data packets to all hosts on a certain network. In order to reduce unnecessary overhead involved in broadcasting, the host can selectively send data only to a specific part of the receivers (either within the domain or between domains). This is the origin of multicast.

![Wireshark packet trace showing RTSP messages between 192.17.1.92 and 192.17.1.63, with OPTIONS, DESCRIBE, SETUP, and PLAY requests highlighted in red boxes]

Figure 2.1.2-2 The packet tracing status of RTSP among the 2 ends using Wireshark[9]

In the TEARDOWN stage, when the playback concludes, the client sends a termination request to the server. Upon receiving this request, the server responds with a "200 OK" status code and disconnects the session. The client then ceases the video playback process, marking the end of the entire live video streaming process under the RTSP protocol between the client and server.

For those interested in the technical implementation, the core code for the RTSP server developed in the affiliated project of this thesis, based on the aforementioned specifications, can be found in the appendix. Due to the extensive nature of the underlying protocol implementation, the provided code includes not only protocol-related message packet settings but also content related to encoding, all of which have been integrated into the final code. The structure of this code is illustrated in Figure 2.1.2-3.

---

[9] WireShark is a very popular network packet analysis tool that can capture data packets in the current LAN environment in real time and perform targeted analysis.

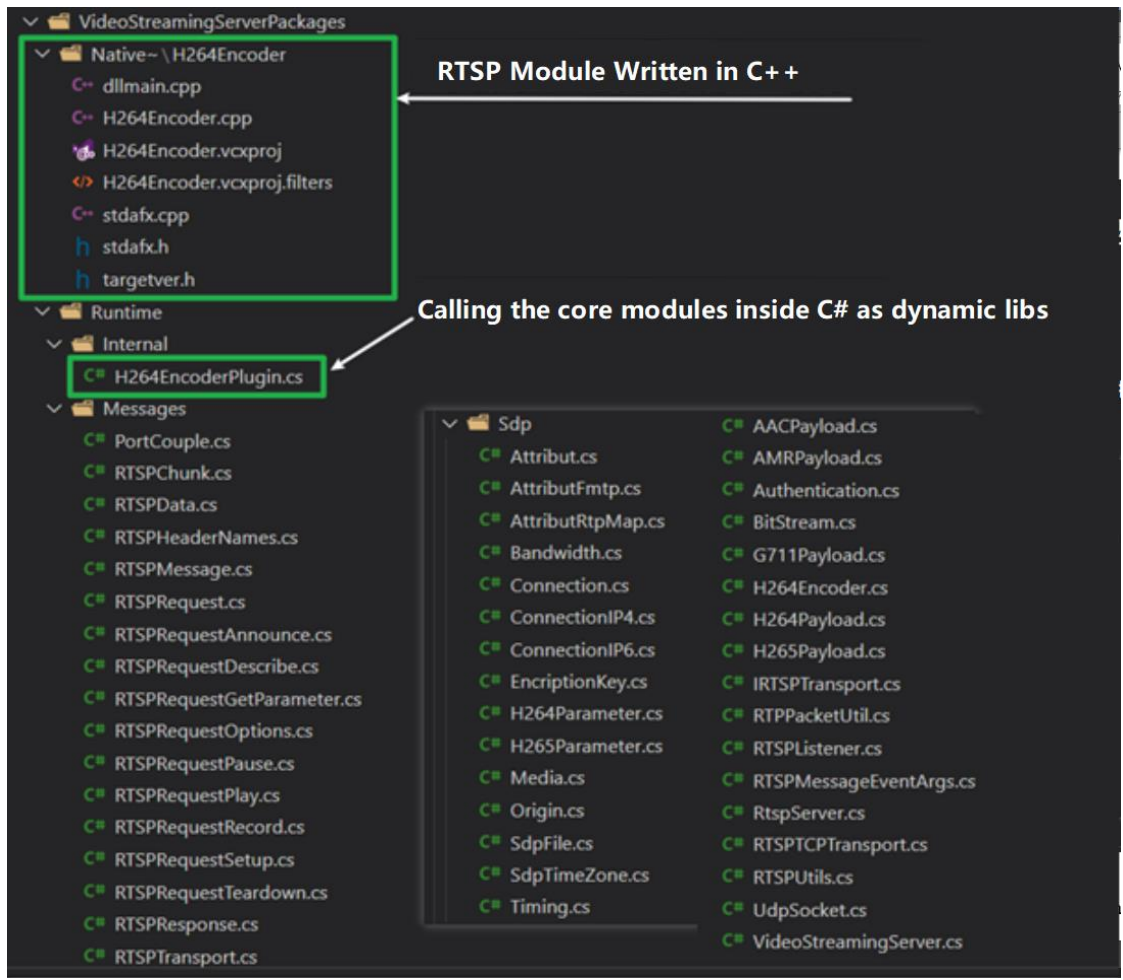

Figure 2.1.2-3 Code structure of the RTSP live streaming.

2.1.3 DEEP INTO RTSP PROTOCOL

By understanding the overall process of the RTSP protocol, I successfully implemented real-time transmission of video streaming media using the protocol. Furthermore, the protocol's architecture effectively addresses network jitter and cumulative delay issues during live video streaming. As outlined in the previous section, the client initiates the communication by sending a request to the server to retrieve interface-related information. Each request is encapsulated as a message packet with a predefined format. In alignment with the RTSP protocol's message structure, the following packet format is utilized for communication between the client and server.

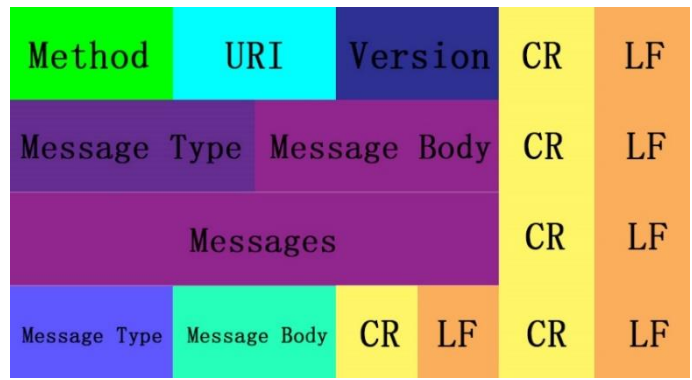

Figure 2.1.3-1 The format of the request message under RTSP protocol

As shown in the Figure 2.1.3-1 above, a request message consists of the "specific method" + "URI" + "RTSP version," followed by one or more headers, where "CR" indicates carriage return, "LF" indicates line feed, and "URI" represents the recipient's address. Upon receiving the client's request, the server responds using a similar packet structure. While the structure remains consistent, only the internal data of the message packet changes to reflect the response.

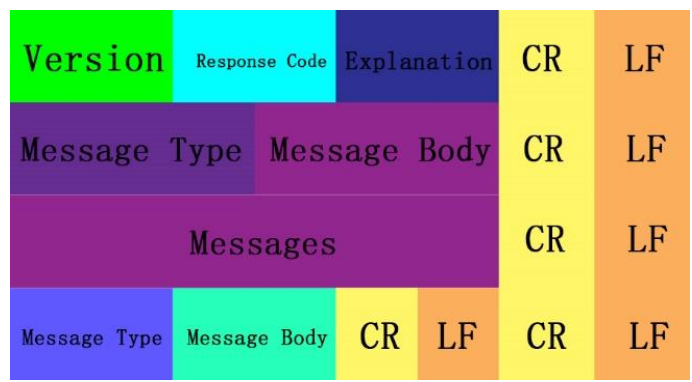

Figure 2.1.3-2 The format of the response message under RTSP protocol

In the response message from the server, the data content in the packet header originally sent by the client is replaced with RTSP version information. The second packet header contains the server's specific response code, indicating the response status, followed by a textual explanation of the status code. Typically, when the response status code is "200," the explanation is set to "OK!" This design minimizes the client's dependence on the server's documents and interfaces, aligning well with the front-end and back-end separation architecture common in modern engineering development models.

While the response code can technically be any real number and its explanation customizable, it is important to note that industry standards predominantly use "200" as the response code with "OK" as the explanation — despite the fact that this explanation holds no practical significance at the program level. During the development process, I adhered to these industry norms to ensure that "the server does not depend on the client, the client does not depend on the server, and the two do not depend on each other." This adherence to standard practices increases the system's robustness, enabling future integration with clients from other manufacturers under general settings. Admittedly, in the early stages of development, I personalized the format of the data packets sent by the server and defined the response codes independently. It wasn't until later, when I was developing the client, that I discovered the commonly used "EasyMovieTexture" plug-in could not successfully parse the custom response code sent by the server. This was because the third-party client plug-in was designed to expect a "200" response code for successful status. Additionally, when there is an inconsistency in the RTSP protocol version numbers between the server and the client, the server responds with a "401" status code, indicating unauthorized access.

During the OPTIONS phase, communication between the client and server occurs via these message packets. In subsequent phases, while the packet content may vary slightly, the overall structure remains consistent. In this phase, the client typically uses specific parameters to obtain and set resource messages from the server, such as GET_PARAMETER[10] and SET_PARAMETER[11]. However, since these parameters are not frequently used in actual development, they will be treated as extended interfaces and omitted from this discussion due to space limitations.

In the DESCRIBE stage, when the server receives a message packet from the client, two possible scenarios arise:

1. **Authentication Required**: This is a common mechanism adopted by commercial software servers like Zoom and Tencent Conference. In the

---

[10] The GET_PARAMETER parameter is used to obtain parameters from the server, generally used to obtain the time range. When there are no relevant request parameters in the sent request, it is used to maintain the RTSP connection.

[11] The SET_PARAMETER method is used to set parameters for the stream address of the client specified by the URI.

affiliated project of this thesis, I followed this commercial approach, implementing an authentication requirement to enhance the security and stability of the live streaming system. If the client message lacks authentication information, the server responds with a "401 Unauthorized" code, signaling that the client needs to provide valid credentials. The client typically sends this authentication information embedded in the URI. Only after the server successfully verifies the authentication details and returns a "200 OK" response can the process continue.
2. **Non-Authentication Required**: If the server does not require authentication, it directly processes the client's request without the need for additional verification steps.

In the third stage, the SETUP phase, the client and server must agree on how the streaming video should be transmitted to ensure correct stream pushing and pulling. During this phase, the client sends a SETUP request, specifying two ports to receive data. The first port, usually an even-numbered one, is designated for receiving RTP (Real-time Transport Protocol) data, which contains the actual video frame data. The adjacent odd-numbered port is used for receiving RTCP (Real-time Transport Control Protocol) data, which contains control information for the video frames. This convention of using even and odd-numbered ports for RTP and RTCP data is a standard practice in the development of live video streaming systems using the RTSP protocol.

In the PLAY phase, which is the fourth stage, the client sends a PLAY request to the server to initiate playback. The client can specify a playback interval in the request. If the connection between the client and server remains stable, the server begins transmitting the audio and video data packets according to the transport method agreed upon in the SETUP phase. At this point, both the client and server enter the Data Stream phase, as depicted in Figure 2.1.2-1, marking the official start of the playback process.

In addition to the PLAY instruction, which initiates playback, there is a corresponding PAUSE instruction. This instruction can be sent through either RTCP (Real-time Transport Control Protocol) or RTSP (Real-time Streaming Protocol) message packets while both ends are in the Data Stream stage. Although both methods can effectively pause playback by stopping the data transmission from the server, they operate differently:

1. **RTCP-based PAUSE Control**: Sending a PAUSE instruction via RTCP packets indicates that the server continues to send data to the client, but the client has control over whether to play it. In this scenario, the server maintains its data transmission state, while the client decides when to process and display the video frames. This method allows for more granular control of playback on the client side without interrupting the server's data stream.
2. **RTSP-based PAUSE Request**: Sending a PAUSE request through RTSP instructs the server to halt the stream transmission to the client. This request causes a temporary interruption of the video stream, effectively pausing both ends of the connection. The client sends the PAUSE command with its own URI, and if the requested URL points to a stream address, only the playback and recording of the stream are interrupted. Using RTSP for pausing is generally preferred for performance reasons, as it directly stops the server from sending data, thus reducing unnecessary network load. In practice, the choice between using RTCP and RTSP for pausing depends on the performance requirements and specific requirements. For optimal performance, the RTSP-based PAUSE request is often utilized, but RTCP control can also be employed based on the application's requirements.

Finally, in the TEARDOWN stage, the client sends a TEARDOWN message to terminate the transmission of audio and video streams. Upon receiving this message, the server responds with a "200 OK" status code and releases any related resources. The completion of the TEARDOWN stage signifies the end of the RTSP connection.

2.1.4 V<span>IDEO</span> C<span>ODEC</span> S<span>OLUTIONS</span>

Live video streaming entails the real-time transmission of high-capacity data with substantial throughput over networks. Key challenges include achieving real-time delivery and minimizing latency. Video encoding and decoding are crucial processes that significantly impact latency control throughout the streaming protocol.

Regardless of the live streaming protocol employed, video encoding and decoding remain central to live video streaming. These processes, which involve compressing each frame of video data for transmission over the network, are critical beyond the scope of the transmission protocol itself. Effective encoding and decoding solutions are essential, but their efficacy can be hindered if the underlying protocol does not support reliable message exchanges. The RTSP offers a robust framework for video encoding and decoding during live streaming, ensuring that the developer's efforts in optimizing these processes are not compromised by external network conditions. This facilitates effective testing and optimization of video encoding and decoding algorithms.

Encoding and decoding can be broadly understood as compression and decompression processes, respectively. Encoding compresses data by eliminating redundancies, enabling efficient transmission over the RTSP protocol. In practical applications, where video content consists of a continuous sequence of images rather than static or fixed images, encoding is further categorized into intra-frame and inter-frame encoding. Intra-frame encoding addresses redundancies within a single image, while inter-frame encoding tackles redundancies across a series of images. The encoding process focuses on the differences between successive images, which serves as the foundation for compression.

This project utilizes several fundamental algorithmic principles in encoding, including transformation, quantization, entropy coding, motion estimation, and motion compensation. Additionally, optimization metrics are incorporated to enhance performance. Each frame of captured image data exhibits strong correlations, which can be exploited by transforming the image from the spatial domain to the frequency domain—a process known as "decorrelation." This transformation significantly reduces the image's spatial footprint. The H.264 video encoding standard typically employs Discrete Cosine Transform (DCT)[12] to perform this transformation. The DCT

---

[12] DCT: Discrete Cosine Transform, mainly used for data or image compression, can convert spatial domain signals to frequency domain, and has good decorrelation performance. DCT transform itself is lossless, which creates good conditions for subsequent quantization, Huffman coding, etc. in image coding and other fields. At the same time, since DCT transform is symmetrical, we can use DCT inverse transform after quantization coding to restore the original image information at the receiving end. DCT transform has a very wide range of uses in the current image analysis and compression fields. Our common JPEG static image coding and MJPEG, MPEG dynamic coding standards all use DCT transform

formula used in the affiliated project of this thesis is straightforward and directly implemented in the encoding algorithm, thereby avoiding the need for extensive mathematical optimization (though various DCT formulas exist, the one presented here is employed in the H.264 video encoding algorithm used in the affiliated project of this thesis):

$$F(u) = c(u) \sum_{t=0}^{N-1} f(i) \cos\left[\frac{(i+0.5)\Pi}{N} u\right]$$

$$c(u) = \begin{cases} \sqrt{\frac{1}{N}}, & u = 0 \\ \sqrt{\frac{2}{N}}, & u \neq 0 \end{cases}$$

Here $f(i)$ is the original signal, N is the number of original signal points, $c(u)$ is the compensation coefficient, $F(u)$ is the coefficients after the DCT.

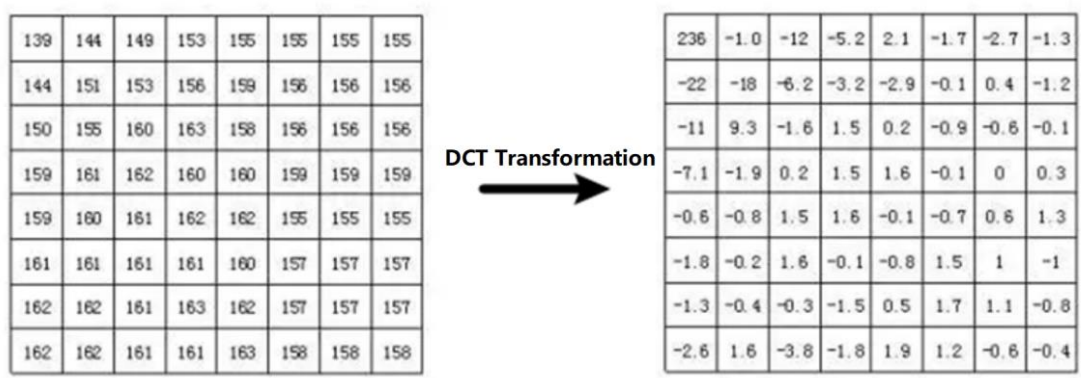

Figure 2.1.4-1 The brief showcase of the process of DCT.

As illustrated in Figure 2.1.4-1, an 8×8 image block is transformed from the spatial domain to the frequency domain following the DCT. The image undergoes significant compression post-DCT; however, to achieve further data reduction, image redundancy can be further minimized through quantization of the DCT coefficients. In the quantization process, the DCT coefficients are divided by a quantization step size, often represented as a constant, and then rounded to the nearest integer to achieve quantization. This process is implemented as follows: the result after quantization is denoted as R. The result F(u) after the DCT transform obtained above is directly used, and the quantization step is denoted as S. The formula is as follows:

$$R = \lceil F(u)/S \rceil$$

Using this formula, we can test the quantized result and compare it with the pre-quantization data. The matrix operation process demonstrated below illustrates a quantization process with a step size of 28, resulting in a final quantization value of 8.

$$\begin{bmatrix} 236.0 & -1.0 & -12.0 & -5.2 \\ -22.0 & -18 & -6.2 & -3.2 \\ -11.0 & 9.3 & -1.6 & 1.5 \\ -7.1 & 1.9 & 0.2 & 1.5 \end{bmatrix} \rightarrow \begin{bmatrix} 9.0 & 0 & 0 & 0 \\ -1.0 & -1.0 & 0 & 0 \\ 0 & 0 & 0 & 0 \\ 0 & 0 & 0 & 0 \end{bmatrix}$$

In practical scenarios, the quantized matrix often contains many zeros in the lower-right corner. From the formula, it is evident that a larger quantization step results in coarser quantization granularity, leading to a higher compression rate and smaller data packet sizes during transmission. Conversely, a smaller quantization step results in finer granularity, preserving image quality but resulting in larger data packets.

The quantized result remains a two-dimensional matrix. To further compress this data, it is typically converted into a one-dimensional format. However, during this conversion, a suitable matrix scanning algorithm must be employed. In computer graphics, various scanline algorithms are used, with raster scanning being one of the most common. In this example, since the non-zero values in the quantized matrix are predominantly located in the upper-left corner, the ZigZag scanning algorithm is used to enhance encoding efficiency. This algorithm aligns with the H.264 encoding standard widely adopted in the industry. After applying ZigZag scanning, the two-dimensional matrix is transformed into a one-dimensional array, as illustrated below:

[9, 0, -1, 0, -1, 0, 0, 0, 0, 0, 0, 0, 0, 0, 0, 0]

After scanning and transforming the matrix into one-dimensional space, it is evident that zeros are predominantly concentrated towards the end of the sequence. This characteristic provides a favorable basis for entropy coding. To accelerate project development, I have employed Context-Adaptive Variable-Length Coding (CAVLC)[13] rather than the more efficient but complex Context-Adaptive Binary Arithmetic Coding (CABAC) [14]. Entropy coding, as a lossless compression technique, involves reintegrating and optimizing the data. Following the entropy coding stage, the entire encoding process is completed.

---

[13] CAVLC stands for Context Adaptive Variable Length Coding, which is an entropy coding method used in H.264/MPEG-4 AVC. In H.264, CAVLC is used to encode the transformed residual blocks in a zig-zag order. CAVLC is a substitute for CABAC.

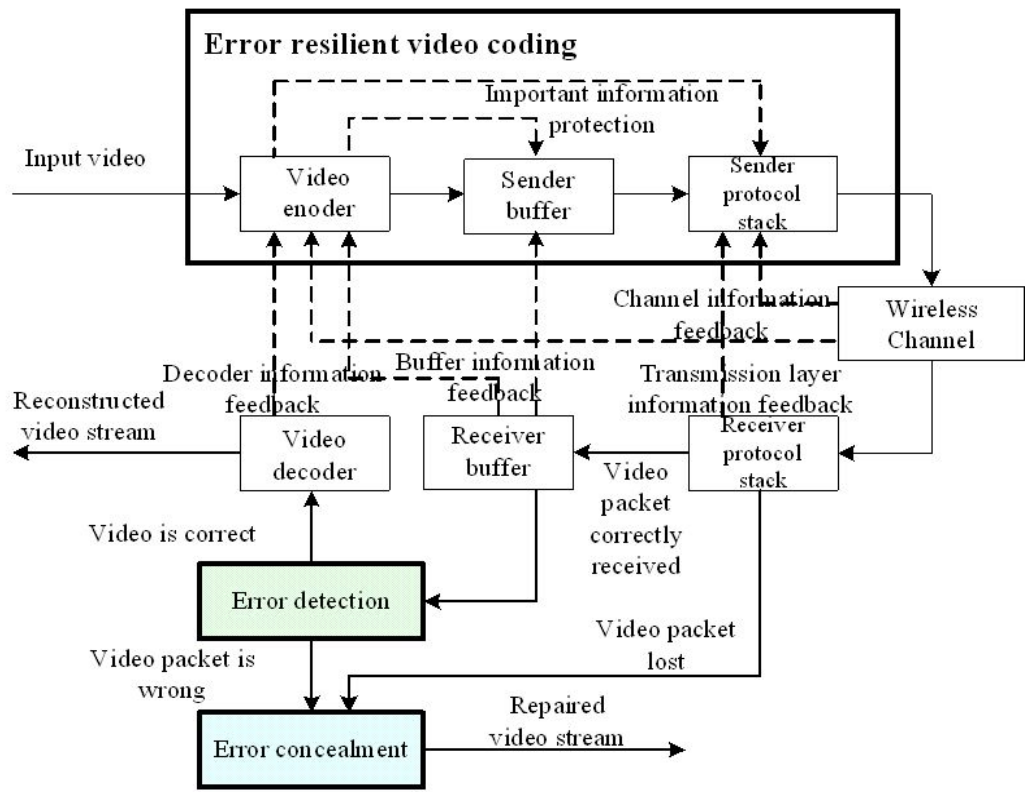

Figure 2.1.4-2 Video encoding and decoding process based on H264.

With this encoding process in place, decoding becomes relatively straightforward. The operations can be directly reversed. The Easy Movie Texture plugin used in the project is compatible with the decoding of video data encoded in H264. Consequently,

---

Although its compression efficiency is not as good as CABAC, CAVLC is simple to implement and is supported in all H.264 profiles.

[14] CABAC stands for Context-based Adaptive Binary Arithmetic Coding. CABAC is one of the two entropy coding methods in the H.264/AVC standard, and its core coding algorithm is arithmetic coding.

implementing the decoding process on the client side, after receiving the video stream and prior to playback, is a straightforward task.

## 2.2 IMPLEMENTATION OF RTSP PROTOCOL INSIDE U3D

The live video streaming/broadcasting module utilizes a purely idealized experimental approach at the research level. Although initial confidence was high when constructing the structure, the challenge of integrating the codec module and RTSP protocol module, developed in pure C# code, into U3D—such that U3D is provided with an application layer interface similar to a game object—emerged as a significant issue during the practical implementation phase following the experimental stage. This chapter outlines the strategies and solutions employed to integrate the research findings from the experimental stage into practical application and presents performance testing results for the live video streaming/broadcasting module after its integration with U3D.

### 2.2.1 DECOUPLING UNDERLYING ALGORITHMS FROM U3D

After thorough process planning, structural design, and algorithm research for the video live module of the video live system, all aspects remain theoretical. Implementing these concepts in practice necessitates the use of the U3D game engine. Utilizing U3D requires adherence to its design principles and API specifications to avoid significant obstacles.

U3D provides developers with C# instead of C++ or C. Although C# is a robust, object-oriented programming language, implementing some of the underlying algorithms discussed requires the efficiency of C++ or even C. C++ is not only efficient but also clean and well-suited for writing encoding and decoding algorithms. To leverage this, I used C++ to develop the encoding algorithms suitable for the affiliated project of this thesis and compiled them into a dynamic library that C# can call. This approach combines the advantages of C#'s object-oriented capabilities in application architecture design with the performance benefits of C++ and C in the underlying system. This results in significant improvements in application performance, particularly in encoding efficiency. The following demonstrates how to call an H264Encoder written in C++ from C#. By utilizing the performance advantages of C pointers, I implemented the encoder as follows:

```cpp
1.    #define PINVOKE_ENTRY_POINT extern "C" __declspec(dllexport)
2.    PINVOKE_ENTRY_POINT H264Encoder* Create(uint32_t width, uint32_t height, uint32_t frameRateNumerator, uint32_t frameRateDenominator, uint32_t averageBitRate)
3.    {
4.    #if ENABLE_TRACE
5.      std::call_once(InitLogOnce, InitLog);
6.    #endif
7.      std::unique_ptr<H264Encoder> encoder(new H264Encoder());
8.      if (encoder->Initialize(width, height, frameRateNumerator, frameRateDenominator, averageBitRate))
9.        return encoder.release();
10.     return nullptr;
11.   }
```

[The encoding method used by the streaming side calling the dynamic library]

In U3D, using the dynamic library approach described above ensures that the functionality is not constrained by the U3D engine itself. This method allows the dynamic library to be directly incorporated into the U3D project. The process is facilitated through standard C# methods for referencing dynamic libraries, as demonstrated below:

```csharp
1.    using System;
2.    using System.Runtime.InteropServices;
3.    internal struct H264EncoderPlugin
4.    {
5.      [DllImport("H264Encoder", EntryPoint ="Create")]
6.      extern public static IntPtr CreateEncoder(uint width, uint height, uint frameRateNumerator, uint frameRateDenominator, uint averageBitRate);
7.    }
```

8. [Basic method of C# importing C++ dynamic library in U3D]

Since C# is derived from C and C++, interoperability between these languages is both common and feasible, enhancing the extensibility of U3D. As U3D engine users, developers can not only "visualize" their algorithmic concepts but also leverage the high-level APIs provided by U3D to achieve their desired outcomes. For specialized functional requirements, developers can even create custom plugins. At the beginning of the affiliated project of this thesis, I developed a standard shortcut plugin for U3D, demonstrating the flexibility of the engine.

While many modern game engines are popular, Unreal Engine is particularly renowned for its open-source nature, exposing all underlying code to users. This openness can be overwhelming for entry-level developers, who might struggle with the sheer volume of code and complex graphical programming interfaces, such as Blueprints. Although Blueprints aim to simplify programming, they can complicate the task for developers accustomed to coding directly, reducing code reusability and efficiency. Once the encoding module was integrated, the focus shifted to developing the video transmission module in C#. This module, developed directly in C#, is not constrained by the U3D engine at the lower level. However, at the application level, it must integrate with U3D's MonoBehaviour to participate in U3D's lifecycle.

By decoupling the underlying algorithm from U3D's official constraints, the streaming structure and RTSP protocol designed by me are seamlessly implemented at the application level. As the saying goes, "When one door closes, another opens." The development process may present challenges, but these should not deter you from your carefully planned approach. The notion of a universal, unchanging development solution does not exist; instead, developers must confront and overcome various challenges posed by their chosen tools. Ultimately, overcoming these challenges leads to successful outcomes and a brighter future.

2.2.2 MAKE THE MOST USE OF U3D

In the development of the live video streaming system client, the U3D game engine serves as a powerful tool. Although many of its interfaces and functions might initially seem irrelevant, they often provide potential solutions to complex problems. This necessitates a development approach tailored to U3D's unique paradigms, rather than relying solely on traditional development experience. For instance, U3D Shaders, which may appear extraneous to the core application, can play a crucial role in solving specific issues. In the proposed live streaming system, the server must convert video data from RGB to NV12 color space before streaming. U3D Shader can efficiently handle this transformation due to its comprehensive math library, which includes built-in functions for fundamental space conversion formulas. For example, in the affiliated project of this thesis, each pixel in every frame needs to be converted from linear space to Gamma space to ensure that the brightness of the final image is perceived correctly by the human eye, regardless of whether the data values are too high or too low. This seemingly redundant operation significantly enhances the quality of the image data received by the client, preventing issues with images appearing too dark or too bright across different devices. U3D Shader's built-in math library includes the function LinearToGammaSpace(color), which can be used directly for this purpose. While it is possible to manually implement the underlying mathematical formulas, understanding the principles behind the built-in function is important. The LinearToGammaSpace function applies Gamma correction using HLSL (High-Level Shader Language), where the constants used in the formula are empirical values. The mathematical formula for this conversion is as follows:

$$sRGB(color) = \begin{cases} 12.92 \cdot color, & color \leq 0.0031308 \\ 1.055 \cdot color^{\frac{1}{2.4}}, & color > 0.0031308 \end{cases}$$

Below is the implementation of the algorithm using shader language in U3D:

```
1.  inline float LinearToGammaSpaceExact (float value)
2.  {
3.      if (value <= 0.0F)
4.          return 0.0F;
5.      else if (value <= 0.0031308F)
6.          return 12.92F * value;
7.      else if (value < 1.0F)
8.          return 1.055F * pow(value, 0.4166667F) - 0.055F;
9.      else
10.         return pow(value, 0.45454545F);
11. }
```

Figure 2.2.2-1 below presents a test image of a scene outside the window, captured during the development phase of the project. The image was processed using the core algorithm described above to transform it from linear space to sRGB space. The image before transformation appears significantly brighter overall; on a sunny day, this can lead to overexposure of the sky and diminished viewing quality for the client user. Such issues are especially pronounced if gamma correction is not applied on the client side. After applying the correction, the image more accurately reflects real-world conditions, enhancing the overall viewing experience.

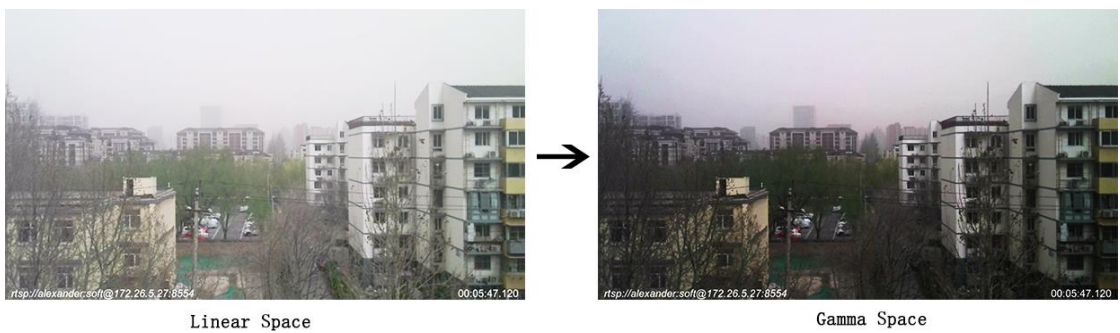

Figure 2.2.2-1 Comparison of the original image and the image after Gamma correction in linear space

By transforming the color space, the data sent to the client is pre-processed to ensure that no additional processing is required on the client side. The server is responsible for maintaining the quality of the image data captured during video streaming. This means that the client only needs to play the received video without further adjustments for issues such as excessive darkness or brightness. Considering the diverse range of client devices that may receive the data, ensuring a consistent visual effect across all clients is essential. To achieve this, standardizing the image data before streaming on the server side is crucial to ensure that any discrepancies in image quality are not attributable to server-side processing. U3D provides numerous APIs that offer effective solutions for developers. In the affiliated project of this thesis, various U3D Shader functions were utilized because they directly influence visual effects. Thus, any visual improvements required during development were addressed using Shader technology. Shader can achieve significant enhancements in image processing when combined with appropriate algorithms.

Despite the challenges encountered, a comprehensive understanding of the U3D API and its core concepts enabled the successful implementation of features that might be challenging or even unattainable through traditional application development methods. Each "scene" within U3D can be viewed as an expansive canvas, allowing developers to explore and implement solutions that leverage U3D's capabilities.

Overall, the integration of protocols such as RTSP into the project showcases the full potential of U3D's tools and demonstrates their ability to address complex development requirements.

2.2.3 PERFORMANCE EVALUATION FOR STREAMING MODULE

With the basic live video streaming module implemented in U3D, the next phase involves testing latency at each end, analyzing the underlying causes of latency, exploring feasible performance optimization solutions, and incrementally improving the system's performance. In the project, I used U3D to display the current streaming time on the canvas to evaluate the specific latency of live streaming. Under optimal conditions, the latency between two different local ports on a dual-core CPU host should be less than or equal to 0.5 milliseconds. Exceeding this threshold indicates significant performance issues within the internal algorithm that necessitate improvement.

Initially, I bypassed encoding and directly employed TCP for push streaming in a peer-to-peer (P2P) setup, with the client pulling the stream. This approach resulted in delays often exceeding 1 to 2 seconds. Moreover, when using a queue-based packet grouping method for push streaming, cumulative delays were observed. As the duration of the live streaming increased, so did the accumulation of delay. On mobile devices, this issue extended beyond simple overheating, indicating a need for a comprehensive overhaul of the underlying algorithm.

Following the standardization of the live video streaming process using the RTSP protocol and H.264 encoding, testing was conducted on a laptop with a dual-core CPU. The streaming end and the receiving end were both active, with the streaming end utilizing the Easy Movie Texture plug-in for playback. Although no delay was visible in Figure 2.1.2-1, precise testing revealed that the measured video delay ranged between 200 milliseconds and 500 milliseconds, with the average maximum local delay not exceeding 500 milliseconds, as shown in Figure 2.2.3-2.

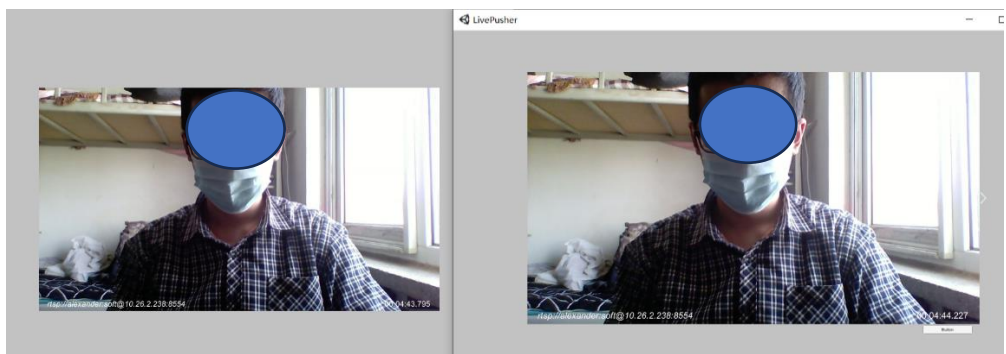

Figure 2.2.3-1 Screenshot for showcasing the video being played by the streaming server (right) on the streaming client (left).

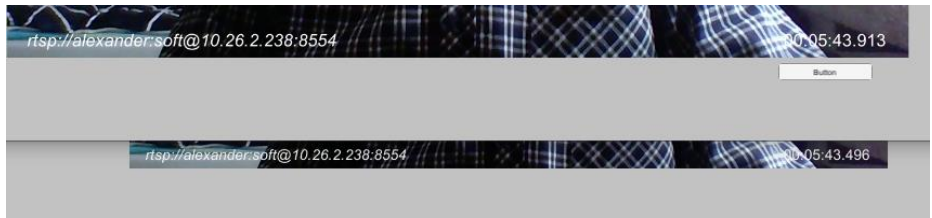

Figure 2.2.3-2 The latency of streaming while the client side (below) plays the streaming video.

To determine whether the nearly 500ms delay is attributable to server encoding and streaming efficiency or client streaming efficiency, I employed PotPlayer, a widely used video player capable of streaming and playing RTSP video streams. Given that the system was originally developed to comply with the RTSP specification, it ensures broad compatibility with streaming media data across various players. Consequently, any player supporting RTSP streaming should be able to handle the data pushed by the server.

After closing all memory-intensive applications on the machine, testing with PotPlayer under optimal conditions reveals that the delay for streaming RTSP live links ranges between 300ms and 600ms, with the maximum delay not exceeding 600ms. This indicates that the fundamental issue with live streaming delay lies not with the streaming end but with the server.

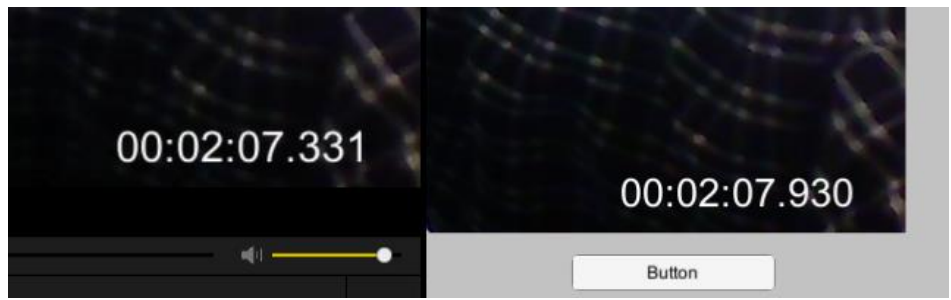

Figure 2.2.3-3 The details of the accumulated delay of the streaming server (left) and the client (right).

To validate this inference further, I employed U3D's built-in Profiler performance testing tool to monitor the project's overall performance. As illustrated in Figure 2.2.3-4, the real-time display of video data captured from external devices on the UI does not adversely affect the performance of the RTSP streaming process. Thus, the impact of local playback on RTSP streaming can be considered negligible.

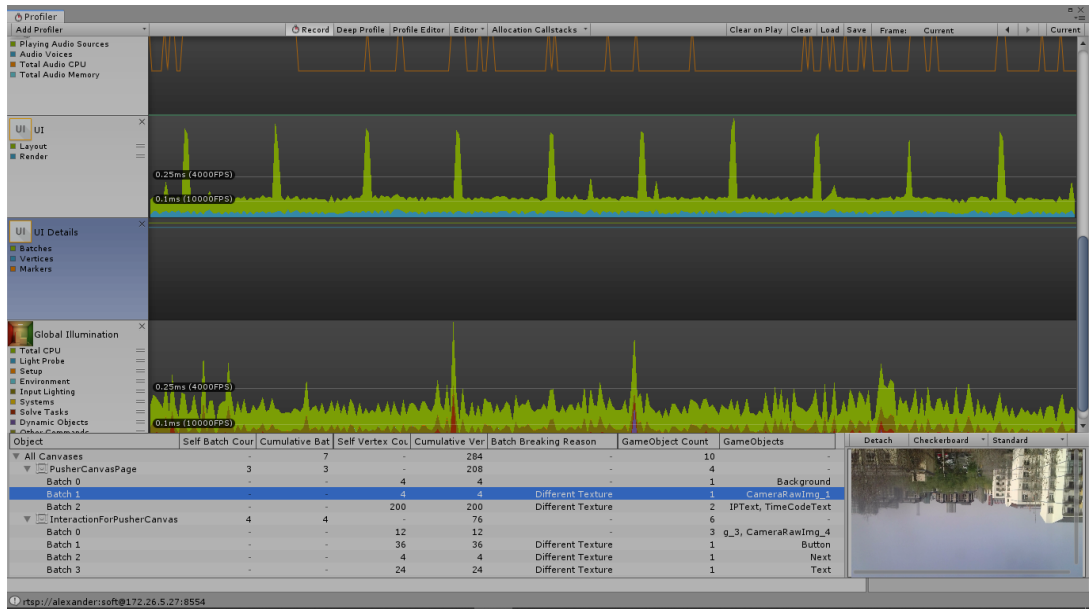

Figure 2.2.3-4 The monitored UI performance shown via U3D Profiler.

Upon further investigation of GPU usage, it was observed that the GPU usage rate was relatively high during local video capture and application of the video to the RawImage component in U3D. This indicates that the performance overhead associated with this part may have an impact on the streaming process. However, since the GPU is responsible for this task, it remains uncertain whether this performance overhead significantly affects the push stream. The performance monitoring results from the GPU are shown in Figure 2.2.3.5:

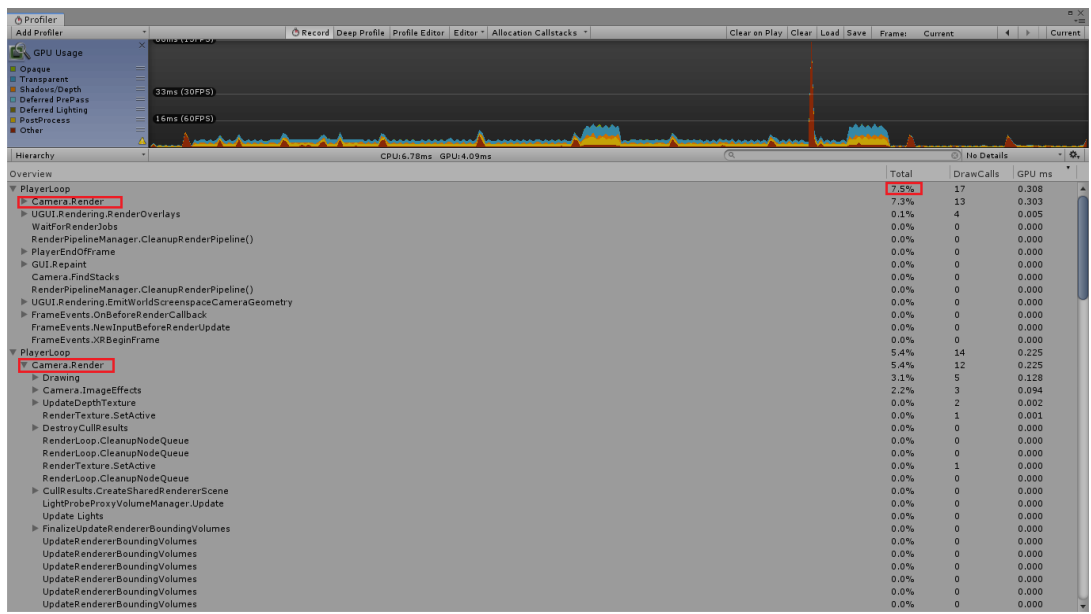

Figure 2.2.3-5 The performance of GPU and its usage shown via U3D Profiler.

With lingering questions, I proceeded to analyze CPU usage. According to Figure 2.2.3-6, while the local video capture and UI presentation process does contribute to CPU usage (with Camera.Renderer accounting for 0.09ms, as depicted in the figure), it is not the primary cause of the delay. The figure shows that the ProcessCommands segment occupies the largest proportion of CPU time. Further analysis revealed that this segment is responsible for video encoding operations, indicating that the efficiency of encoding each frame on the server — both before and during streaming—directly impacts the final delay.

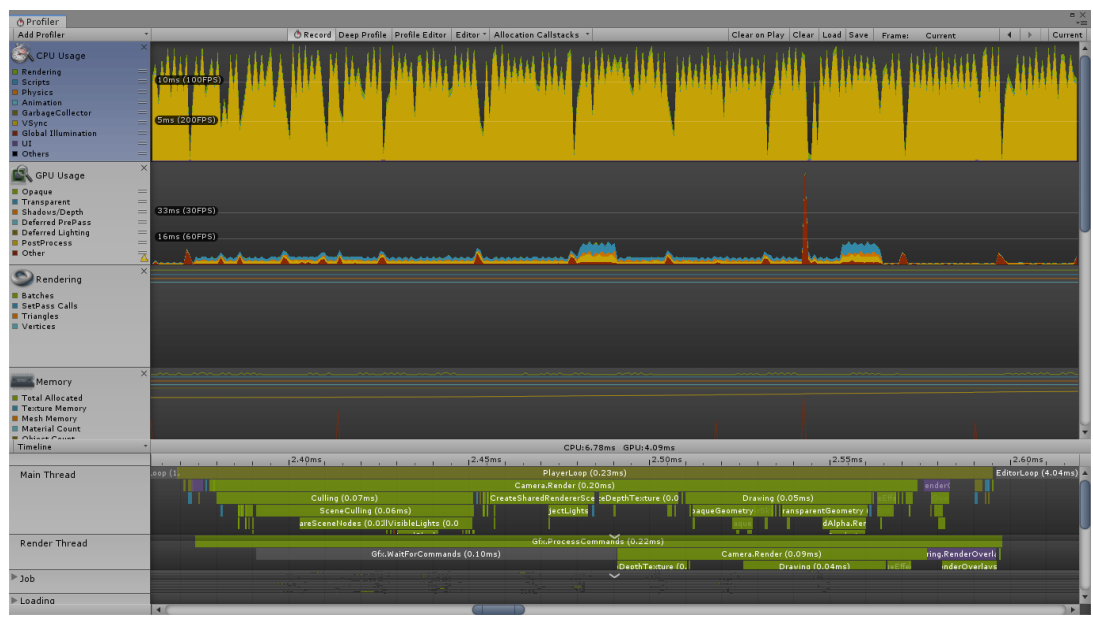

Figure 2.2.3-6 CPU performance and CPU consumption of streaming module.

The testing conducted at this stage serves as a crucial foundation for enhancing the performance of the video live module and reducing video latency. It also sets higher expectations for the subsequent development of the entire project. As other modules are developed, it is essential to avoid introducing additional performance overhead and to minimize or even eliminate performance bottlenecks. The design plan for the client application integration stage, detailed in Chapter 4, addresses these concerns.

To optimize performance, I propose a dual approach: reducing RTP packet size and improving encoding efficiency. This approach alone is not recommended as it only addresses the size of individual data transmissions without fundamentally solving the issue.

The more effective solution involves improving encoding performance by using YUV color space as input instead of RGB. YUV color space can significantly reduce redundant information, thereby enhancing encoding efficiency. The conversion from RGB to YUV is straightforward and can be performed using the following formula:

**RGB to YUV Conversion**

$$Y = (0.257 * R) + (0.504 * G) + (0.098 * B) + 16$$

$$Cr = V = (0.439 * R) - (0.368 * G) - (0.071 * B) + 128$$

$$Cb = U = -(0.148 * R) - (0.291 * G) + (0.439 * B) + 128$$

Figure 2.2.3-7 Formula of RGB to YUV conversion[15].

Below, Figure 2.2.3-8 presents the inside structure of the video encoder.

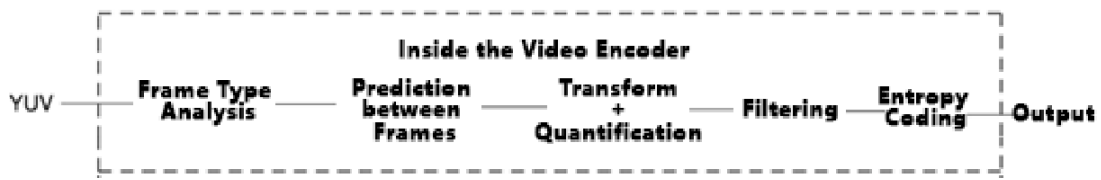

Figure 2.2.3-8 The inside structure of the video encoder.

---

[15] The RGB to YUV conversion formula used in the project is as follows:

Y = (0.257 * R) + (0.504 * G) + (0.098 * B) + 16

Cr = V = (0.439 * R) - (0.368 * G) - (0.071 * B) + 128

Cb = U = -(0.148 * R) - (0.291 * G) + (0.439 * B) + 128

Accordingly, converting from YUV to RGB requires the following conversion formula:

B = 1.164(Y - 16) + 2.018(U - 128)

G = 1.164(Y - 16) - 0.813(V - 128) - 0.391(U - 128)

R = 1.164(Y - /16) + 1.596(V - 128)

## 2.3 SUMMARY OF THE STREAMING MODULE

The live video streaming system based on U3D represents a comprehensive full-stack solution centered around live video streaming. In this chapter, the research on the RTSP protocol and video encoding and decoding serves as a critical foundation for the project's functionality and operational stability. Utilizing the surveyed materials, development experience, and test results, I have established the basic framework for a one-to-many live video streaming system.

However, as the project is still in its underlying development phase, the application layer remains rudimentary, and the client application has yet to evolve into a prototype of the final application. The next chapter will introduce an innovative functionality and its integration with live streaming, building on the research and development findings discussed here. This innovation will be crucial to the final application and its implementation.

We will now proceed to the next chapter, where we will explore multi-channel video capture technology and examine the specific solutions for integrating the live streaming module with the multi-channel video capture module.

# 3. MULTI-CHANNEL VIDEO STREAMING

## 3.1 GETTING STARTED WITH MULTI-CHANNEL CAPTURING

In this chapter, we will delve into the definition of multi-channel video, the application of multi-channel video capture technology within the industry, and the specific methods for capturing multi-channel video data in U3D. Additionally, related research will be discussed to preliminarily clarify the significance and value of multi-channel video capture for the affiliated project of this thesis.

### 3.1.1 DEFINITION OF MULTI-CHANNEL VIDEO CAPTURING

In the industry, multi-channel video refers to a series of streaming videos from multiple video stream sources. For instance, channels such as CCTV1 to CCTV10 represent ten separate live streaming data channels. In this context, the term "multi-channel" is inspired by the concept of multi-channel video and generally refers to the video data lines from multiple external devices that are captured locally by the host before being streamed. For example, if five external cameras are connected to the host, there are five captureable video data lines. These video data lines are independent of each other at the hardware transmission level, with unique properties such as resolution, frame rate, and color space potentially varying between devices.

In the affiliated project of this thesis, the captured video data from these multiple video device lines are integrated using the U3D API to achieve consistency, resulting in a unified live stream that can be streamed to one or more clients. While the server captures video data from multiple external devices in real time, it continues to stream over a fixed-width network line. The format of the live streaming/broadcasting, based on the RTSP protocol, remains a single-channel stream with unchanged encoding and decoding target data size, as well as the streaming packet size. The only difference lies in the enriched content of the streaming video. The proposed system allows for the seamless addition of external video devices to enhance live video content without altering the data volume. The server can push the content from all external video devices to the client without changing the overall data size. Consequently, the number of video lines captured by the server during streaming is not limited by the software, meaning that the live video content can be richer. Furthermore, the server user has the flexibility to selectively push any of the multi-channel videos at any time, which enhances both the flexibility and stability of the live stream.

Since the data volume pushed by the server remains constant, the client's stream pulling is decoupled from the server's stream pushing. This decoupling ensures that an increase in video content does not result in increased data reception at the client side, thereby avoiding issues such as live streaming delay, video quality degradation, and dependency on server-side streaming forms. Multi-channel video capture, therefore, refers to the process of integrating data from multiple external camera devices at the software level, bringing video devices with different attributes into a consistent state.

### 3.1.2 APPLICATION OF MULTI-CHANNEL VIDEO CAPTURING

Multi-Channel video capture is widely utilized across various industries. One of the most common applications is in postgraduate exam interviews, especially during the COVID-19 pandemic. In this scenario, candidates take exams in front of a computer connected to two video capture devices. One camera monitors the candidate's behavior from behind, while the other, positioned in front of the candidate, is used for video communication with the interviewer. Similarly, in the medical field, remote surgeries often rely on multi-channel video capture technology, where a computer connected to multiple cameras facilitates video streaming of the procedure.

As illustrated in Figure 3.1.2-1, OBS Studio[16] — a popular software in the industry— demonstrates multi-channel video capture by accepting video input from multiple driver-free cameras. Even when multiple camera inputs are active, OBS can capture the window of a single application running on the desktop or host and merge it into a single stream for live streaming. The live streaming is conducted using the RTMP protocol. OBS Studio is highly regarded for its open-source nature and user-friendly interface, supporting H264 (X264) and AAC encoding, as well as unlimited scenes and video sources. Additionally, it supports real-time RTMP streaming to platforms such as Twitch, YouTube, DailyMotion, and Hitbox.

---

[16] OBS Studio: The full name is Open Broadcaster Software, which is a free and open source video recording and live streaming software. It has multiple functions and is widely used in video acquisition, live broadcast and other fields. It is sponsored by YouTube and Facebook.

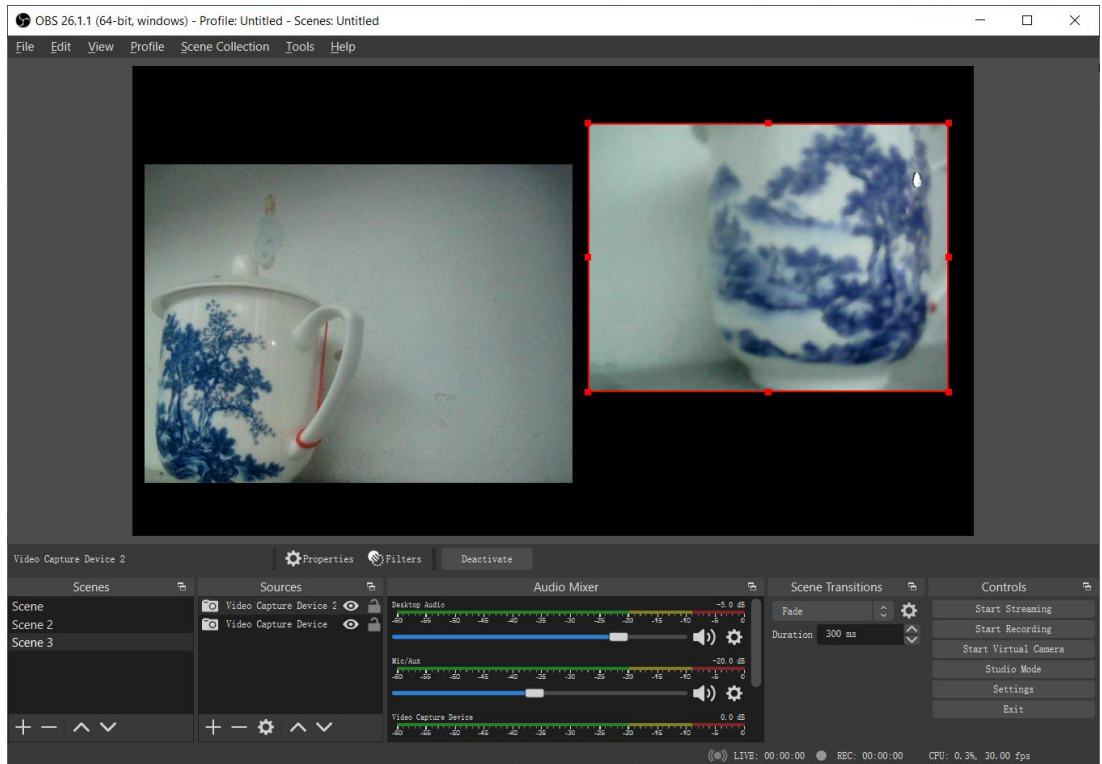
Figure 3.1.2-1 OBS capturing video data from 2 external cameras.

Under current hardware conditions, most operating systems support the integration of multiple external devices. In a star-shaped physical configuration, a single USB[17] 2.0 port can theoretically support up to 127 external devices. Given that most hosts have at least two USB ports, the 127-device limit per USB 2.0 port is more than sufficient for basic live video streaming technology. For special usage scenarios requiring a higher number of peripherals, the hardware setup can be optimized by upgrading the USB Hub's power supply and enhancing the host's hardware performance, allowing for stable connections with potentially thousands of USB driver-free cameras. However, the affiliated project of this thesis targets general civilian use cases where it's common to connect fewer than 10 external cameras via a USB Hub.

In the affiliated project of this thesis, scenarios involving more than 10 external cameras were not considered during the experimental, development, and testing phases. Therefore, this article remains neutral on the system's versatility in such extreme cases, although theoretically feasible solutions will be discussed. Assuming that hardware manufacturers can fully address the challenge of unlimited USB external camera expansion, software integration becomes straightforward. The operating system, equipped with the necessary peripheral drivers, can easily incorporate one or more external devices into the host.

---

[17] USB: Abbreviation for Universal Serial Bus, an external bus standard.

At the software level, key considerations include reading specific information from external devices, capturing sensory data, and exercising control over these devices. Fortunately, most APIs provide interfaces for accessing data captured by external devices. Even for image data perceived from the external environment, specific APIs can retrieve various attributes for each frame. This information is crucial for processes such as image processing, compression, and video encoding and decoding, directly influencing the final quality of the live video stream.

3.1.3 REASONS FOR CHOOSING MULTI-CHANNEL STREAMING

From a technical perspective, multi-channel video capture involves connecting multiple external devices to a host to simultaneously obtain video data from multiple sources. In single-application scenarios that do not involve live streaming, its value is often straightforward. For example, in a classroom, a lecturer may use a projector to display test papers or teaching materials onto an electronic screen; in vehicle monitoring, a driver uses onboard cameras to view traffic conditions from multiple directions in real-time; or in 3D real-time scene reconstruction, multiple cameras are used to recreate a scene in three dimensions. More accurate reconstruction typically requires depth maps, often obtained through depth cameras. However, during my internship in the Sony XR[18] technical research room, I learned that the industry is exploring the use of machine learning to achieve 3D reconstruction using ordinary cameras.

Other examples include cable TV monitoring, where security personnel monitor various areas of a facility through wired connections, and comprehensive observation of a patient's internal organs in an operating room. While the industry has been working towards achieving similar results through wireless means — effectively making them live video scenarios—real-time transmission of data from multiple cameras requires a stable and efficient network environment, minimal latency (comparable to wired transmission), and low cost. These requirements are challenging to meet. The research in this article on multi-channel video capture technology is not merely a reiteration of existing wired multi-peripheral video capture methods but an effort to extend and expand these capabilities at the network level, aiming to meet the aforementioned demands.

---

[18] XR: Short for X Reality or Extended Reality. The "X" here means the inclusion of all related technologies, commonly known as extended reality. In a broad sense, it includes augmented reality (AR), virtual reality (VR), mixed reality (MR) and other related technology fields that have not yet been foreseen but may exist in the future. XR can be defined as a container to reduce public confusion.

When multi-channel video capture technology is applied at the network level and extended further, it opens up various new application values and can even contribute to advancing areas such as 3D scene reconstruction or AR applications. Wireless multi-channel video capture technology can transmit video data from multiple cameras to the client in real-time, allowing for the integration of this data on the client side to "reconstruct" the real world to an ideal state. As this technology becomes more stable and mature, it will greatly benefit fields such as medicine, where accuracy is critical. For instance, this article focuses on live video technology while integrating multi-channel live video capture as its main innovative feature. The research value lies in providing diverse solutions for future live video formats, meeting the demand for richer live video content, and offering a solid foundation for expanding into other fields such as medicine, VR[19], and AR[20]. Specifically, various AR and VR applications developed using U3D can be built upon this framework without requiring significant restructuring. This set of applications can be directly used for further development to achieve desired effects. Whether for three-dimensional reconstruction or live video, the amount of data captured by multiple cameras is directly proportional to the richness of the final content. Therefore, the significance of multi-channel video capture technology for live streaming and video-related software applications is particularly crucial.

---

[19] VR: Short for Virtual Reality, virtual reality (technology) is a computer simulation system that can create and experience a virtual world. It uses computers to generate a simulated environment to immerse users in the environment.

[20] AR: Short for Augmented Reality, augmented reality (technology) is a technology that cleverly integrates virtual information with the real world. It widely uses a variety of technical means such as multimedia, three-dimensional modeling, real-time tracking and registration, intelligent interaction, and sensing. It simulates computer-generated virtual information such as text, images, three-dimensional models, music, and videos, and applies them to the real world. 2 types of information complement each other, thereby achieving "enhancement" of the real world.

## 3.2 DEEP INTO MULTI-CHANNEL VIDEO CAPTURING

In this section, I will introduce the specific principles of multi-channel video capture technology, including the technical standards employed during the multi-channel video capture process and its various applications. We will discuss in detail the underlying technical methods required for effective multi-channel video capture, along with the interfaces provided by different devices and the hardware-level development options available. Additionally, a brief analysis will be conducted on the multi-channel video capture process at the intersection of underlying software and hardware interfaces.

### 3.2.1 UVC-based Video Capturing Strategy

The industry generally uses the UVC (USB Video Class) standard to further process data captured by multiple external cameras. This standard is not only applicable to Windows clients but is also widely used in Linux and Android systems, making it a relatively mature standard for video line access.

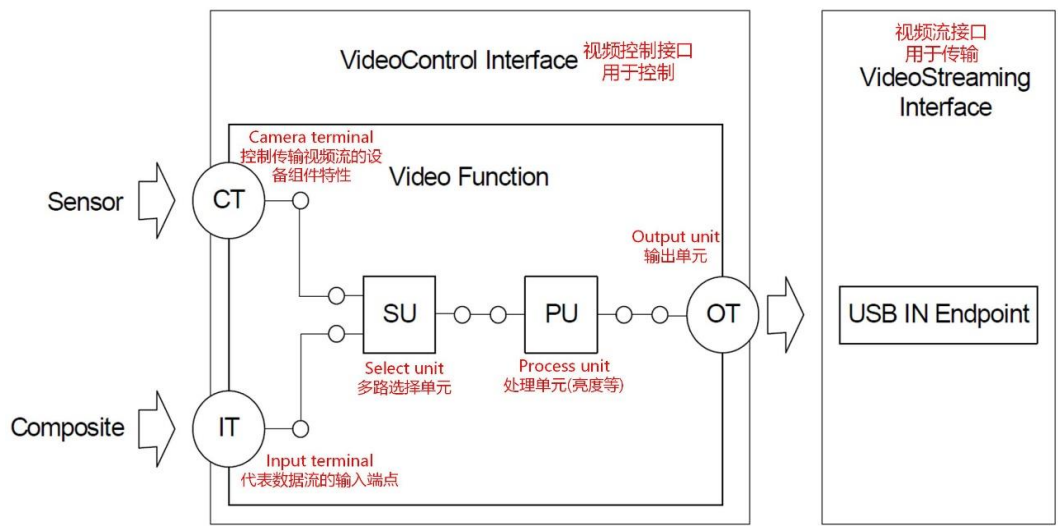

Figure 3.2.1-1 VS interface and VC interface inside the structure of USB.
[Figure Inserted from Original Thesis Source, 2021.7]

In essence, UVC devices are multi-interface devices, differing from standard USB flash drive connections. As is shown in Figure 3.2.1-1, UVC devices typically have at least two interfaces: the Video Control Interface (VC) and the Video Stream Interface (VS). The UVC protocol mandates that any device with functional UVC capabilities must include a VC Interface and one or more VS Interfaces. The VC interface is primarily used for configuration and control, setting the UVC device into different operational states, while the VS interface handles the transmission of video data streams. The full functionality of UVC devices relies on the coordination between the VC and VS interfaces. The protocol also requires that UVC devices use an Interface Association Descriptor (IAD) to describe the interface set that includes both the VC and VS interfaces. Figure 3.2.1-2 shows the overall structure of a USB device, showing the specific roles of different interfaces and their positions within the system. Under the UVC standard, video capture begins at the hardware level, with video data from multiple external camera devices being connected to the software level in its native form. Fortunately, the Windows operating system and most development tools encapsulate the underlying capture processes of hardware devices, allowing developers to work directly with high-level APIs. However, it is essential to have a general understanding of the overall structure described in this section to capture video data from multiple external devices more effectively during development. This structure also allows developers to capture information from various external devices, including but not limited to somatosensory cameras, scanner cameras, and thermal imaging cameras.

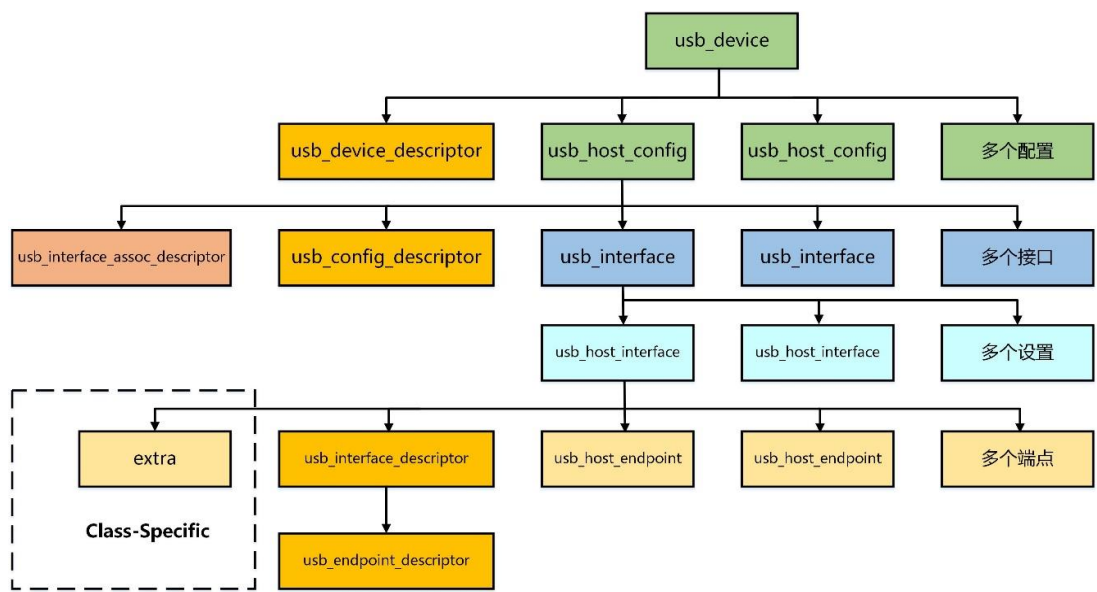

Figure 3.2.1-2 Overall structure of the capturing device and interfaces of USB.

During development with U3D, I utilized pre-packaged APIs designed for UVC interfaces to manage multiple external devices. However, whether using U3D or other APIs, the underlying implementation still involves providing integrated solutions for the VC and VS interfaces tailored to specific operating systems. [Figure Inserted from Original Thesis Source, 2021.7]

3.2.2 VIDEO CAPTURING STRATEGY VIA U3D

The video capture strategy implemented using the U3D API relies on UVC standard support at the lower levels. Essentially, it involves detecting multiple external cameras based on the UVC protocol within a specific system environment and obtaining real-time video data. Each set of video data captured by the U3D API is represented as a WebcamTexture, which inherits from Texture, offering more than just basic texture functionality. Typically, when attaching camera-captured video data to the U3D canvas (Canvas), an image (Image), or a raw image (RawImage), WebcamTexture can be used as a standard Texture. It is highly flexible and provides pixel-level control, allowing operations on individual pixels. For instance, the YUV color format conversion used in the proposed system is handled through the interfaces provided by WebcamTexture.

3.3 SUMMARY OF MULTI-CHANNEL VIDEO STREAMING MODULE

"Multi-Channel video capture" is a term defined by me, referring to the process of acquiring data from multiple external devices using drivers or internal standards provided by operating system manufacturers. In the affiliated project of this thesis, multi-channel video capture technology serves as a fundamental tool for enhancing the richness and novelty of live video content. By leveraging this technology, the potential of live video is significantly expanded, allowing for more diverse applications beyond traditional live streaming. Despite its maturity on major operating systems like Windows, Linux, and macOS, multi-channel video capture technology faces challenges on Android devices. Current Android interfaces support single-line video but lack support for simultaneous acquisition from multiple peripherals. This limitation is partly due to hardware constraints, but also involves software challenges such as managing multiple device drivers, addressing power supply issues, and processing multiple video streams concurrently. As Android hardware and software interfaces advance, it is anticipated that these issues will be resolved, making multi-channel video capture technology a viable solution across various platforms. Until then, its practical implementation on mobile devices remains limited, pending advancements in hardware and software capabilities, similar to how Apple devices have already supported simultaneous dual-camera functionality.

For the live video streaming system based on U3D, the development of the multi-channel video capture module relies heavily on U3D's underlying API and the concept of "scene." The "scene" functionality enables the simultaneous collection and utilization of video data from multiple external devices. Without this capability, capturing video from multiple external cameras would be challenging. During development, I tested multi-camera support in mainstream browsers and found the results unsatisfactory. As shown in Figure 3.3.1, supporting multiple external camera devices on a single browser page requires specialized browser development, a task demanding extensive technical expertise. While the affiliated project of this thesis focuses on public needs, the challenge of adapting general browsers for multi-camera support is recognized. Nevertheless, with increasing demand for multi-channel live video streaming and broadcasting, exploring such innovative approaches will become increasingly valuable. Notably, some international institutions are already working on custom browsers to support simultaneous camera data acquisition on the same page.

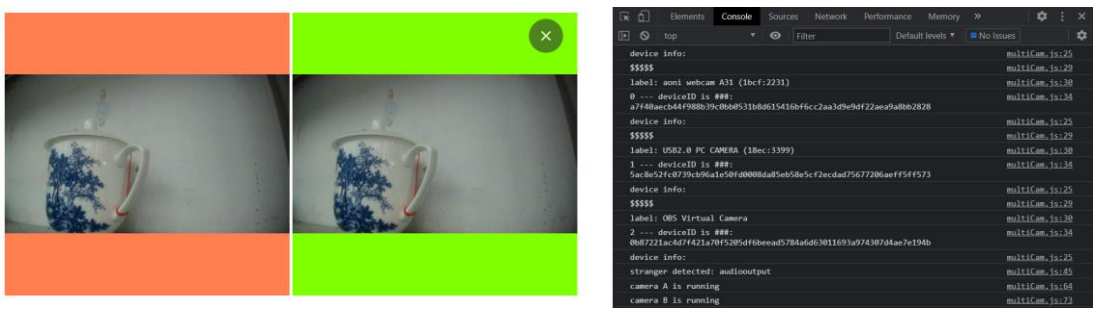

Figure 3.3.1 In the experiment, I found that Google Chrome browser cannot use multiple cameras as inputs simultaneously on a single webpage.

From Figure 3.3.1, it is evident that although both camera devices are detected and operational, Google Chrome does not support accessing multiple camera video streams simultaneously on a single page. Similarly, in testing with Firefox, while it does not support simultaneous recording from multiple cameras, it prompts users to select which camera to use upon opening the page.

Figure 3.3.2 provides details from Firefox, showing the detected camera devices. It lists the IDs of the two external cameras, demonstrating how Firefox identifies and differentiates between multiple cameras, even though it still restricts simultaneous access to them.

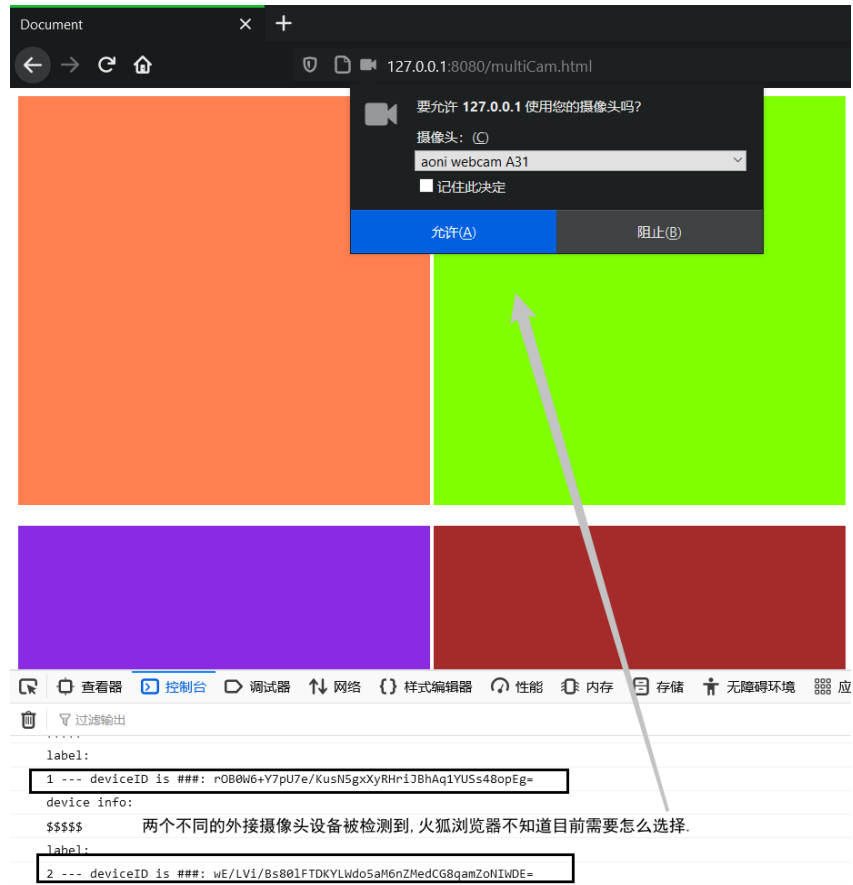

Figure 3.3.2 In the experiment, I found that using Firefox browser, it will detect multiple cameras and pops up all the choices to let the user choose only a single camera device as the main input. [Figure Inserted from Original Thesis Source, 2021.7]

Although multiple cameras are detected, the browser still prompts that only one camera can be selected for data access. In Figure 3.3.3, it is shown that the official browser currently recognizes two external cameras along with one OBS Studio virtual camera, which is used to capture the computer desktop. Despite detecting these devices, the browser restricts data access to only one camera at a time.

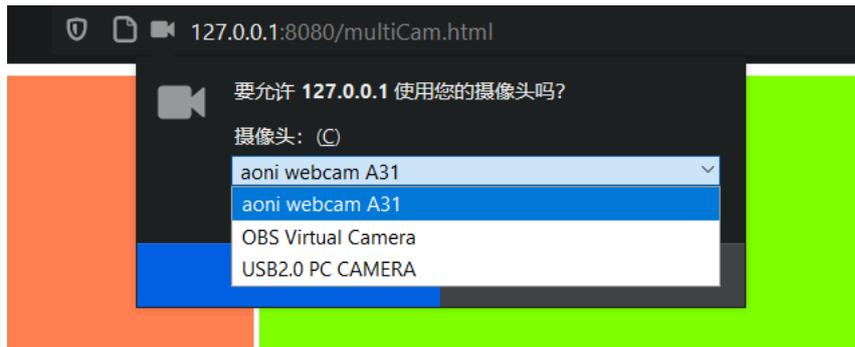

Figure 3.3.3 Firefox browser detects all camera devices connected to the computer and prompts the user to force the user to select only a single camera.

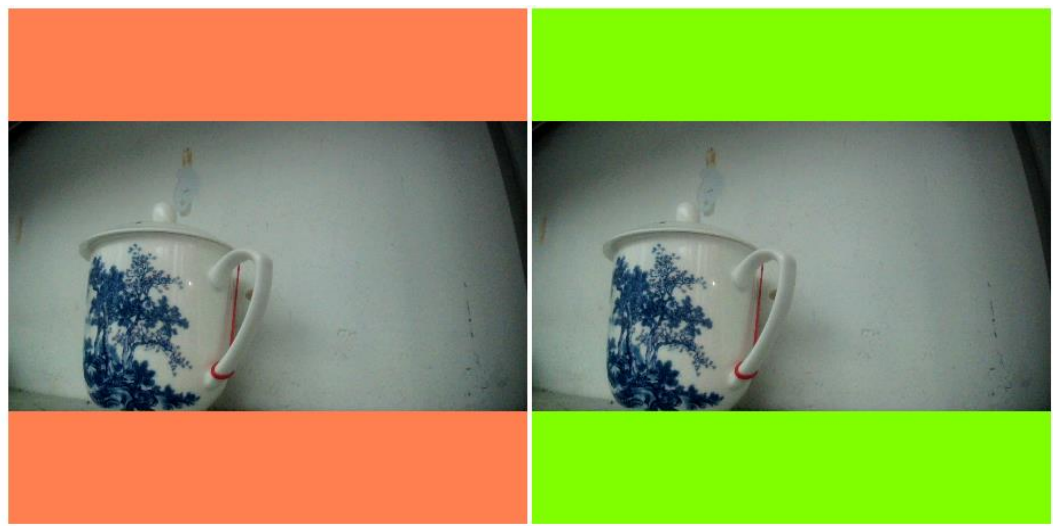

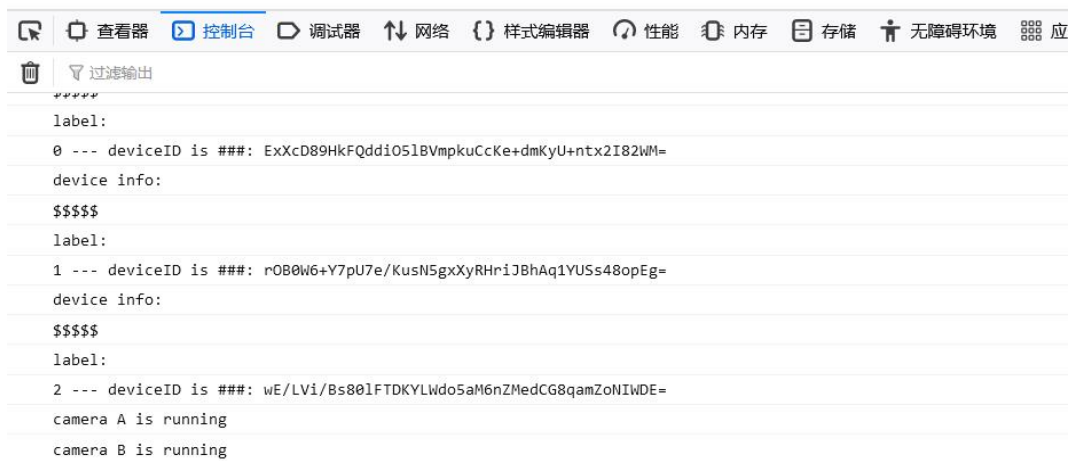

Figure 3.3.4 The experiment on the Firefox browser indicated that it only supports 1 single video capturing device as its input at run time

When developing the webpage, I attempted to use two separate boxes to display video data from different cameras. However, in practice, the browser only accepts video data from a single capture device at a time, resulting in two identical images being shown, as illustrated in Figure 3.3.4. To view the video from the other camera, the device must be selected again on a newly opened page. Figure 3.3.5 shows the image captured by the second external camera.

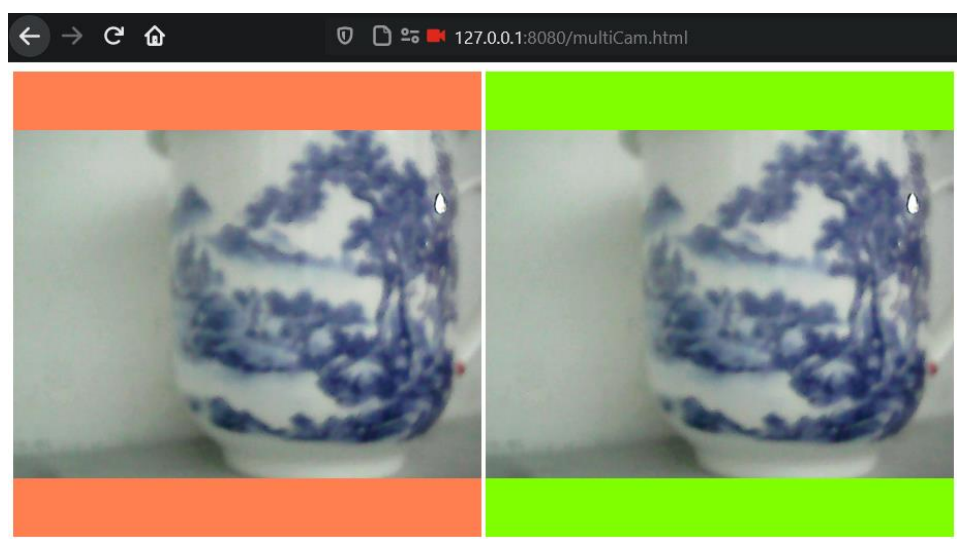

Figure 3.3.5 Experiments shows that the captured image from another camera can only be displayed in another new webpage in Firefox browser.

The maturity and stability of the underlying multi-channel video capture technology ensure a reliable content foundation for the multi-channel live video streaming and broadcasting system. The RTSP protocol provides essential protocol support, while the H264 encoding algorithm establishes a standard for video data packets. With the foundational research largely complete, the focus will now shift to implementing and designing the specific content derived from these research findings. The forthcoming chapters will transition from theoretical research to practical application, detailing how these underlying technologies are integrated into condition-oriented functions. The various functional modules explored in the research will be utilized independently or further optimized in real-world applications. As German philosopher Ludwig Feuerbach noted, "Theory comes from practice and must return to practice." The subsequent sections will illustrate the practical application of the research results within a live video streaming system, culminating in the development of a U3D-based client for both one-to-many and one-to-one multi-channel live video streaming.

# 4. DESIGN & IMPLEMENTATION OF CLIENT APPLICATION

## 4.1 OVERALL DESIGN OF THE CLIENT APPLICATION

In this chapter, I will outline the foundational aspects of the client component within the U3D-based live video streaming system. It will detail the objectives, necessary infrastructure, and technical elements required for the client application's implementation. Additionally, it will provide an initial exploration of the general design concepts before the client development begins. The design approach considers the entire application as a cohesive system, with each component contributing to the overall functionality. An overview of the system's architecture is depicted in Figure 4.1.1.

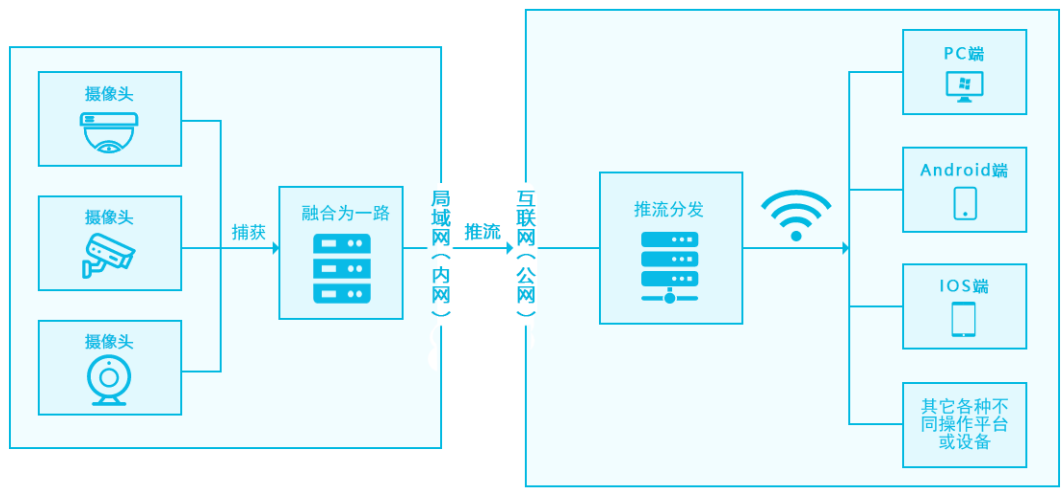

Figure 4.1.1 Overall software architecture based on RTSP.
[Figure Inserted from Original Thesis Source, 2021.7]

Based on this structure, we further developed the client application, defining it as a comprehensive system capable of both pushing and pulling streams. The core design principles are as follows:

Firstly, utilizing the interface provided by the U3D engine for accessing external camera device information, we retrieved specific video frame data from multiple external cameras. This data was then displayed on separate Raw Images within the U3D environment. Each Raw Image operates independently, ensuring that the video frame data in one Raw Image is not affected by external factors or interference from other video frames, thereby preserving the integrity of the original data.

Secondly, we employed a canvas to integrate the video frame data from multiple Raw Images. This canvas was strategically arranged within the interface and placed in world space. This setup allows the U3D engine's camera to capture every element on the canvas and map its data onto the Render Target using a suitable color format. The resulting data represents the actual video content, which is subsequently processed through encoding, compression, packaging, and transmission to be delivered to the client.

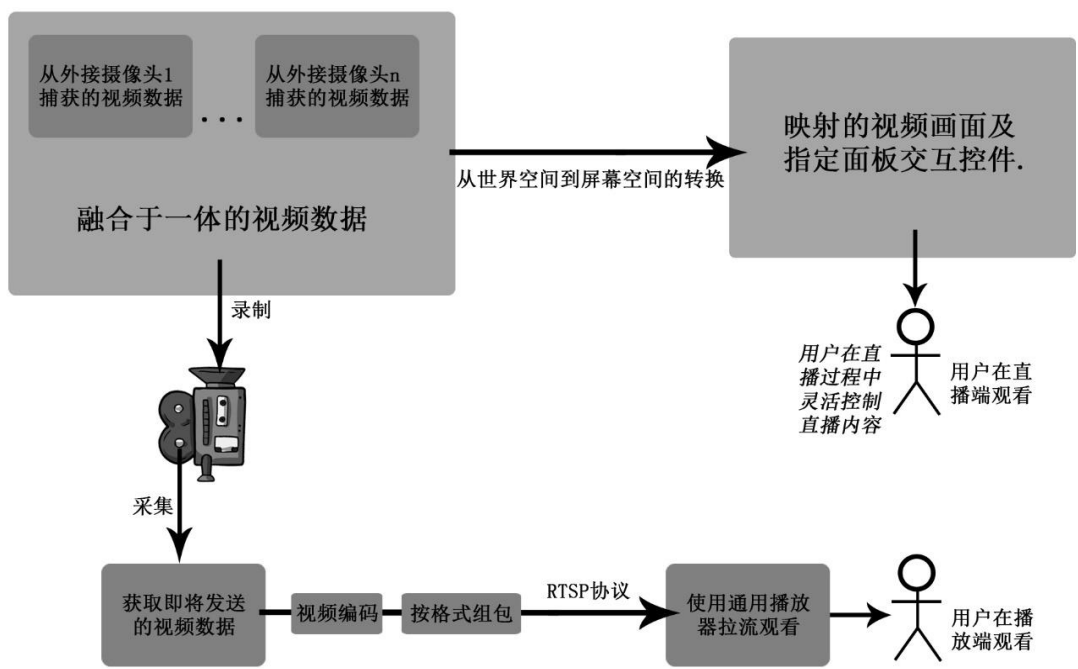

Figure 4.1.2 Core framework of the client-side application of multi-channel live video streaming and broadcasting [Figure Inserted from Approved Chinese Patent, 2023.5].

We then employ a canvas within the screen space to project the video frame data onto the Render Target, thereby displaying the captured and transmitted video content on the local screen for the live streaming end to view in real time.

Finally, we implement control mapping to associate the multiple Raw Images, identified in the initial stage, with various interactive controls in screen space using an orthogonal projection mechanism. This setup allows users at the live streaming end to zoom in on the video content from specific external devices by clicking on the video feed. Additionally, it provides users with effective control over an unlimited number of video capture devices.

4.1.1 OBJECTIVES

As live video streaming application software, the client application is designed to provide a set of fundamental low-latency multi-channel video streaming tools through its video push and pull modules. Both pushing and pulling functionalities are integrated within this application, with each function being proprietary to the system's functional modules. As discussed previously, various technical solutions and feasible methods have been explored to enhance the universality of each module. This research involved studying and experimenting with live video streaming technologies based on the RTSP protocol and achieving compatibility with industry-standard H264 encoding for video streaming. Consequently, the proprietary features of each functional module have been generalized to the maximum extent during implementation. The client's streaming player is capable of pulling and playing back any mainstream live video streaming data. Simultaneously, the video data pushed by the client's streaming module can be reliably received by any industry-standard streaming player.

During the testing phase of the client player, pull analysis was conducted on live data protocols including RTSP, RTP, RTMP, and HTTP-FLV. This testing confirmed the basic pull and playback functionality of live video streams, demonstrating the general-purpose capability of the client. For testing the playback of push stream data, two well-known industry players — PotPlayer and VLC Video Player — were utilized to verify correct playback results, ensuring the standardization of push stream data. Both PotPlayer and VLC Video Player successfully and stably played the video data pushed by the client.

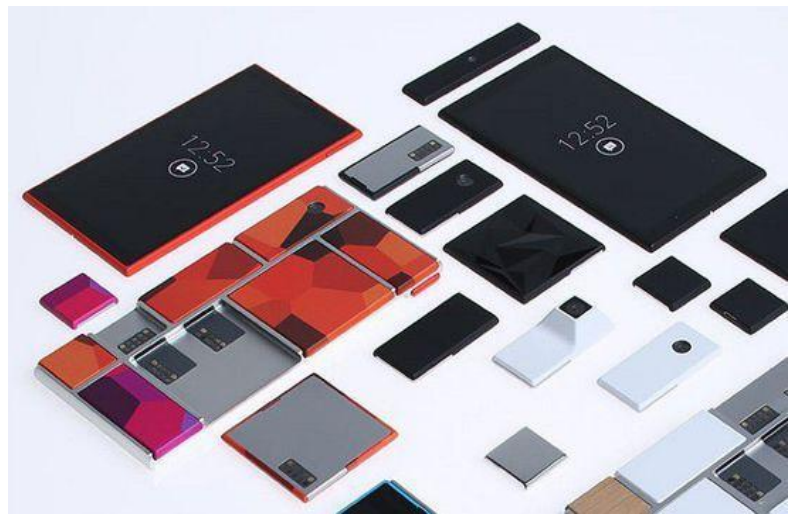

Figure 4.1.1-1 The concept of modular phone proposed by Google.

In addition to the generalization of the push and pull modules discussed earlier, the client application also aims to achieve the independence of other modules, adhering to fundamental principles of software engineering development and design. This involves decoupling various components to maximize reusability. The independence of each module, as outlined in the chapter on multi-channel video capture technology, allows the multi-channel video capture module, the pull module, the push module, and other components of the client application to be disassembled and separated from each other without impacting the core functionality of the application. The development approach for this client application was inspired by the modularity concept model introduced by Google, as illustrated in Figure 4.1.1.2. This concept emphasizes modularity in mobile phone design, aiming for each module to function as an independent technical element. In line with this inspiration, the application software was designed to ensure that each module operates independently while contributing to the overall functionality, thereby enhancing the flexibility and reusability of the system.

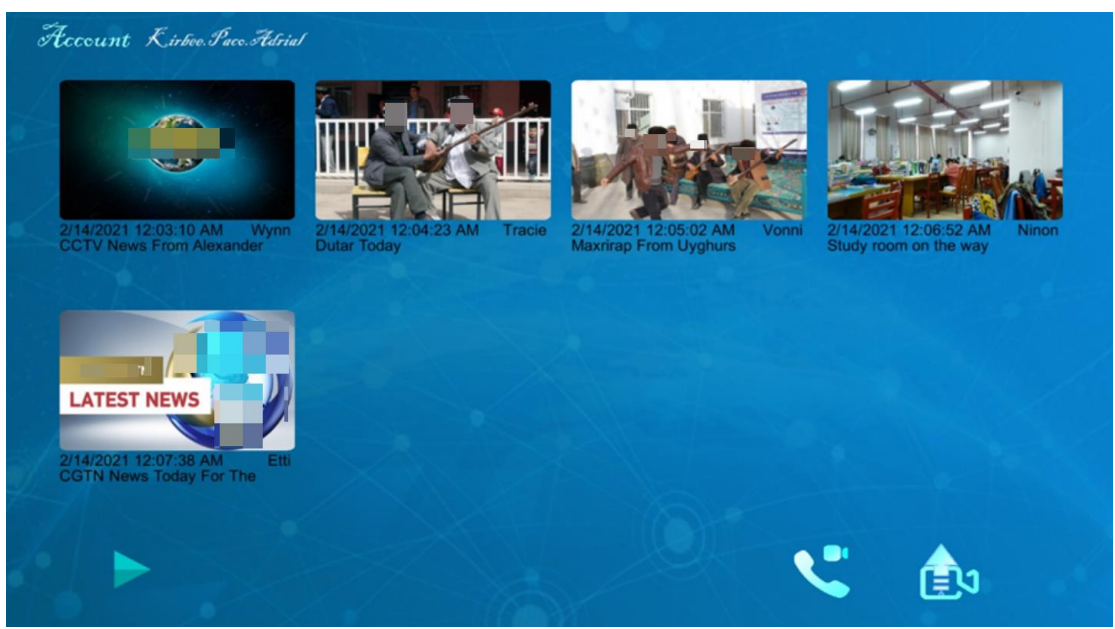

Figure 4.1.1-2 Main panel of the client application.

Simultaneously, the client must fulfill status detection requirements. Upon connecting to the server, it should implement fundamental status detection functions and continuously update the current status in real time to ensure stable operation.

As a practical software application, the client must be user-oriented, with each module designed to provide a user-friendly experience. Upon entering the main interface, users can choose between a standard login mechanism or a guest login option. This project supports both approaches: users who wish to access advanced features may register an account and log in, while those requiring basic functionality can use the application without logging in. The backend should facilitate guest access, ensuring that users can fully utilize the live video streaming system's features even without a registered account.

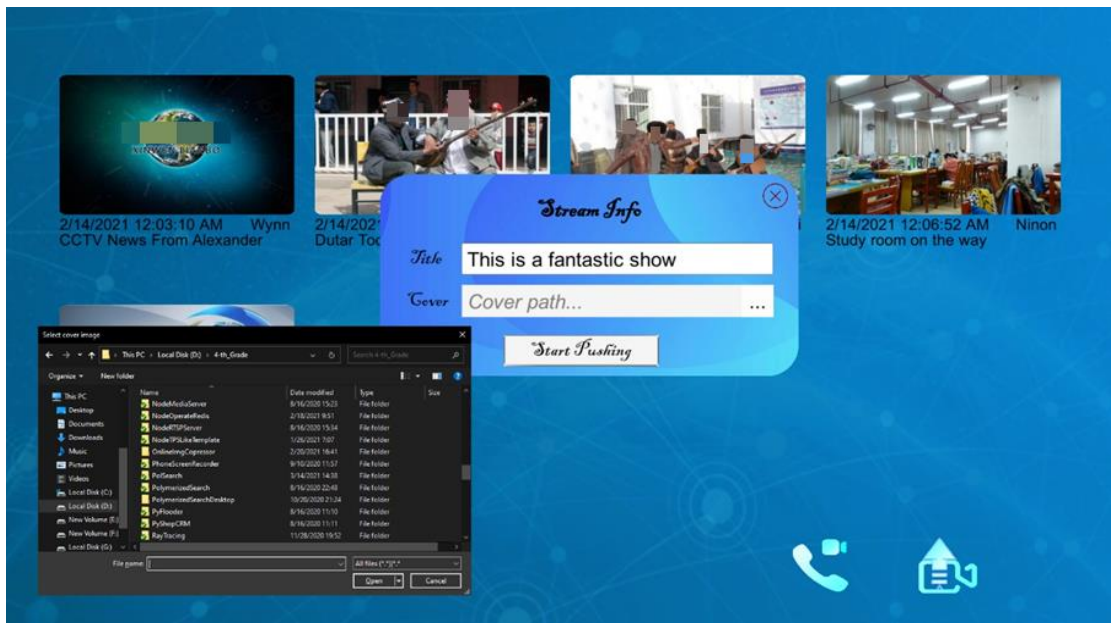

Figure 4.1.1-3 Registration panel before streaming at the client side.

When a user opens and logs into the live video application, whether as a visitor or a registered user, they will be presented with a list of public live streaming rooms and information about current live streaming. By selecting any public live streaming room from the interface, the user can immediately enter the room to view the live streaming content. To utilize the live streaming functionality, users must register and log in. Once logged in, users can start a live streaming by clicking the "Live Broadcast" button. They will need to enter a title and upload a cover image for the streaming. After setting up, the video content being streamed will be visible to all users of the system.

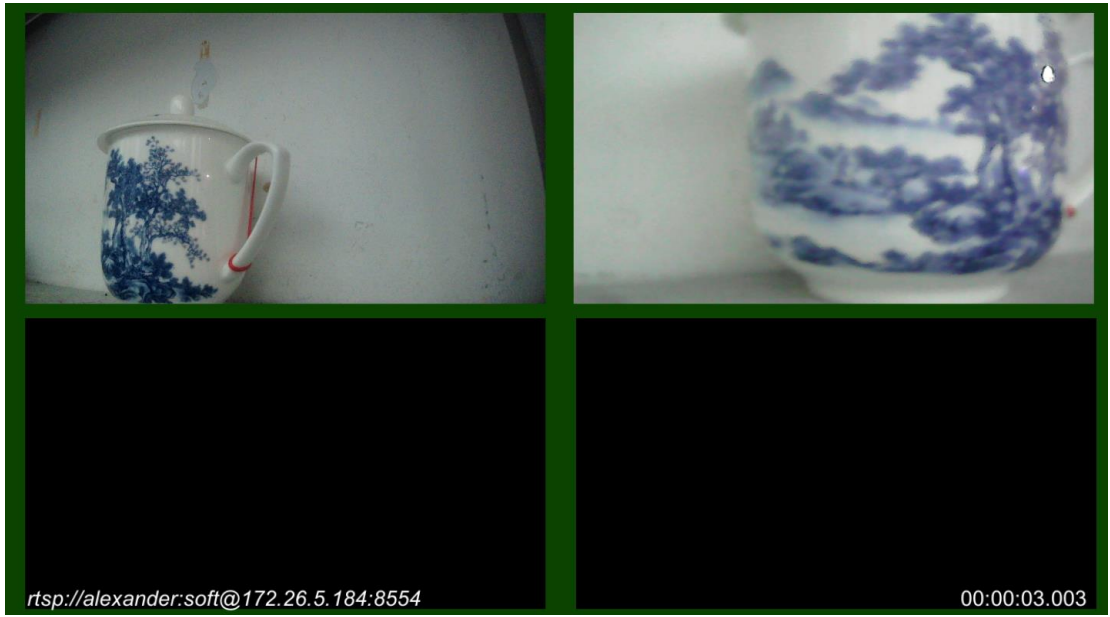

Figure 4.1.1-4 Public live streaming status at the client side.

In addition to those functions, the application also supports user-to-user calls when users are not actively streaming. If an online user is in an idle state, other online users can initiate a call to them. When User A attempts to call an idle User B, User B's client will receive a notification displaying the details of the incoming call. User B can then choose to accept or decline the call based on this notification. If User B accepts the call, a one-to-one video communication session will commence between User A and User B. If User B declines the call, both users will revert to their idle states. This functionality establishes a basic one-to-one video call feature within the application.

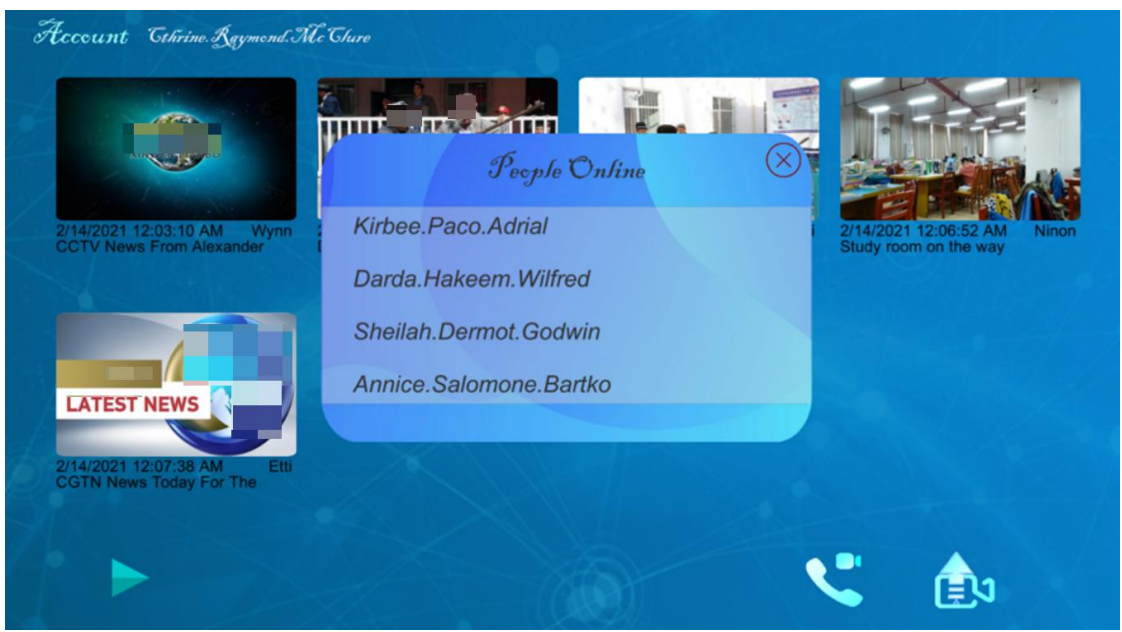

Figure 4.1.1-5 Online user list at the client side.

When a user goes online, their status must be updated in real time. The primary statuses include "on the phone," "watching a movie," "live streaming," "receiving a call," and "idle." The user's interaction with any functional module may alter their status. Consequently, the status must be updated dynamically to ensure that other users can view the most current status when accessing the online user list and initiating calls to specific users.

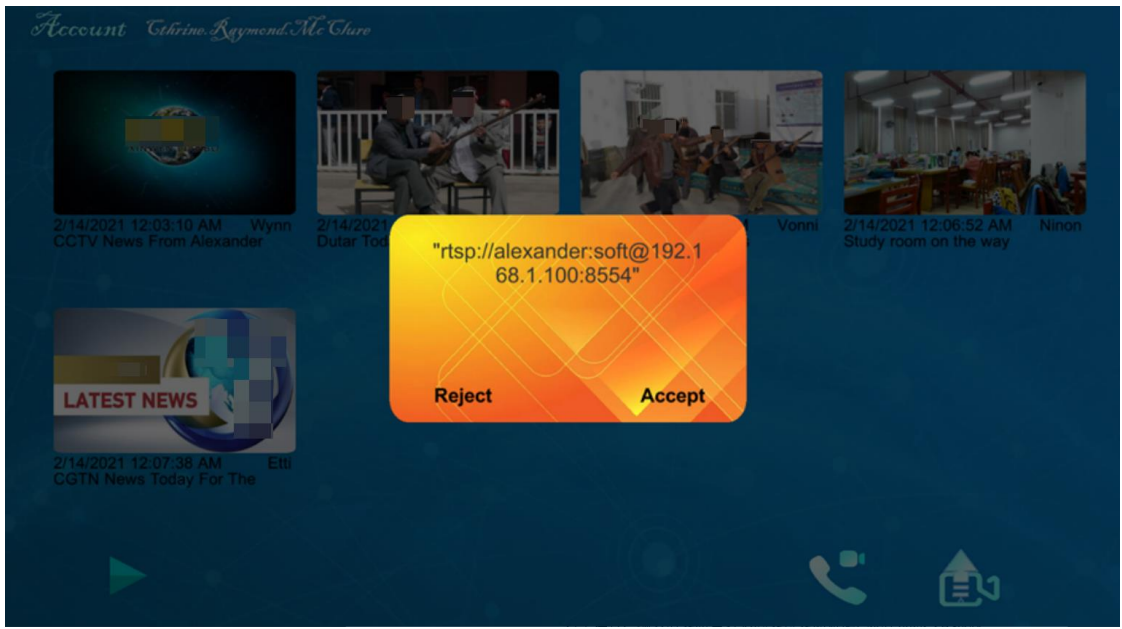

Figure 4.1.1-6 The client is being called by another user at idle state.

To ensure stable real-time status updates, a heartbeat mechanism is incorporated. This mechanism monitors connectivity and facilitates reconnection in case of disconnection.

In summary, the final application goals for the client application are as follows:

| Object | Functional Modules | Objectives | Basic Principles |
|---|---|---|---|
| Client | Encoding module | The encoding result should be as small as possible and as stable and efficient as possible. | Modular: The two ends are separated from each other in development and |
| | Streaming module | The streamed video data pushed to the server can be directly | |

| | | played by main-stream video players in market | maintenance and do not affect each other; the interfaces are isolated; each module is reusable and combinable to achieve better maintainability. |
|---|---|---|---|
| | Decoding module | Can stably and smoothly parse video data under RTMP, RTSP, M3U8, HTTP-FLV and other protocols | |
| | Playback module | Can play all kinds of live video data in the industry stably and smoothly | |
| | Multi-Peripheral Support Module | Multiple external video capture devices work independently without affecting each other | |
| | | Supports both the access to a single video capture device and the access to multiple external video capture devices | |
| | | The access to external video capture devices does not affect the performance of client applications | |
| | Heartbeat module | Equipped with a heartbeat control mechanism to achieve disconnection and reconnection when network conditions are unstable or relatively poor | |
| | Communication function | Ability to make one-to-one calls between online users | |
| | Public Live Streaming Room | One-to-many live data is visible to all client users | |
|  | other | The module of the external video capture device is independent of the live video streaming module and does not affect each other. In the absence of an external video capture device, the various modules of the live video streaming application can work normally. The push stream module can still push streams normally (the push stream data is a black background), and the pull stream can be used normally and play various live video data. |  |
| Back-end | Underlying database | The data table is relatively standardized, with as little redundancy as possible, stable performance, and SQL characteristics |  |
|  | Database Cache | High performance, easy to use |  |
|  | HTTP Services | MVC[21]Standardized, suitable for long-term maintenance and expansion |  |
|  | WebSocket Service | It also supports one-to-one and one-to-many communication with clients, and supports |  |

---

[21] MVC: is a development model, which can also be regarded as a development framework or specification. It is the abbreviation of Model, View, and Controller. M refers to the business model, V refers to the user interface, and C refers to the controller. The purpose of using MVC is to separate the implementation code of M and V, so that the same program can use different forms of expression. Among them, the definition of View is relatively clear, which is the user interface, Model is the data layer, and Controller is the control layer.

|  | | connecting sessions between different clients. | |
|---|---|---|---|
|  | Load balancing module | Supports classifying requests and forwarding them to multiple dedicated servers based on the classification results | |

Table 4.1.1-7 Objectives for the client-side application

The establishment of these objectives defines the functional requirements for the system. During the subsequent development phase, the design and implementation of the final application will be executed strictly in accordance with these goals to deliver a user-oriented live video streaming client application.

4.1.2 CLIENT APPLICATION STRUCTURE

The structure of the client application can be divided into four functional modules: the Message Panel, the Main Panel, the Player Module, and the Live Tool Module. Among these, the Player and Live Tool Modules occupy a central role in the video call functionality. Based on the specific implementation technologies, the system can be further categorized into modules such as the External Device Data Capture Module, Stream Pulling Module, Stream Pushing Module, Heartbeat Module, Public Data Acquisition Module, and Data Push Module. Although the core of this simple live video streaming application revolves around the Stream Pulling and Stream Pushing Modules, all these components are integral to the system's overall functionality.

In developing the Player Module, the Easy Movie Texture plug-in was utilized to parse and play various types of live video data. This third-party plug-in for U3D offers excellent version compatibility and serves as an efficient and universal video data playback solution. By leveraging this plug-in, developers avoid the need to build a video player from scratch, allowing them to focus on the design and implementation aspects of the application.

After integrating the plug-in and setting up the player file path interface with a reasonable design, a "universal video player" based on U3D was created. This video player supports playback of various locally stored video formats, including mov, mp4, and flv. Moreover, Easy Movie Texture enables the pulling and playback of network video data, handling all pre-playback processes—such as protocol parsing, video data acquisition, decryption, decoding, and packaging—efficiently and effectively.

Following the implementation of the RTSP protocol-based live video streaming module during the experimental phase, it was connected to the client, resulting in a basic RTSP-based video streaming tool. The video data pushed by this tool can be decoded and played using mainstream streaming players.

These two modules—Player and Live Tool—constitute the core of the live video streaming system. The remaining components, including the Message Panel and the Main Panel, form the peripheral structure, focusing on enhancing the user's experience when not engaged in live streaming.

The streaming module includes data acquisition functions for multiple external video capture devices. The acquisition process depends on the integration between the Main Panel and the streaming module, with the streaming module managing the activation and deactivation of multiple external video devices. The Main Panel directs specific operations to these devices.

The Main Panel collects user data, live streaming information, and other related messages from the server. This module is highly dependent on the server for data retrieval. Without the server, no data is available for rendering on the front-end interface, rendering the module's functions unusable. The heartbeat function, implemented in this panel, maintains the connection by sending periodic data to the server to ensure the connection's integrity. If the client fails to receive a response from the server, it indicates a lost connection. During such disconnections, the client can only perform local operations, and data cannot be shared with the server or saved to the database. To minimize redundant work, the client should prompt the user about the disconnection, as connection issues are often due to the client's network environment. Users are advised to check their network hardware and settings, and if necessary, the client should attempt reconnection a limited number of times, as specified by the developer. This practice aligns with recent trends of limiting reconnection attempts rather than allowing unlimited retries.

The system's structure also includes the Message Panel, which interacts with the Public Data Acquisition Module and the Data Push Module. Operations affecting the database are immediately reflected in the public data module, which other clients can query from the server. The Message Panel maintains information on live streaming titles, timings, client statuses, and user levels (such as visitor or registered user). It serves as a bridge between users and between users and the server, facilitating data exchange essential for executing operations and requirement logic.

In summary, while the modules of the client application are interrelated, they operate independently, each contributing to the overall functionality of the U3D-based live video streaming system. These modules, each fulfilling distinct functions, collectively form an integrated and efficient system. Figure 4.1.2-1 presents the code structure of the client-side application layer, excluding the plug-in code developed by myself.

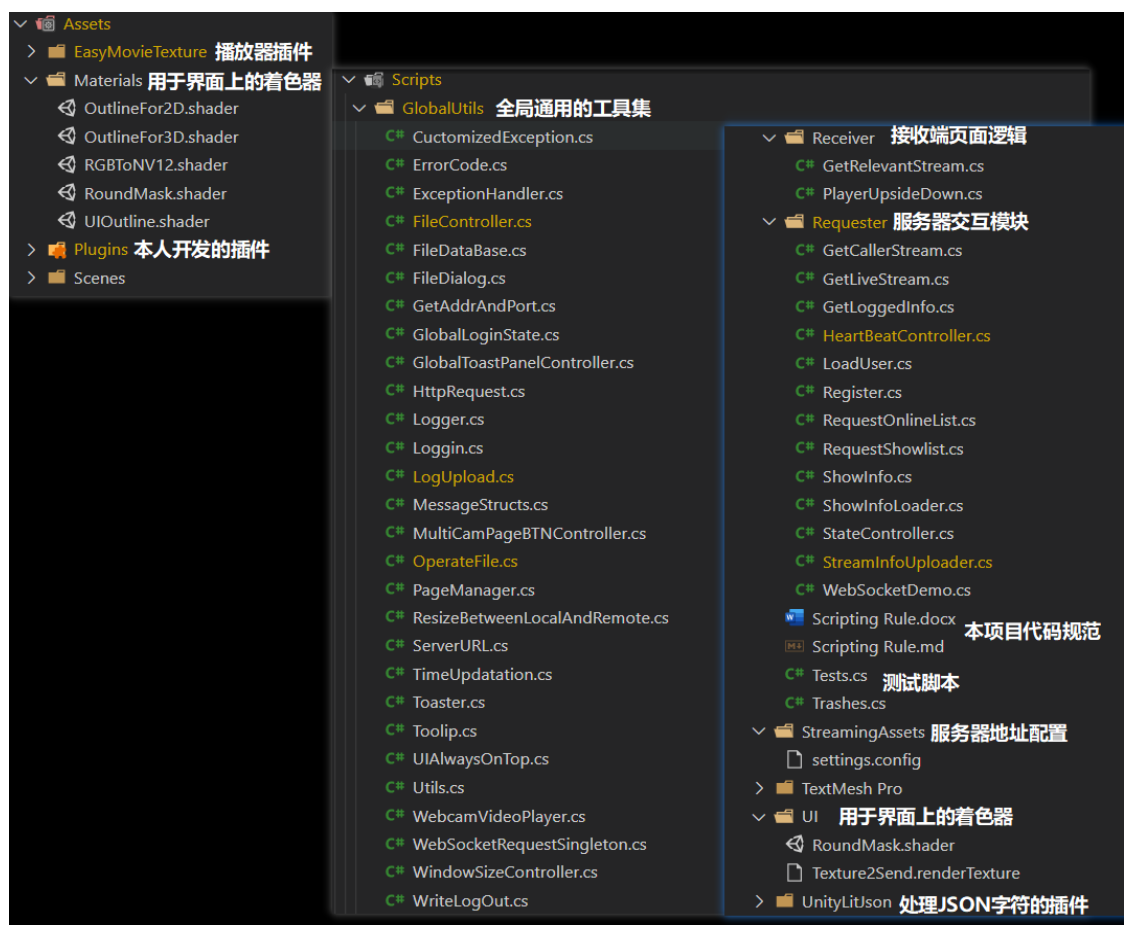

Figure 4.1.2-1 Schematic diagram as a code structure of the client-side application.
[Figure Inserted from Original Thesis Source, 2021.7]

The code across different modules demonstrates significant reusability, particularly concerning interface components. This modular design approach facilitates efficient updates and modifications at the code level, reducing the need for repetitive design efforts by art planners. For example, the config.settings file provided with the project is used specifically for configuring server address information in the packaged version. This design ensures a clear separation between the front-end and back-end, allowing the front-end to function independently of back-end constraints. Adjustments to back-end configurations can be made directly within the packaged application by modifying this configuration file.

Concerns regarding the accidental deletion of the configuration file are acknowledged. Although such scenarios cannot be entirely prevented, I have implemented measures to mitigate this risk. The configuration file is stored in a location that is not easily accessible to users and is accompanied by clear documentation and specifications. This approach helps ensure that users are unlikely to inadvertently interfere with the file under non-critical circumstances, thereby preserving its integrity.

## 4.2 TECHNICAL ELEMENTS

As a live video streaming application, the specific technical modules employed by the client have been detailed in previous sections. The technical elements integral to these modules play a crucial role in their development, outlining the specific technical aspects and providing clear guidance for implementation. This section aims to further elucidate these technical elements to better understand their application and development.

Firstly, in the streaming module of the live video client application, I utilized research on the RTSP protocol and H264 encoding to integrate these results into U3D. The live streaming library was incorporated into U3D, where it functions as a component within the U3D game scene. By creating a game object and attaching the component to it—whether it is an empty game object or one with a geometric shape—the script associated with the component can be compiled and executed during runtime in U3D, achieving the desired streaming effect. Initially, the streaming module and other modules were developed in a componentized manner, allowing for the direct application of pre-developed components to achieve basic streaming functionality.

Secondly, the Easy Movie Texture plug-in was employed in the streaming module to facilitate the integration of various technical elements, ultimately forming a versatile video player. The author connected the public interface from the MediaPlayerCtrl script to the Input control under the sub-panel, enabling users to view live content by inputting the live video URL. Subsequently, to enhance user experience and integrate with the back end, live video data URLs were retrieved from the back end and presented in the main panel. Each URL (which contains more detailed live video information beyond what is discussed here) was bound to a specific button control and instantiated using the *Instantiate(GameObject gameObject[ DEFAULT_VALUE, GameObject parentObject])* method provided by the U3D API.

It is important to note that in U3D, all objects that can be displayed to users are categorized as GameObjects. U3D also includes Components and its own Systems. Unlike Unreal Engine 4, U3D's system-level functions are not open source and are considered a "black box" to developers. However, GameObjects and Components are highly flexible and can be customized to meet specific rerequirement. The author developed a library of components during the early stages of development, which provided the foundation for the project. The advantage of components in U3D is their reusability, which significantly enhances the cohesion of project modules and improves the overall reusability of different project components.

In developing the main interface, I focused on maximizing user fluency by providing relevant prompts when acquiring message data for the main panel, as guided by the requirements. The underlying system utilizes the HTTP protocol, specifically employing HTTP GET requests to retrieve public data from the backend for display on the main panel. This public data includes video URLs and other live video information. The data provides clients with the title, cover, opening time, and duration of each online public live room, enabling users to select and view the live streaming based on this information.

For user interactions within the client, such as user status detection, information updates, registration, and login operations, HTTP POST requests are employed. These requests carry data to the backend for inspection and updates in the database, following the specified message content. Although database update operations are commonly used, the affiliated project of this thesis employs a method of appending new data, which significantly enhances database performance and security. The communication format between the front-end and back-end uses the widely adopted JSON format, which facilitates message processing on both ends. It is worth noting that while numerous JSON data parsing libraries exist, not all are suitable for large-scale projects involving complex data combinations. The author initially underestimated this aspect, leading to various errors and slow development and debugging processes when using an inadequate third-party JSON parsing library. The JSON data parsing library used on the Node.js backend is relatively stable, benefiting from Node.js's robust development ecosystem. Node.js offers powerful data parsing and processing libraries for various data types, including XML, JSON, and TXT. However, options for JSON parsing libraries on U3D-based front-ends are more limited. The industry generally uses the JSON parsing library provided by C#, specifically LitJSON.

In the affiliated project, I utilized a third-party plugin, U3D LitJSON, which is a secondary encapsulation of LitJSON. However, the encapsulated version differed significantly from the original, resulting in numerous development and debugging issues. Consequently, I opted to use native C# string methods for JSON parsing. Although this approach is somewhat more cumbersome, it provides a more intuitive and controllable process, avoiding issues associated with immature third-party libraries and ensuring greater reliability in development.

Almost all functions related to the main panel involve interaction with the backend. Data for the main interface cannot be hard-coded on the client side; instead, it must be stored in a remote database to ensure data consistency across different clients. This facilitates video live communication and viewing of designated live video streaming between clients using the live video streaming system. Additionally, storing data on the remote server ensures that data is updated in real time. In the heartbeat module, the client communicates with the backend once per second. The client sends its latest status to the backend. While this periodic status update may not affect the status change directly (since the client actively sends status updates when it needs to change), it helps evaluate the current network environment. This ensures ongoing condition-related interactions between the client and the server. The heartbeat module, together with the public data pull and data push modules, supports the system's functionality.

For obtaining information from multiple peripheral video capture devices, I utilized the WebcamTexture interface provided by the U3D API. The GetWebcamTexture() method is implemented to retrieve specific information about external devices, such as color mode, video screen, and device name. The author considers U3D's WebcamTexture API as a crucial technical element for this module. A loop is designed to traverse all camera device information, retrieve pixel data from each frame, compile it into a Texture, and project it onto the specified RawImage. This image is then integrated into the Canvas in World Space, forming an effective method for managing multiple peripheral video capture devices in the project.

In summary, the technical elements employed by the client are as follows:

1. Server-side streaming technology based on the RTSP protocol.
2. Video encoding technology based on H.264.
3. Real-time decoding and playback technology for online or offline video data, facilitated by the Easy Movie Texture Plugin.
4. Network communication technology based on HTTP/WebSocket protocols.
5. Integration for multi-peripheral video capture based on U3D's WebcamTexture and RawImage combination.

These technical elements contributed to the successful development of the client software, resulting in a stable live video client application that operates on the Windows operating system, built on the U3D platform.

## 4.3 IMPLEMENTATION OF THE CORE MODULE

Phase 1 - Setting up the recording area

[0001] Download and install the U3D engine (any version) from the U3D official website;
[0002] Create a new U3D project, create a new scene, and enter the scene;
[0003] Make sure that the newly created scene has two basic game objects: MainCamera and Directional Light;
[0004] Create a new Canvas in the Hierarchy panel, and make sure to generate an Event System game object at the same time. This game object is mainly used to monitor events on the Canvas and is extremely important for user interaction.
[0005] In the Inspector panel, set the rendering mode of the Canvas to World Space Canvas;
[0006] In the Inspector panel, set the size of the Canvas to 1920 in the X direction and

1080 in the Y direction;

[0007] Create an empty object in the Canvas sublayer and name it RTSPServer to mount the RTSPServer script;

[0008] Create an empty object in the Canvas sublayer and name it WebcamVideo Player to mount the external device data receiver script;

[0009] Create a new Camera in the Canvas sublayer and set its Tag to WorldSpaceCamera. Set the Camera to world space. Select the Orthographic option in the Inspector panel to set it to orthographic projection, or select Perspective to set it to perspective projection.

[0010] Create a new UI-Image object in the Canvas sublayer and name it BackgroundImage. It is used to set the background of the Canvas and the objects above it. Set it to a certain color to distinguish it from the UI controls in the Canvas sublayer (the gray part in this screenshot is this UI control);

[0011] Create a new empty object in the Canvas sublayer and name it MultiCamHolder to hold multiple camera devices;

[0012] Create a new UI-Raw Image object in the sublayer of MultiCamHolder and name it MultiCam, which is HolderPage1. Set it to a certain color to distinguish it from the background image control (the green part in this screenshot is this UI control). It is used to present the video data captured by the camera. This object is the first page. If there are many external camera devices, you can create multiple MultiCamHolderPages according to your needs, and map the interactive content according to the page layout and mapping relationship in MultiCamHolderPage1;

[0013] Create four UI-Image objects in the sublayer of MultiCamHolderPage1 and name them CameraRawImg_1 to CameraRawImg_4;

[0014] Arrange CameraRawImg_1 to CameraRawImg_4 on MultiCamHolderPage1 in an orderly manner. The arrangement method is arranged in a grid pattern. From the left, there are only CameraRawImg_1 and CameraRawImg_2. From top to bottom, there are CameraRawImg_1 and CameraRawImg_3.

[0015] Create a new UI-Text in the Canvas sublayer and set its default string to RTSP Address in the settings panel to display the RTSP streaming address;

[0016] After completing the above operations, the final UI layout should be as shown below:

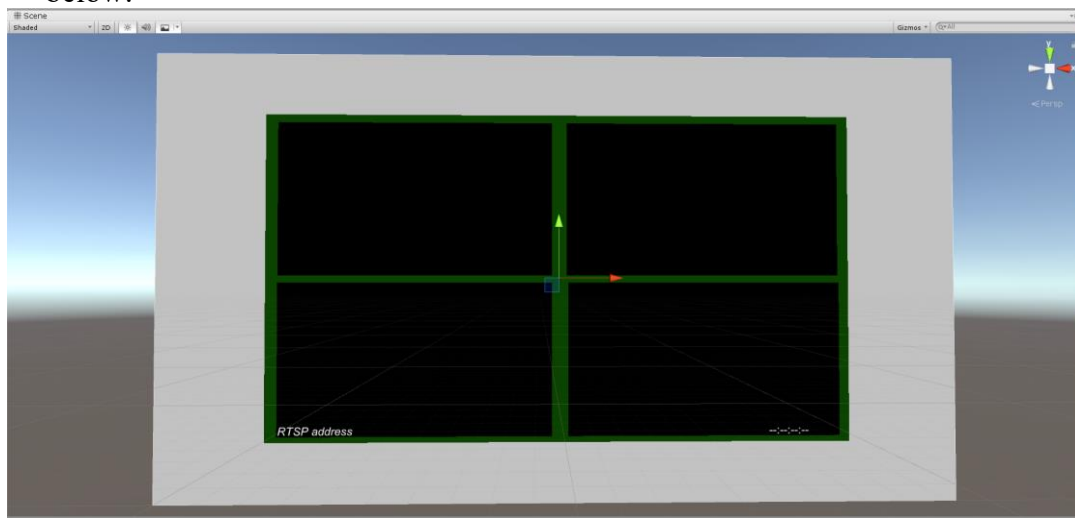

[0017] After executing the above operations, the layout of the game objects under the Hierarchy panel is as follows (in this example, one page of multi-camera receiving controls × 2 is used, a total of two pages, so you can see that there are two MultiCamHolderPages, which can be increased automatically when the number of external cameras increases):

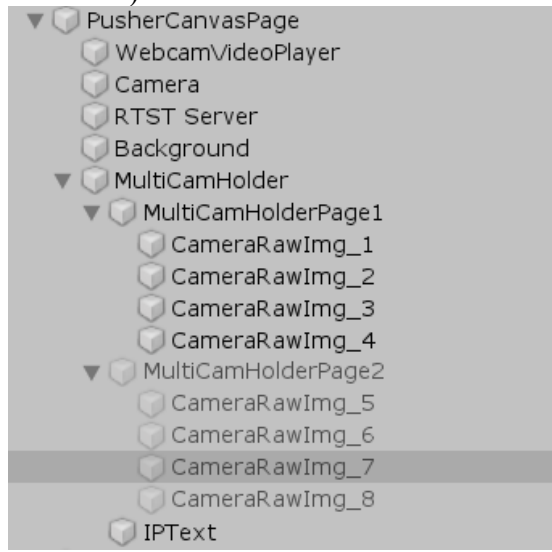

[0018] The Camera created in step "[0007]" is aligned with the area of MultiCamHolderPage1, which is the area where the initial stream will be pushed. The Camera will record this area and push the stream according to the RTSP protocol. If it is not aligned with this area, the content of the push stream will change due to the change of the area recorded by the Camera.

Phase 2 - setting up the interactor

[0019] Create a new Canvas in the Hierarchy panel and name it InteractionCanvas. This Canvas mainly receives the user's screen interaction operations;
[0020] Set the rendering mode of the InteractionCanvas to Screen Space Overlay in the Inspector panel;
[0021] In the Inspector panel, set the size of the InteractionCanvas to 1920 in the X direction and 1080 in the Y direction;
[0022] Copy MultiCamHolderPage1 and its child objects (CameraRawImg_1 to CameraRawImg_4) created in Canvas and place them in the InteractionCanvas sublayer;
[0023] Add Button components to CameraRawImg_1 to CameraRawImg_4 in the InteractionCanvas and in the MultiCamHolderPage1 sublayer. Add Button components in the Inspector panel.
[0024] Adjust the Color property of the Image controls on CameraRawImg_1 to CameraRawImg_4 in the InteractionCanvas, which are in the MultiCamHolderPage1 sublayer, to be transparent, and set their alpha to 0.
[0025] Create two UI-Button controls in the InteractionCanvas sublayer and name them Next and Previous. This is because two pages are used in this example to carry the contents of 8 cameras. Each page presents the contents of 4 cameras. Different

pages can be switched through these two buttons.

[0026] Place the Next and Previous buttons in the appropriate locations in the InteractionCanvas to facilitate page switching.

[0027] After performing the operations described in "Step 2", the control hierarchy of the InteractionCanvas on the Hierarchy panel is as follows:

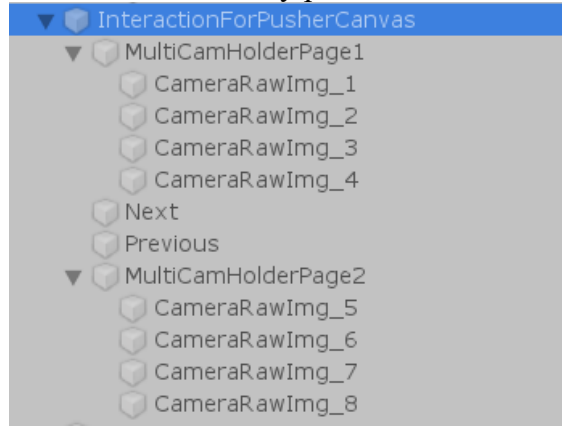

[0028] After executing the operation of the "Second Phase", hide the controls in the first phase and only display the controls in the second phase. The content displayed in the scene editor should be as follows (only one or two controls are visible, which are Button controls used to turn pages):

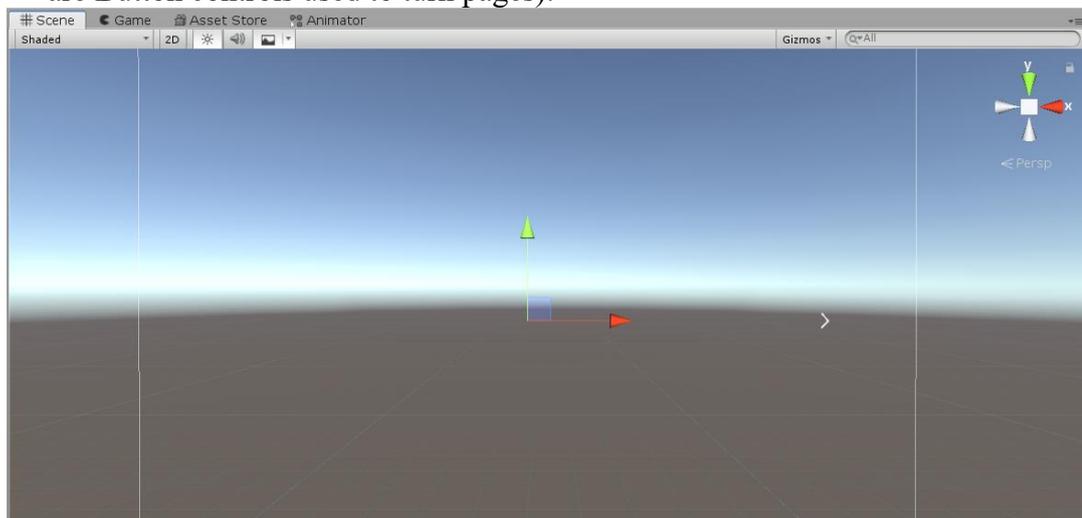

The arrow pointing right in the figure is the Next page button, and the Previous page button is hidden and controlled by code.

Phase 3 - introducing the live streaming library

[0029] The RTSP video live library used in the proposed system has been open sourced by me. It can be downloaded from the Github official website and placed in the root directory of the affiliated project of this thesis. The download link is: https://github.com/Ezharjan/UnityVideoStreamingPackage.git, follow the instructions in the README file in this repository (details are as follows);

[0030] Replace the automatically generated Packages folder in this U3D project with

the Packages folder in this folder;

[0031] Place the VideoStreamingerverPackages folder in the root directory of the U3D project;

[0032] At this point, the import work is basically completed. Check again to ensure that there is a "com.U3D.ig.video-streaming.server": "file:../VideoStreamingerverPackages/" statement in the manifest.json in the Packages folder and that the file path described by the value pointed to by its key is correct;

[0033] After completing the operations described in "Step 3", the structure of the project folder should look like the following:

| 名称 | 修改日期 | 类型 |
|---|---|---|
| Assets | 2021/5/4 9:13 | 文件夹 |
| Library | 2021/5/4 10:38 | 文件夹 |
| Packages | 2021/5/4 9:13 | 文件夹 |
| ProjectSettings | 2021/5/4 9:13 | 文件夹 |
| SimpleLiveServer | 2021/5/4 9:13 | 文件夹 |
| StreamingAssets | 2021/5/4 9:13 | 文件夹 |
| Temp | 2021/5/4 10:38 | 文件夹 |
| VideoStreamingServerPackages | 2021/5/4 9:13 | 文件夹 |

[0034] Find the Shader folder in the VideoStreamingerverPackages folder and place the RGBToNV12.shader shader file in it in the Assets folder of the U3D project for subsequent use;

[0035] After completing the "third step", enter the U3D Editor of the U3D project to observe, and you can see the hierarchical relationship in its Project panel as shown in the following figure (note whether the VideoStreamingerver folder and the files inside it are missing):

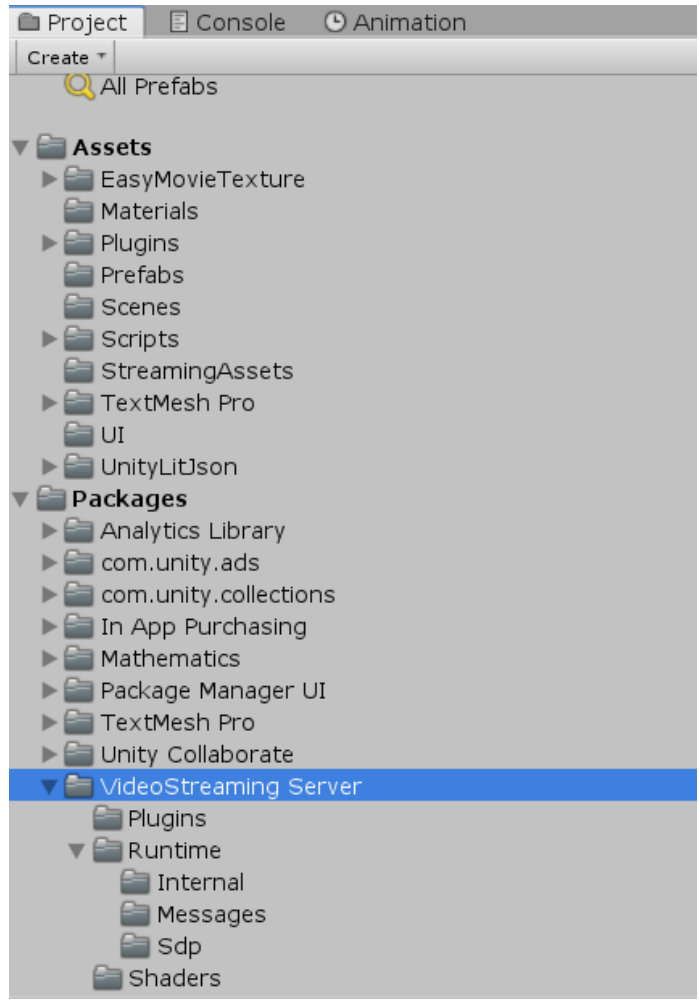

Phase 4 - Using the Live Library

[0036] In the Project panel of the U3D Editor, find the VideoStreamingerver script in the Runtime folder inside the VideoStreamingerver folder, and mount it as a component on the empty object named RTSPServer created in step [0007] in the first stage.

[0037] In the VideoStreamingerver component on the RTSPServer empty object, reference the specified content on its public interface. The "IP Text" interface references the UI-Text control named "Addr" created in step "[0015]" of "The First Link"; in its RGB To NV12 Shader interface, reference the RGBToNV12.shader shader file copied in step "[0034]" of "The Third Link"; set the live streaming port number, user name, password, and frame rate information used by the RTSP live streaming server (the frame rate used in this example is 30 frames per second); at this time, a Render Target interface is left without any object referenced;

[0038] In the U3D Editor's resource panel, right-click to create a Render Target, and set its length and width to 1920 and 1080 respectively in the Inspector panel. This value is the pixel size of the video data that will be pushed out. No matter how many external cameras are connected or how many screens are in the Canvas, it

will not be affected. It (Render Target) is like a film, and the amount of data that is pushed out in the end is only the size of this film.

[0039] Set the color mode of the Render Target created in step [0038] to BGRA32 mode and check sRGB mode;

[0040] Go back to step "[0037]", check whether the reference relationship on the RTSPServer object is correct, add the Render Target interface that was just missed, and reference the Render Target created in step "[0038]" on this interface;

[0041] After executing steps [0036] to [0040], the components and their reference relationships on the RTSPServer game object should be as shown below:

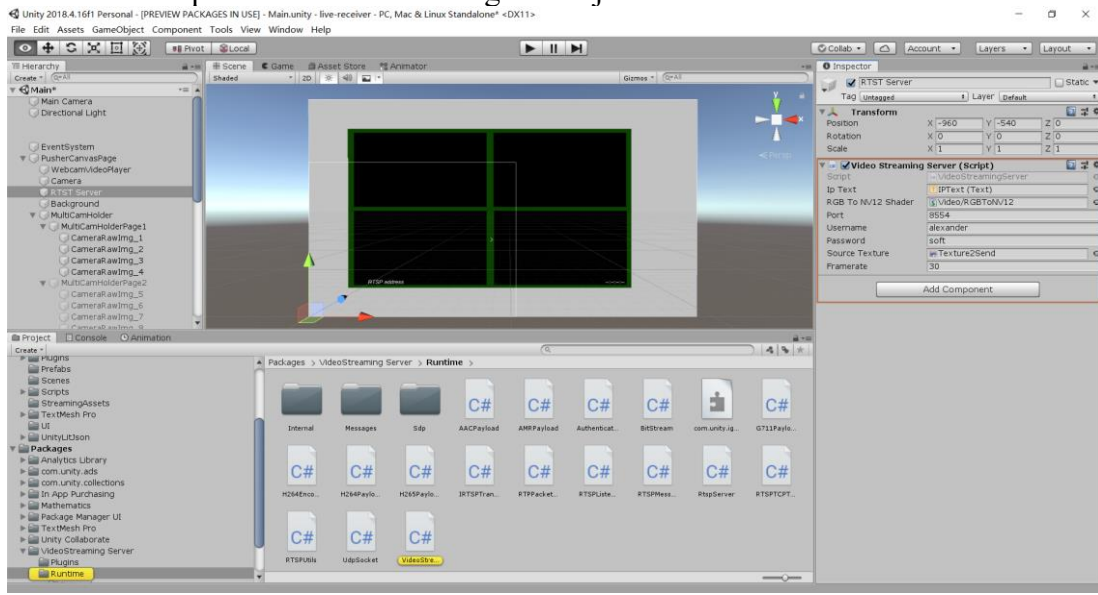

[0042] The Render Target object created and set in step "[0038]" is mounted on the "Target Texture" interface of the Camera created in step "[0009]"; after running the program, the content stored in the Render Target is the image data of the video area captured by the Camera, which is equivalent to the Render Target being a film of the Camera. The film will be transmitted and distributed to the client in the form of live streaming impact data through H264 encoding and RTSP protocol, thereby forming a basic live streaming mechanism;

[0043] Write the following script to iterate over the images captured by all external cameras connected to this device when the program is running:

```
1.  using System.Collections;
2.  using System.Collections.Generic;
3.  using U3DEngine;
4.  using U3DEngine.UI;
5.  /*by Alexander*/
6.  public class WebcamVideoPlayer : MonoBehaviour
7.  {
```

```csharp
8.  public List<GameObject> rawWebCam = null;
9.  private WebCamTexture[] webCamTextures;
10. private WebCamDevice[] webCamDevice;
11. private Color32[] color32Data;
12. private bool isAuthorized = false;
13. private int webCamCount = 0;
14.
15. void OnEnable()
16. {if(!isAuthorized)
17. {StartCoroutine("GetCameraDevice");
18. }else
19. {//StartWebCam(0);
20. StartAllWebCams();}}
21. void OnDisable()
22. {StopAllWebCams();// crucial to stop wihle going out
23. }
24. void OnDestroy()
25. {CleanWebCams();
26. }
27. IEnumerator GetCameraDevice()
28. {yield return Application.RequestUserAuthorization(UserAuthorization.WebCam); //Wait for user to allow access
29. if(Application.HasUserAuthorization(UserAuthorization.WebCam))
30. {webCamDevice = WebCamTexture.devices;//Get the device first
31. webCamTextures=new WebCamTexture[webCamDevice.Length];
32. for(int i =0; i < webCamDevice.Length; i++)
33. {if(!webCamDevice[i].name.ToLower().Trim().Contains("virtual"))// ignore virtual cameras
```

```
34. {webCamTextures[i] = new WebCamTexture(webCamDevice[i].name);
35. webCamTextures[i].requestedHeight = 1080;
36. webCamTextures[i].requestedWidth = 1920;
37. rawWebCam[i].gameObject.GetComponent<RawImage>().texture =
    webCamTextures[i];
38. StartWebCam(i);
39. rawWebCam[i].GetComponent<RawImage>().color
    = new Color(255f,255f,255f,255f);// whiten image for clearence
40. }}
41. isAuthorized = true;
42. }}
43. void StartWebCam(int i)
44. {if(i < 0 || i > webCamTextures.Length -1)
45. {Debug.LogError($"The camera with index {i} is outside the valid range, please
    check your index!");
46. }
47. else
48. {if(webCamDevice[i].name.ToLower().Trim().Contains("virtual"))
49. {Debug.Log($"The camera with index {i} is virtual, please check your index!");
50. }
51. else
52. {webCamTextures[i].Play();
53. color32Data = new Color32[webCamTextures[i].width *
    webCamTextures[i].height];
54. webCamCount++;
55. }}}
56. void StartAllWebCams()
57. {for(int i = 0; i < webCamTextures.Length; i++)
58. if(!webCamDevice[i].name.ToLower().Trim().Contains("virtual"))// ignore
    virtual cameras
```

59. webCamTextures[i].Play();

60. }

61. void StopAllWebCams()

62. {for(int i =0; i < webCamTextures.Length; i++)

63. if(!webCamDevice[i].name.ToLower().Trim().Contains("virtual"))// ignore virtual cameras

64. webCamTextures[i].Stop();

65. }

66. void CleanWebCams()

67. {StopAllWebCams();

68. rawWebCam =null;

69. StopCoroutine("GetCameraDevice");

70. }}

[0044] Save the script written in step "[0043]" and mount it as a component on the WebcamVideo Player empty object created in step "[0008]", and set its UI-Image capacity in the Inspector panel according to the number of devices currently required to be connected or the number of CameraRawImg controls. In this example, since there are 4 controls per page and a total of two pages, the value here is set to 8; after setting the number, reference the first page CameraRawImg_1 to CameraRawImg_4 and the second page CameraRawImg_1 to CameraRawImg_4 controls of the MultiCamHlder sublayer in each blank item, and reference them in sequence;

[0045] After performing the operations described in step [0044], the component properties on the WebcamVideoPlayer object and its relationship with the game objects in the scene should be as shown below:

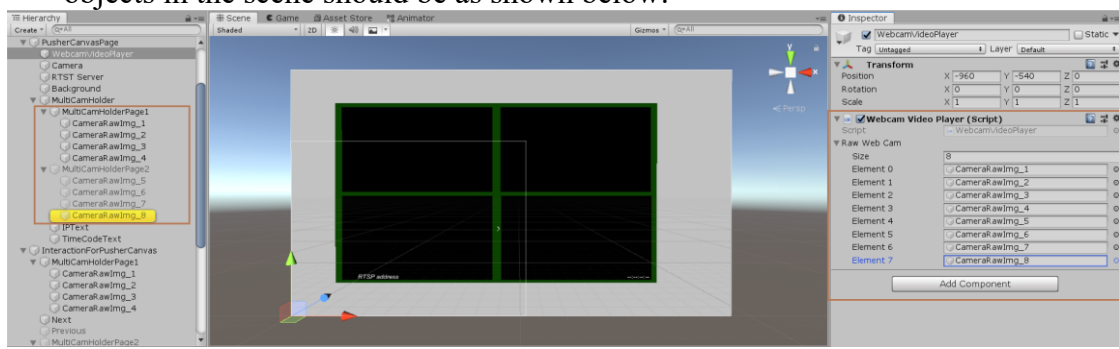

Phase 5 - connecting the interactive area and the recording area

[0046] Adjust the MainCamera at the top level of the scene to focus its view on all areas of the Canvas created in step [0004];

[0047] Set the projection mode of MainCamera to perspective projection or orthographic projection;

[0048] Add the following script as a component to the 8 CameraRawImgs under InteractionCanvas (1 to 4 on the first page, 1 to 4 on the second page, a total of 8) to control the current size (full screen or minimized) of each CameraRawImg window:

```
1. using System.Collections.Generic;
2. usingU3DEngine;
3. usingU3DEngine.UI;
4. /*by Alexander*/
5. public classWindowSizeController:MonoBehaviour
6. {
7. public RectTransform thisRect =null;
8. public GameObject parentGameObj =null;
9. publicList<Button> btnList;// to be optimized
10. Vector2 ownSize;
11. Vector3 ownPosition;
12. boolisMaximized =false;
13. privatevoidStart()
14. {StoreOwnInfo();
15. }
16. privatevoidStoreOwnInfo()
17. {ownPosition = thisRect.anchoredPosition;
18. ownSize = thisRect.sizeDelta;
19. }
20. privatevoidMaximize()
21. {thisRect.anchoredPosition =newVector3(0f,0f,0f);
```

```
22. thisRect.sizeDelta = new Vector2(1920f, 1080f);
23. isMaximized = true;
24. }
25. private void ResizeBack()
26. { thisRect.anchoredPosition = ownPosition;
27. thisRect.sizeDelta = ownSize;
28. isMaximized = false;
29. }
30. public void ResizeWindow()
31. { if (isMaximized)
32. { ShowAllActiveWindow();
33. ActivateAllButtons();
34. ResizeBack();
35. }
36. else
37. { LeaveOnlyOneActiveWindow();
38. LeaveOnlyOneActivButton();
39. Maximize();
40. } }
41. private void LeaveOnlyOneActiveWindow()
42. { RawImge[] rawImages = parentGameObj.GetComponentsInChildren<RawImage>();
43. foreach (RawImage rawImage in rawImages)
44. rawImage.enabled = false;
45. thisRect.gameObject.GetComponent<RawImage>().enabled = true;
46. }
47. private void ShowAllActiveWindow()
```

```
48. {RawImage[] rawImages =
    parentGameObj.GetComponentsInChildren<RawImage>();
49. foreach(RawImage rawImage in rawImages)
50. rawImage.enabled = true;
51. }
52. private void LeaveOnlyOneActivButton()
53. {foreach(Button btn in btnList)
54. btn.enabled = false;
55. this.gameObject.GetComponent<Button>().enabled = true;
56. }
57. private void ActivateAllButtons()
58. {foreach(Button btn in btnList)
59. btn.enabled = true;
60. }}
```

[0049] Find the Button component on the 8 CameraRawImgs (1-4 on the first page, 1-4 on the second page, 8 in total) under InteractionCanvas and reference the WindowSizeController script created in step "[0048]". Find the OnClick event column in the time list and add the ResizeWindow() method of WindowSizeController as the response event. Make sure that each CameraRawImg is filled in like this.

[0050] Find the script component just created in step "[0048]" on the 8 CameraRawImg (1~4 on the first page, 1~4 on the second page, 8 in total) under InteractionCanvas, and add the real picture CameraRawImg mapped in the Canvas sublayer to each CameraRawImg (1~4 on the first page, 1~4 on the second page, 8 in total), and mount the MultiCamHolderPage under the Canvas sublayer on the Parent Game Obj option (set each sub-item, there are two MultiCamHolderPages, set under the Interaction sublayer of two different pages); enter "4" on its Btn List according to the situation shown on each page, and reference all the CameraRawImg used for interaction in the Interaction Canvas sublayer, 4 on one page, 8 in total;

[0051] After executing this step, click any CameraRawImg control in the InteractionCanvas sublayer. The settings on its Inspector panel should correspond to the following (each correspondence is related to its serial number):

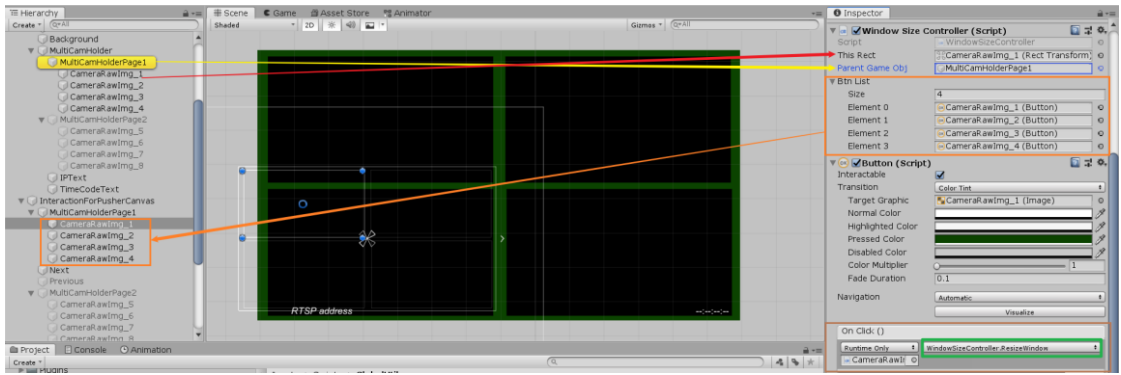

[0052] At this point, all development has been completed. Click the run button of U3D Editor or directly package the application on the local computer, open the packaged software and use it according to the prompts. The effect is that the user can see the video devices captured by all external cameras currently presented on his host; the screen where all video capture devices are located can be maximized or minimized by clicking; at the same time, the current push stream address can be seen in the lower left corner of the screen, and the audience uses the push stream address to watch the video content posted by this user; the live streamer can arbitrarily connect multiple video capture devices during the live streaming, and the video data of multiple devices will be distributed, but because the size of the data content sent is always 1920×1080 pixels, the connection of any number of cameras will not affect the delay of the live video streaming.

[0053] The idea of this design can be applied to any live video streaming, virtual reality, and entertainment game software applications developed based on U3D.

4.4 CLIENT R&D SUMMARY

As a software application directly interacting with users, the client plays a crucial role in the system. During its development, I enhanced the rendering performance of the internal user interface (UI) using scene technology provided by U3D, while also ensuring the reusability of UI elements. This approach led to the creation of an online video live client based on U3D. Figure 4.4 presents the core invention block diagram of the client for which the invention patent was ultimately filed. The short video demonstration for the evaluation of the communication latency is demonstrated [here](here).

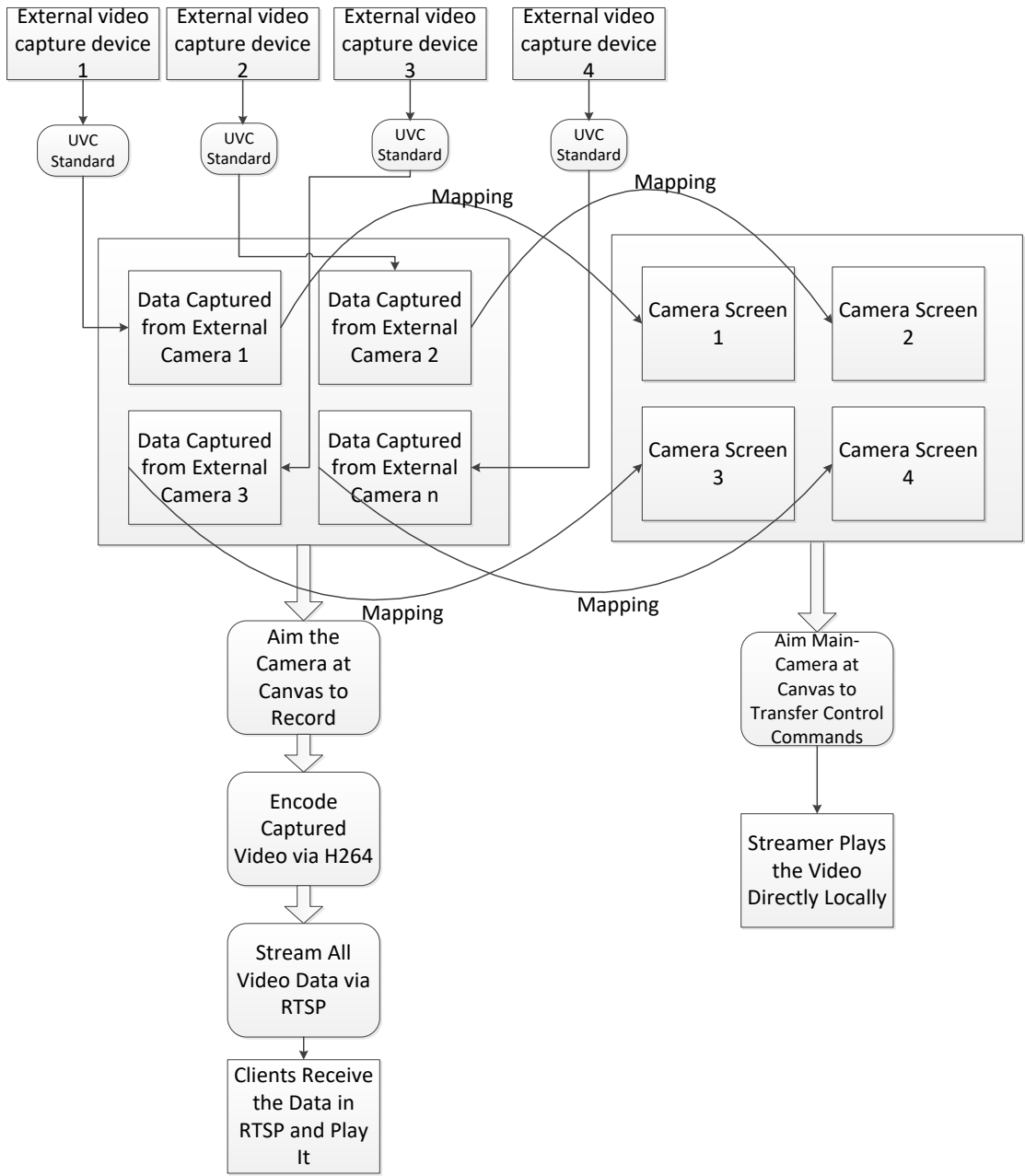

Figure 4.4 Core structure of the low-latency multi-channel streaming module proposed live video streaming system.

However, for the client to operate effectively and efficiently over extended periods, robust backend support is essential. In the following chapters, I will provide a detailed explanation of the backend architecture design and the implementation solutions for its various modules.

# 5. BACKEND ARCHITECTURE DESIGN

## 5.1 BACKEND INFRASTRUCTURE OVERVIEW

As a network-based application, the client software relies heavily on server-side processing for handling all public data. The server retrieves user data from the database and communicates it to the client. Thus, the backend, which includes the server and the database, is crucial for the effective functioning of the client application. This section introduces the backend architecture of the live video streaming system based on U3D and provides an overview of the interaction processes between the database and the server, as well as between the server and the client. This explanation aims to clarify the fundamental role of the backend in the overall project. The backend of the affiliated project of this thesis consists of several key components:

1. HTTP Server: Handles simple message-based communication.
2. WebSocket Server: Manages long-lived connections for real-time interactions.
3. Database: Utilizes MySQL for data storage.
4. Database Cache: Employs Redis as an intermediate caching layer.
5. Message Forwarding Server Cluster: Uses Nginx for distributing messages.
6. Static Resource Server Cluster: Also utilizes Nginx for serving static resources.

### 5.1.1 HTTP Server for Instant Communications

As a fundamental service protocol, HTTP facilitates interaction between the frontend and backend by responding to requests made by the frontend. When implementing the HTTP protocol in any project, it is essential to establish a consistent message format between the frontend and backend, specifying the required content. This content typically includes the response code, request result, and response message string. Although the response message string may not be practically useful for end-users, it serves as a valuable tool for developers, allowing them to quickly and intuitively understand the results of their requests without needing to consult backend documentation.

In two-way communication, maintaining a consistent message format is crucial for ensuring a robust application environment. However, the importance of this consistency also increases the cost of server maintenance, particularly regarding security. A compromised server could lead to severe consequences if hackers gain access to sensitive data and key information through malicious instructions.

In the affiliated project of this thesis, the HTTP service protocol is utilized for handling all simple message data. As described in the previous chapter, the client sends GET or POST requests to retrieve specific content from the database provided by the server. The backend processes several types of requests from the client, including:

1. User Registration: Recording the user's email address, username, and password.
2. Duplicate Email Registration: Informing the user if the email address is already registered.
3. Duplicate or Invalid Username Registration: Notifying the user if the username is already taken or invalid.
4. Incorrect Login Credentials: Alerting the user if the username or password is incorrect.
5. Successful Login: Updating the user's status in the database upon successful login.
6. User Online Status: Updating the user's status to "online" when they come online.
7. User Offline Status: Updating the user's status to "offline" when they go offline.
8. Busy Status: Updating the user's status to "busy" when they enter viewing mode, receive a call, are in a video call, or are streaming video.
9. Public Live Broadcast Information: Pushing information about live streaming rooms, including cover images, to online clients to inform users of the current streaming room status.
10. Online Personnel List: Providing the current list of online personnel and related live streaming information to users for making live video calls.
11. Guest Login: Providing a random nickname and unique user identification code for guest users to access all software functions.

The above operations illustrate the HTTP server's role in processing client requests. Notably, these requests are handled individually without the need for a persistent connection between the server and client. Hence, HTTP is selected for one-to-one processing of client requests.

5.1.2 WEBSOCKET SERVER FOR PERSISTENT COMMUNICATION

In real-time online applications, the HTTP protocol alone is insufficient to achieve long-lasting connections with excellent performance. While HTTP can handle short-lived connections effectively, maintaining a long connection requires the client to continuously send requests to the server. Shortening the interval between requests improves communication accuracy but significantly increases the number of requests, placing a substantial performance burden on the server, especially in large-scale projects. Although this approach was traditionally used, it has largely been supplanted by more efficient solutions.

A better solution might well be the protocols designed for persistent and long-lived connections. While TCP and UDP are transport layer protocols that can achieve long connections, they come with their own set of complexities. To address these issues, the WebSocket protocol was developed. As an application layer protocol built on top of TCP, WebSocket is compatible with HTTP services and facilitates long connections, thus providing an effective solution for maintaining persistent communication.

In the affiliated project of this thesis, the WebSocket server is employed to manage long connections, particularly for the client's heartbeat and online call modules. This approach is inspired by major network communication tools that use WebSocket to maintain a reliable connection between clients and servers. Many instant messaging applications utilize the WebSocket protocol for this purpose. To ensure that users can make online calls as if they were on a phone call, it is essential for users to be online. The heartbeat module plays a critical role in this. It enables the client to periodically send requests to the server to confirm its online status. If a request is not responded to in a timely manner, it indicates that the client may be offline. The software, therefore, includes mechanisms for user prompts and automatic reconnection attempts. This long connection approach ensures that User B remains reachable for calls from User A, provided User B is online and maintains stable communication with the WebSocket server. When User B comes online, the server records their status and assigns a unique identity code. This code is then made public in the online personnel list, allowing all users to view currently available contacts. Users can connect in real-time by selecting a specified user from this list. The actual process involves selecting a user name, which corresponds to a unique UUID[22] (Universally Unique Identifier) in the underlying code. The server maps usernames to UUIDs, ensuring that each UUID is unique and preventing duplicate entries in the mapping table.

Before a user is called, the WebSocket server checks the status of the recipient user from the database using the recipient's UUID. It then sends the calling user's UUID and video streaming URL to the recipient based on their current status. The recipient user can receive these messages while idle, and the content of the message will dictate the appropriate action. Specifically, if the UUID in the message is new and the URL is updated, this indicates that the recipient is being called by another user with the specified UUID. The local client then displays a prompt on the main interface to inform the user of the incoming call.

The client handles the user-facing interface, and the WebSocket server ensures accurate and timely delivery of call-related prompts. Given that user actions during a call—such as answering, accepting, or rejecting—are not easily tracked, the use of WebSocket for maintaining a long connection efficiently captures all user actions and establishes a reliable two-way connection. At the service level, when User A initiates a call to User B, the call remains active for a specified duration (e.g., 15 seconds, configurable via a settings file). This duration is inspired by traditional telephone calls, which typically have a set call cycle. If User B does not answer within this timeframe, the call request is terminated. This mechanism is similar to the disconnection and reconnection strategy mentioned earlier, where attempts continue until a successful connection is made.

Although the WebSocket server is compatible with the HTTP protocol, it is beneficial to separate these services to improve scalability and modularity. In the affiliated project of this thesis, the HTTP and WebSocket services are treated as distinct entities to enhance the efficiency and maintainability of each module. This separation aligns with best practices for large-scale commercial systems, even though the overall project size is relatively small. Such an approach ensures that failures or modifications in one service do not adversely impact the other, thereby optimizing user experience.

During the development of the WebSocket service, TypeScript was used as the development language, which was subsequently compiled into efficient JavaScript code.

---

[22] UUID: Universally Unique Identifier, a software construction standard and part of the Open Software Foundation in the field of distributed computing environments.

Although the use of Node.js for the WebSocket server is less common in the industry, I have documented and refined this code into a reusable framework. The core framework, referenced in this paper, is currently closed source but is planned to be open-sourced in the future.

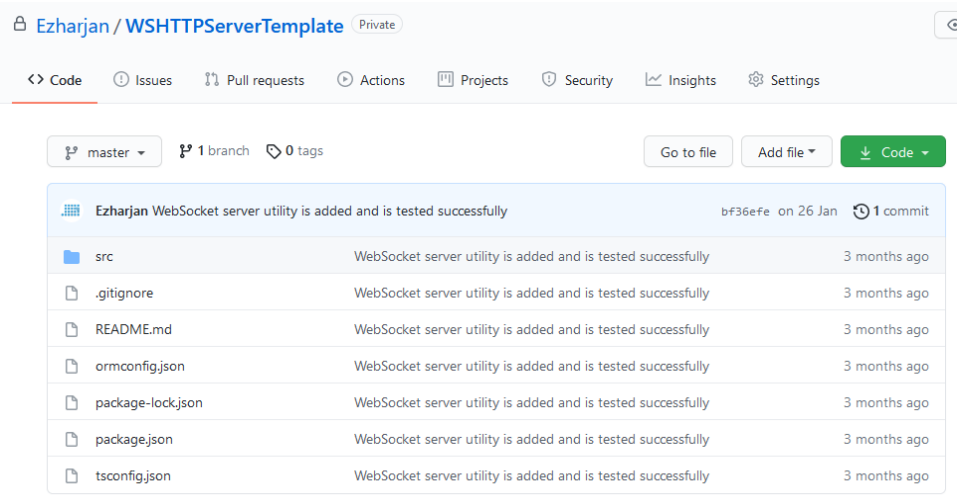

Figure 5.1.2-1 The WS-HTTP server framework is now open source and publicly available on GitHub.

The framework utilized for the affiliated project of this thesis depends on several third-party libraries to facilitate WebSocket and HTTP protocol services. These libraries include @types/body-parser, @types/express, @types/node, @types/ws, body-parser, express, module-alias, mysql, reflect-metadata, typeorm, typescript, and ws. This framework represents a general server development architecture combining WebSocket, HTTP, TypeScript, Express, and MySQL. It enables developers to efficiently develop WebSocket services and perform operations on MySQL databases using Node.js. As illustrated in Figure 5.1.2-2, the framework is organized into an EFR (Entity, Function, Routes) structure.

The "EFR" (namely Entity, Functions and Routes) structure employs TypeScript's decorator functions, which are not available in JavaScript, to efficiently manage MySQL database operations through the Entity layer. The Function layer implements specific functionalities invoked by the top-level Routes interface. Notably, Friend in this structure functions similarly to Function, specifically handling WebSocket-related services.

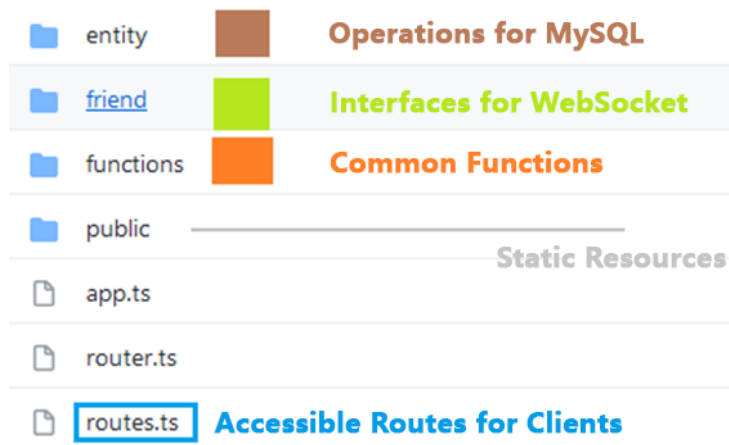

Figure 5.1.2-2 Entity, Function, and Routes structure of the backend framework.

Readers should not be confused by the "public" folder depicted in Figure 5.1.2-2. That directory is designated for storing static resources. When users request specific static resources, the WS-HTTP server retrieves the appropriate files from this directory. In many large-scale projects, a dedicated static server is employed to serve these resources. Similarly, the affiliated project of this thesis utilizes such a mechanism. Consequently, I incorporated this directory into the framework during the refinement process. In practical scenarios, static resources are served by an Nginx load balancing server. For further details, please refer to the sections discussing the load balancing server.

5.1.3 WEBSOCKET DUAL-END MESSAGE STRUCTURE

The project employs two servers: a WebSocket server and an HTTP server. The WebSocket server is essential for real-time communication of online status with each user, which is critical for establishing one-to-one live video streaming. To maintain a user's online status, the client periodically sends a heartbeat packet and status messages to the WebSocket server. In response, the WebSocket server acknowledges receipt with a confirmation message, ensuring both parties that the connection is active. Simultaneously, the WebSocket server updates and records the client's status based on these frequent messages. When another client attempts to connect, the WebSocket server uses the stored user status to facilitate the connection. Through investigation, it has been observed that many commercial WebSocket servers establish a stable connection with clients using this approach. The heartbeat packet is more than a simple connection check; it is a structured message containing the client's status. Upon receipt, the server sends a confirmation message, focusing on the client's status rather than just the connection status. This mechanism not only meets client needs but also addresses server requirements effectively. Early TCP-based servers used heartbeat packets primarily for connection verification, which was resource-intensive. A more efficient approach is to determine connection status based on the receipt of server messages. If the client does not receive messages from the server, it is assumed that the connection has been lost. Conversely, if messages are received, they should be correctly processed and recorded by the server, assuming adequate server capacity and normal operation.

The status message follows a specific format and structure. Each message must adhere to this general format, especially for one-to-one video calls. The structure of the message is as follows:

{ UserState, Calling, Called, CallResult, CallerUUID, CalleeUUID, StreamAddress}

The keys in this structure represent various request contents, and the server processes them accordingly. The complete message structure and its options are detailed below:

1. HeartBeat: {true, false},
2. UserState：{Online, Chatting, Offline, Pushing},
3. Calling: {true, false},
4. Called: {true, false},
5. CallResult: {Accepted, Rejected, StandStill}
6. CallerUUID: {/* caller's nickname */},
7. CalleeUUID: {/* caller's nickname */},
8. StreamAddress:{/* requester's streamming address */}

However, the client sends a heartbeat packet to the server to communicate its current status. This is distinct from the message structure previously described, necessitating two separate cases: one for heartbeat requests and another for other types of requests. For heartbeat requests, the meaning and format of the key-value pairs used in the UserState field are as follows:

UserState：Online, Chatting, Offline. Pushing.

In this context, the statuses are defined as follows:

- *Online*: Indicates that the user is currently online and available to receive one-to-one live video streaming requests from any user in their contact list.
- *Chatting*: Signifies that the user is either currently engaged in a video call or is in an active one-to-one live video streaming state.
- *Offline*: Represents the state when the user has activated "Do Not Disturb" mode (typically used while watching videos).
- *Pushing*: Refers to the state when the user is conducting a one-to-many live video streaming.

Upon receiving the status message packet from the client, the server immediately updates the database cache with this information. Given the frequent changes in user status, the information is not directly entered into the primary database but is cached in a Redis non-relational database for efficient access and updates. Concurrently, the server sends a confirmation message to the client. The format of the confirmation message is as follows:

$$StateCheck: 200$$

Here, StateCheck: 200 corresponds to the HTTP Response Code 200, indicating that the connection status between the client and server is normal. During the initial design phase, the decision between incorporating StateCheck into the HeartBeat key or using a direct echo mechanism was considered. Ultimately, the echo mechanism was chosen to avoid redundancy in the message structure. When the message received by the client matches the message sent, it confirms that the message is a heartbeat acknowledgment from the server. However, to prevent interference with client status updates, a dedicated StateCheck key was introduced, enabling the server to return the appropriate network status response code to the client.

In the message structure, the contact list serves the purpose of streaming the information of all online users to all clients. This allows clients to send one-to-one video call requests to other users. The contact list reveals the nickname information of each user, and the interface associated with each user's nickname includes the following details:

$$Nickname, UUID, StreamAddress.$$

In the proposed system, StreamAddress represents the user's current IP address encapsulated in the RTMP streaming address format, while UUID denotes the user's unique identification number on the server. This unique identification number is essential for the server to accurately send and forward requests.

The system employs a mechanism similar to that used in WeChat video calls for one-to-one live video interactions. For instance, if User A initiates a one-to-one live video request to User B, User A first retrieves User B's streaming address and UUID from the contact list. User B's UUID is used to send a request to the server, which then forwards the request to User B. Meanwhile, User A utilizes the streaming address to play User B's video stream locally. If User B accepts the call, User B begins streaming, and User A gains permission to view the stream. The complete message structure sent by User A to the server to request a live video streaming with User B is as follows (where User A is the Caller and User B is the Callee):

Calling: UUID_Of_Caller, StreamAddress_Of_Caller, UUID_Of_Callee.

From the message structure above, it is evident that a group of people requesting a one-to-one video call must provide not only the UUID of the recipient but also its own UUID and streaming address. Upon receiving the request, the server maps the UUID_Of_Callee from the request to the corresponding client in the client table and then sends the Caller's UUID and streaming address to facilitate communication between the two parties. When User B (the Callee) receives the video call request, they use the Caller's streaming address (provided by User A) to view the Caller's video stream. User B's decision on whether to accept or decline the call is sent to the WebSocket server, which uses the Caller's UUID to process the result. In the context of sending a one-to-one video call request, it is essential that none of the five keys in the message structure are left empty for non-heartbeat requests. Each value must adhere to the agreed-upon range to ensure consistency in the backend's message processing interface.

The use of a unified message structure simplifies subsequent parsing and processing on both ends. This section of development required considerable time and effort. Many message structures either become overly complex due to an overemphasis on initial design considerations or lack the flexibility needed for future expansions. The following section defines the universal message structure adopted in the final project, which is used for communication between both ends.

```
1.    enum UserState {
```

```
2.      Online = "Online",
3.      Chatting = "Chatting",
4.      Offline = "Offline",
5.      Pushing = "Pushing"
6.  }
7.  enum CallResult {
8.      Accepted = "Accepted",
9.      Rejected = "Rejected",
10.     StandStill = "StandStill"
11. }
12. function DataStruct(heartBeat: string, responseCode: string, userState: string, isCalling: string,
13.     isCalled: string, callResult: string, callerUUID: string, calleeUUID: string, streamAddress: string) {
14.     const result: string = `{
15.         "HeartBeat": "`+ heartBeat + `",
16.         "ResponseCode": "`+ responseCode + `",
17.         "UserState": "`+ userState + `",
18.         "Calling": "`+ isCalling + `",
19.         "Called": "`+ isCalled + `",
20.         "CallResult": "`+ callResult + `",
21.         "CallerUUID": "`+ callerUUID + `",
22.         "CalleeUUID": "`+ calleeUUID + `",
23.         "StreamAddress": "`+ streamAddress + `"
24.     }`
25.     return result;
26. }
27. export {
```

```
28.        UserState as UserState,
29.        CallResult as CallResult,
30.        DataStruct as FillDataStruct
31. }
```

Having navigated the complexities of message structure design, future challenges of a similar nature will be more manageable due to the insights gained. The message structure developed has proven robust and has required no fundamental changes since its inception. Consequently, the structure outlined in the above code has been consistently utilized in subsequent implementations.

### 5.1.4 MySQL Database for Data Storage

In the previous section, I discussed a general server development framework I developed, based on a WebSocket + HTTP + TypeScript + Express + MySQL architecture. The framework's compatibility with MySQL, as demonstrated through its Entity and dependencies, may raise questions for some readers. Specifically, why not utilize MongoDB or other non-relational databases that are often considered more popular with Node.js?

Firstly, it is true that MongoDB offers greater adaptability, higher flexibility, and is frequently used in Node.js-based backends. Despite these advantages, factors such as a shortage of high-end talent in related fields, the complexity and cost of migrating traditional architectures, and the associated risks have led many industries to continue using established technologies like MySQL. Node.js, while capable and supported by major industry players, is still not universally adopted as a pure server technology.

The choice to use MySQL in the affiliated project of this thesis was driven by several considerations. MySQL's relational database model offers significant advantages in handling complex relationships inherent in real-world scenarios. Although MongoDB and other non-relational databases can provide superior performance in certain contexts, the relational capabilities of MySQL were deemed more suitable for the project's requirements.

Additionally, TypeScript's role in this framework is noteworthy. By transforming JavaScript into a strongly typed language, TypeScript enhances maintainability and aligns well with MySQL's ORM databases. This combination is anticipated to become increasingly relevant in the industry.

This affiliated project of this graduation thesis attempts to employ advanced selection schemes and integrate them based on a forward-looking perspective. The goal was to address a wide range of factors, both current and future, creating a Node.js backend development framework characterized by high scalability, ease of use, and maintainability. This effort reflects not only a bold technological exploration but also a commitment to meeting diverse needs. As Fred Brooks famously stated, "There is no silver bullet in software engineering."

After opting to use MySQL, I encountered a limited selection of TypeScript libraries for interacting with MySQL. Following extensive testing of various libraries, I ultimately chose the TypeScriptORM library. This library leverages TypeScript's experimental feature—decorators—to simplify database entity operations and offers a more robust public interface for managing complex relationships.

In the affiliated project of this thesis, MySQL is employed to store user login information and live streaming-related data. This includes, but is not limited to, the user's live streaming URL, IP address, MAC address, the path to static resources for streaming covers, and live streaming start times. The database schema comprises three main tables: the user table, the user network information table, and the live streaming information table.

- User Table: This table stores user-specific data.
- User Network Information Table: This table records the user's current network data, including online status, network segment, and port number used for live streaming.
- Live Streaming Information Table: This table logs details of the user's public live streaming. It ensures that all public streaming and broadcasting activities are recorded and can be accessed by users requesting live streaming room data, thereby maintaining the transparency and reliability of the streaming rooms.

To optimize database utilization, it is crucial to design appropriate foreign keys between these tables. Effective use of foreign keys helps manage the relationships between tables and reduces redundancy. Beginners often face challenges when storing disparate information across multiple tables without establishing proper connections, leading to significant database redundancy. Although some degree of redundancy is acceptable, avoidable redundancy resulting from poor database design should be minimized.

Relational databases necessitate a higher level of design expertise compared to NoSQL databases, which often provide a more straightforward experience for beginners. Designing foreign keys between tables, such as between the user table and the user network information table, helps link related data, thus enhancing query efficiency and reducing redundancy.

As illustrated in Figure 2.1.4-1, the schema comprises three independent tables, each featuring a Primary Key (PRI) but lacking foreign keys. Notably, the repeated "name" field across all three tables indicates redundancy. While redundancy in the database content itself is a concern, it falls outside the immediate scope of database design and will not be discussed further here.

Figure 5.1.4-1 Structure of 3 user-related tables in the database of the proposed live streaming system.

Such design reflects my intention to scale the system as a large-scale commercial project. In such extensive projects, multiple databases are employed, including a middle-tier database used as a cache to enhance server-side database efficiency. This contrasts with traditional approaches that typically rely on a single database, with additional databases used primarily for backup purposes. As projects grow in complexity and diversify their functionalities, there is an increasing need for performance optimization. This necessitates the inclusion of various new modules designed to enhance backend performance. The middle-tier database serves this purpose by caching frequently accessed data and reducing the load on the primary database.

In the current design, which comprises only 3 tables as a basic example showcase, the user network information table will eventually be migrated to the middle-tier database. This approach will facilitate improved performance and scalability. While the present table design does not include foreign keys, future expansions of the database schema will incorporate these keys to further enhance database operation efficiency.

A middle-tier database, such as Redis, plays a critical role in modern large-scale projects. Its function is to provide fast access to frequently used data, thereby reducing latency and improving overall system performance. In the following sections, we will explore the role of Redis as a middle-tier database and its significance in contemporary project architectures.

### 5.1.5 REDIS FOR STORING KEY-VALUE PAIRS

In the previous section, I discussed the MySQL database and its fundamental advantages as a traditional solution for persistent storage of relational data. We will address the limitations of MySQL and present an effective solution to mitigate these drawbacks in this section. The concept of the "middle layer" has significantly diversified traditional backend architectures.

Traditionally, all data for frontend applications was rendered using server-side rendering (SSR). This approach meant that all user-visible information was processed by the backend and sent to the frontend for display. The frontend merely acted as a presentation layer, with minimal processing. While SSR centralizes functionality on the server, it also imposes performance constraints as the server bears a heavier load.

Recent advancements have introduced frontend rendering mechanisms, allowing the server to focus on enhancing database performance. The middle layer concept enables the frontend to directly request data from an intermediate server, thus improving efficiency. Consequently, non-relational databases such as MongoDB, CouchDB, Hypertable, and Redis have gained prominence as middle-layer solutions. By incorporating a middle layer between the frontend and the server, as well as between the server and the database, backend performance can be significantly improved, especially for CRUD operations.

Among the various middle-tier databases, Redis is widely recognized for its performance advantages and is extensively used. In the affiliated project of this thesis, Redis serves as the middle-tier database, enhancing both the efficiency of data operations and the security of the underlying MySQL database. Specifically, Redis is utilized to update user online statuses in real time, leveraging the WebSocket connection maintained with the client. The project includes a heartbeat mechanism that updates the user's network status in real time. As the number of users grows, relying solely on MySQL for real-time updates becomes a performance bottleneck. Although MySQL manages atomic operations at the field level, preventing table-wide locks, the performance overhead of frequent status updates can be substantial. Non-relational databases like Redis, which are optimized for key-value pairs and do not require complex data relationships or persistent storage, offer a more efficient solution. By simplifying data access and reducing overhead, Redis improves the overall performance and scalability of the backend architecture.

Redis is a key-value pair storage database implemented in C. It is one of the few open-source databases with robust support from a wide developer commU3D. Redis is versatile, supporting various use cases such as caching, event publishing and subscription, and high-speed queues. It also accommodates a range of data types, including strings, hashes, lists, sets, and sorted sets (zsets).

During the experimental phase of the affiliated project of this thesis, I encountered limitations with Redis on the Windows operating system, which restricted its full functionality. This experience highlighted the rapid evolution of technology: by the later stages of the project, Redis had become fully supported on Windows, making it as functional as on Linux. Ultimately, I opted to install and use Redis on Windows for development purposes. Although this approach is unconventional — enterprise-level projects typically use Linux due to its security and performance advantages — I chose it for ease of development. The method of use remains consistent across platforms.

The client updates its status every second, quickly forming a key-value pair relationship between the user's UUID and their current status. Integrating Redis into the framework was straightforward, thanks to the numerous third-party libraries available for non-relational databases in Node.js. After evaluating various options, I selected the most effective solution for development, completing the middle-layer implementation. Redis, functioning as a middle-tier database or database cache, addresses performance and security concerns in backend architecture. In a single-node server setup, the architecture typically features direct interaction between the database layer and the client, as depicted in Figure 5.1.5-1. This setup exposes the database directly to the client, potentially compromising both performance and security. Incorporating a middle-tier database effectively mitigates these issues, enhancing both performance and security by adding an intermediary layer between the client and the database.

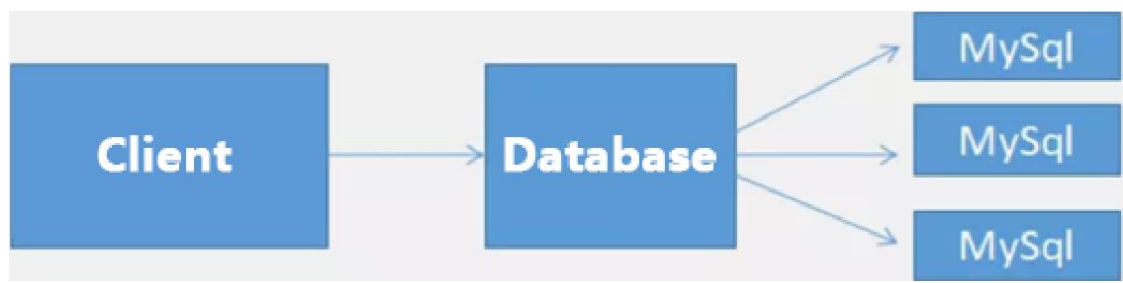

Figure 5.1.5-1 Traditional front-end and back-end architecture.

The concept of the middle layer has emerged with the expansion and increasing complexity of projects and useful operations. Directly using MySQL for handling massive amounts of data can lead to performance degradation, resulting in slow data reading and writing. Additionally, as previously mentioned, in the event of a server attack, the database is vulnerable, compromising backend security beyond basic defensive measures. The introduction of a middle layer addresses these issues effectively. In large-scale projects, architects can leverage caching to manage extensive data volumes more efficiently. Figure 5.1.5-2 presents the typical architecture of modern large-scale enterprise projects, where a middle-tier database, serving as a cache layer between the server and the MySQL database, enhances the system's security and performance. By incorporating Redis as a high-performance cache, the overall efficiency of the service is significantly improved, providing robust protection for the MySQL database while optimizing data handling and processing.

Figure 5.1.5-2 Common-used front-end and back-end architectures for modern large-scale enterprise projects.

In the architecture depicted in Figure 5.1.5-2, the database layer interacts directly with the cache. When data is present in the cache, it is promptly returned to the client. If the data is not available in the cache, a query is made to the MySQL database. This approach significantly alleviates the load on the database, enhances the efficiency of database operations, and simultaneously provides an additional layer of protection.

Figure 5.1.5-3 Redis Server running on a public network Cent OS server.

In the affiliated project of this thesis, Redis demonstrates its significant value, particularly in real-world scenarios. For instance, consider a situation where users are participating in a red envelope giveaway during a live streaming. Handling multiple users attempting to grab red envelopes simultaneously necessitates high performance, which traditional database locks may struggle to provide. In such cases, Redis's atomic operations offer an effective solution. The process involves storing the red envelopes in a list within Redis. Suppose there are 10 red envelopes; 10 numbers are pushed into the list, each representing a red envelope. As users attempt to grab a red envelope, a number is popped from the list, signifying that the user has successfully obtained a red envelope. When the list is depleted, it indicates that all red envelopes have been claimed.

The atomicity of the pop operation ensures that even with numerous concurrent requests, each operation is handled sequentially.

In large-scale real-world scenarios, requests to grab red envelopes are typically managed on the front end to prevent overload on the back end. Although these requests are intercepted and ordered on the front end, they are sequentially sent to the middle-tier database. This is where Redis's caching capabilities prove highly effective.

I installed Redis on my public network server (Linux CentOS) and tested its performance in real-world scenarios. The results showed that Redis achieved a reading speed of 110,000 operations per second and a writing speed of 81,000 operations per second, vastly outperforming MySQL's performance. This performance improvement is substantial. Figure 4.1.5.4 below illustrates the actual performance metrics of Redis when processing 1,000 request locks.

```
Last login: Sat May  1 21:48:25 2021 from 223.104.38.68
[root@iZ2ze7o9yr29yyxgrh17ooZ ~]# redis-benchmark -n 10000  -q
PING_INLINE: 91743.12 requests per second
PING_BULK: 95238.10 requests per second
SET: 96153.84 requests per second
GET: 96153.84 requests per second
INCR: 94339.62 requests per second
LPUSH: 95238.10 requests per second
RPUSH: 86206.90 requests per second
LPOP: 78740.16 requests per second
RPOP: 92592.59 requests per second
SADD: 97087.38 requests per second
HSET: 96153.84 requests per second
SPOP: 94339.62 requests per second
LPUSH (needed to benchmark LRANGE): 94339.62 requests per second
LRANGE_100 (first 100 elements): 54945.05 requests per second
LRANGE_300 (first 300 elements): 22321.43 requests per second
LRANGE_500 (first 450 elements): 17857.14 requests per second
LRANGE_600 (first 600 elements): 14792.90 requests per second
MSET (10 keys): 82644.62 requests per second
```

Figure 5.1.5-4 Results of Redis performance test on CentOS public server.

So why is Redis so fast? Firstly, Redis operates entirely in-memory. When persistence is required, data must be manually written to disk, specifically to the MySQL database in the affiliated project of this thesis. Secondly, Redis is single-threaded, which avoids the overhead associated with context switching in multi-threaded environments. Additionally, Redis employs simple data structures and straightforward data operations. It also utilizes a distinct underlying model compared to other databases. Unlike traditional systems that rely on general system calls and functions—which introduce latency—Redis integrates its own virtual machine mechanism, minimizing overhead. Finally, Redis implements a multi-channel I/O multiplexing model, utilizing non-blocking I/O operations. These factors contribute to its popularity and performance in the industry.

The discussion in this section provides a brief overview of the role of caching within the context of the affiliated project of this thesis. As data volumes increase, developers frequently need to implement master-slave replication technologies to achieve read-write separation. This topic extends beyond the scope of the current discussion. Since this article does not serve as a tutorial on related technologies, master-slave replication is not covered. Readers are encouraged to explore this topic independently. The subsequent sections will focus on the load balancing module utilized in the backend of the affiliated project of this thesis.

5.1.6 NGINX-BASED CONTENT DISTRIBUTION NETWORK

A Content Delivery Network (CDN) is a strategically deployed system encompassing distributed storage, load balancing, network request redirection, and content management. In the affiliated project of this thesis, the CDN's load balancing and network request redirection functionalities enable the client to automatically select the optimal network node to interact with the HTTP server based on the user's IP address, which indirectly reveals the user's geographical location. While the affiliated project of this thesis does not primarily rely on the HTTP protocol, the CDN's benefits become more apparent as the number of users and the scale of the project increase. The use of a CDN for forwarding, intercepting, distributed storage, and load balancing will significantly enhance performance. As illustrated in Figure 5.1.6-1, the early-stage content distribution structure includes a CDN layer positioned above the server. This layer interacts directly with the user, effectively isolating the server to some extent. When a user sends an HTTP request to access a website, there are multiple ways to reach the website: through the domain name or directly via the CDN node domain name. However, the client typically interfaces with a single top-level domain name. When the client requests live cover information from the static resource server, the intermediate request forwarding server identifies the request and directs it to the appropriate CDN node. This approach mitigates issues such as content duplication on a single server, server overload, and security vulnerabilities.

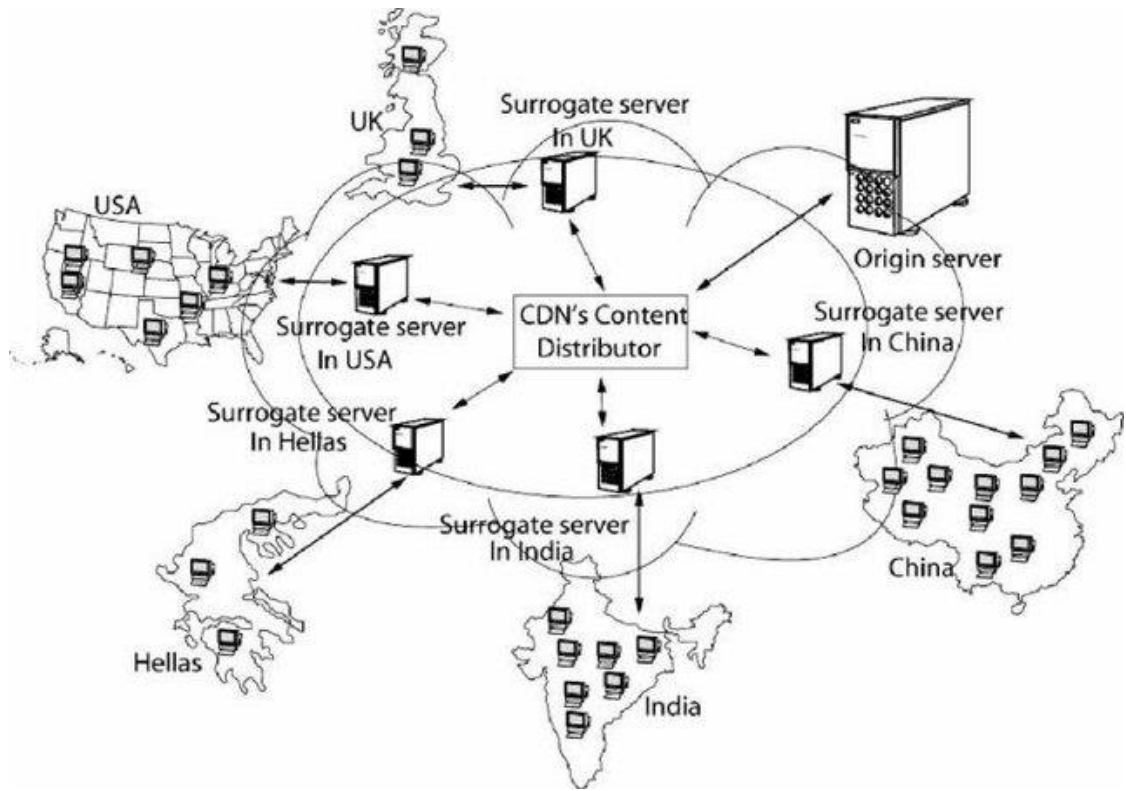

Figure 5.1.6-1 Content distribution network in the whole system. [10]

During the design phase of the affiliated project of this thesis, I opted to place static resources, specifically cover images, on a dedicated image hosting server. This strategy ensures that live cover data is managed more efficiently, with the image hosting server exclusively used for storing cover image information. Additionally, video data recorded during live streaming is stored on a separate static resource server designated for video streaming data. This server is configured with robust security measures to protect against unauthorized access.

Furthermore, a server cluster was implemented to handle client data requests. This cluster comprises multiple specialized servers, each responsible for specific functions. Some servers handle requests for video resources, others manage requests for image/cover resources, and additional servers are dedicated to processing database queries. By segmenting the backend infrastructure into function-specific servers, each server can utilize hardware optimized for its particular service requirements, enhancing both performance and security. A dedicated request forwarding server cluster manages the initial layer of request routing, thereby facilitating the creation of a comprehensive CDN network.

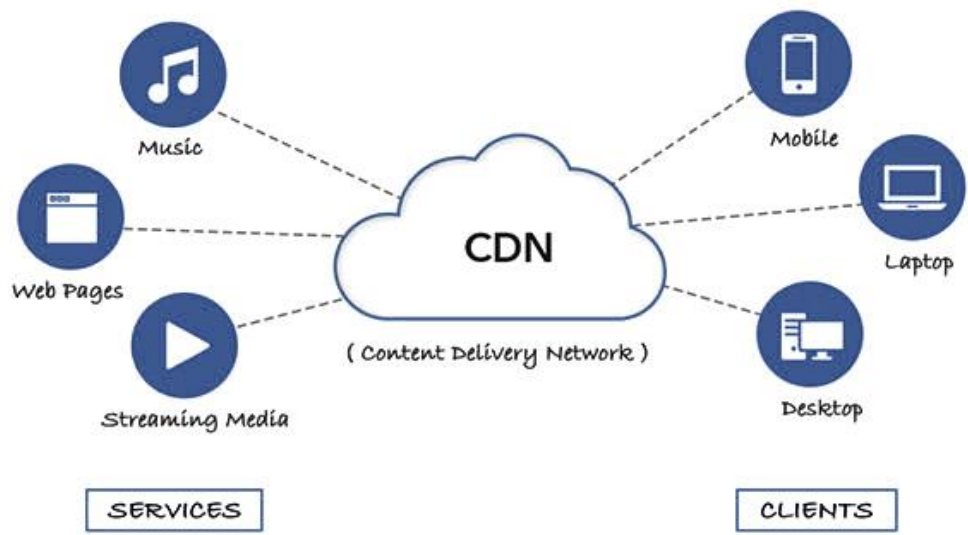

Figure 5.1.6-2 Using CDN to distribute different client requests to dedicated servers.

This mechanism is commonly employed in large-scale projects, and its advantages will be elaborated upon in the "Performance Optimization" chapter. It is important to note that this approach involves considerable financial and material resources, making it less feasible for smaller projects. The design chosen for the affiliated project of this thesis is intended to illustrate a theoretical expansion of the system, considering future scalability. In the affiliated project of this thesis, load balancing within the content distribution network is implemented using the Nginx server. An Nginx-based layer is established on the server side to intercept client requests. The Nginx server screens these requests and forwards them to the appropriate servers based on their specific content, thereby facilitating the content distribution network infrastructure. It is worth noting that large-scale CDN clusters are more common in industry practices. The design implemented in the affiliated project of this thesis represents a simplified version based on the existing conditions and resources.

5.2 IN-DEPTH IMPLENTATION OF BACKEND MODULES

This chapter begins with the fundamental approach to setting up the server and details the specific implementation steps, potential issues, and strategies for addressing each component of the backend system used in the affiliated project of this thesis. There are numerous methods for implementing each part of the backend, and listing all possible approaches for constructing an HTTP server would be impractical. The author will focus on the selection of a specific solution tailored to the project's characteristics. However, this focus does not imply that alternative approaches are infeasible or inapplicable. Given the rapid advancement of technology, top-level technical methods may soon be superseded by newer innovations. This chapter aims to provide a foundation by explaining underlying principles and detailing specific implementation methods, thereby enabling readers to develop a backend system suited to their own projects using the guidance provided.

5.2.1 BUILDING HTTP SERVER

Since the affiliated project of this thesis utilizes Node.js as the backend, I employed the Express framework to construct HTTP services. This choice aims to expedite the development of servers suitable for diverse commercial production environments. Express, a third-party library, can be incorporated into the project via NPM by executing the appropriate installation command. Once integrated, setting up the HTTP server is straightforward: simply configure the port settings in the main entry point and establish a listener to complete the setup.

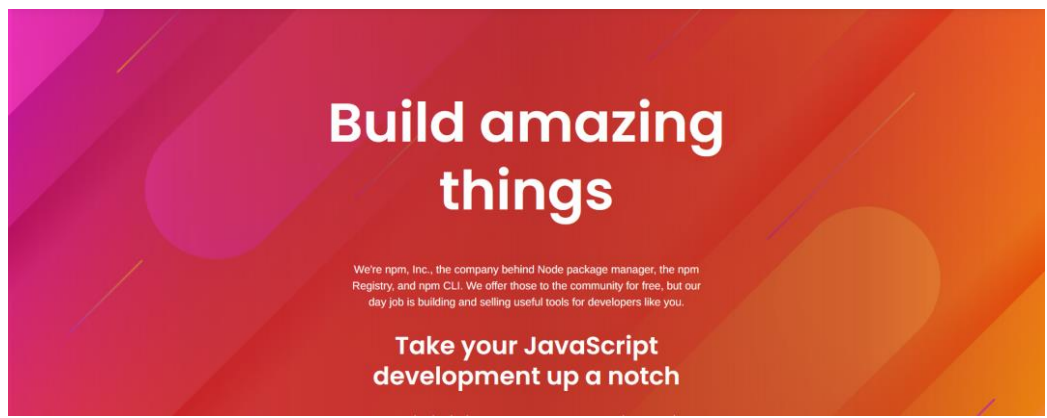

Figure 5.2.1-1 Home page of the NPM official website.

In the main entry script, additional third-party libraries can be incorporated as needed to extend the functionality of the project. The standard method for including these libraries in a Node.js project is through NPM commands. For unfamiliar or obscure code, developers can consult the NPM official website, which typically provides comprehensive documentation, usage instructions, and basic examples to facilitate rapid adoption. If further understanding is required, developers may also review the source code of the specified library. While it is possible to modify the source code to address specific needs, I advise against making such changes without contributing to the library's official repository (e.g., via a pull request). Altering the source code without such contributions can lead to inconsistencies, particularly in collaborative environments.

The final code included in the main entry script is shown below. In addition to the core Express library, the script integrates an ORM library for MySQL operations, a file name reference conversion tool, a data parsing library, and a file upload processing library. These libraries were selected based on their popularity and widespread use in the industry. The official documentation for these libraries is extensive, offering developers a more efficient development process.

```
1.  import "reflect-metadata";          // TypeORM library
2.  import "module-alias/register";     // importing the script via its alias (@/)
3.  import express from 'express';      // express
4.  import bodyParser from 'body-parser'; // request data
5.  import router from '@/router';      // router
6.  import objMulter from './objMulter' // the library for processing the context
7.  let port = 8080;
8.  const app = express(); // initialize express
9.  app.use(express.static('./public')); // This is just the root folder here the TS project needs
10. app.use(objMulter.any());
```

```
11.  app.use(bodyParser.urlencoded({ extended: true }));

12.  app.use('/', router);   // add routes

13.  app.listen(port, () => {

14.      console.log(`Application has started on Port ${port}`);

15.  });
```

After completing the main entry file, it is essential to expose external interfaces for client requests. In the Express framework, this is done similarly to Python's Flask library. By defining specific handlers for GET, POST, PUT, and DELETE requests within the public interface script, you can specify the paths for external requests. Below is a more distinctive approach utilized in the affiliated project of this thesis.

```
1.   import express from 'express';

2.   import visitIndexPage from './functions/visitIndexPage';

3.   import GetRandomName from './functions/getRandomName';

4.   import SaveUser from './functions/saveUser';

5.   import GetOnlineUser from './functions/getOnlineUser'

6.   interface callbackInterface {

7.       (request: express.Request, response: express.Response): void;

8.   }

9.   type Routes = {

10.      path: string;

11.      type: "GET" | "POST" | "DELETE" | "PUT";

12.      callback: callbackInterface;

13.  }[];
```

```
14.    export const routes: Routes = [

15.      { path: "/home", type: "GET", callback: (requst, response) => { response.send("Welcome to home page.") } },

16.      { path: "/index", type: "GET", callback: visitIndexPage },

17.      { path: "/test_user", type: "GET", callback: SaveUser },

18.      { path: "/random_name", type: "GET", callback: GetRandomName },

19.      { path: "/get_online", type: "GET", callback: GetOnlineUser },

20.      { path: "/client_log", type: "POST", callback: StoreUploadedFiles },

21.      { path: "/img_base", type: "POST", callback: StoreUploadedFiles },

22.    ];
```

To enhance clarity and ensure that the final usage is as convenient and efficient as defining routes in Think PHP5, I have created a dedicated script for handling callback functions. The code provided above is implemented following the setup of this callback function script. This approach clearly delineates the processing methods and specific paths for each interface, facilitating easier maintenance and future expansion. The internal details for configuring the callback functions are outlined below.

```
1.    import express from 'express';

2.    import { routes } from './routes';

3.    let router = express.Router();

4.    // Circuit to get the array introduced by Routes and listen to it through router

5.    routes.map((value) => {

6.      const { path, type, callback: callback } = value;

7.      if (type === "GET") { router.get(path, callback); }

8.      if (type === "POST") { router.post(path, callback); }
```

```
9.      if (type === "DELETE") { router.delete(path, callback); }

10.     if (type === "PUT") { router.put(path, callback); }

11. });

12. export default router;
```

Due to space constraints, here I omit the specific implementations of each interface's function body in this thesis. Instead, I provide only the fundamental top-level design to give readers a clearer and more practical understanding of how to build an HTTP server.

5.2.2 BUILDING WEBSOCKET SERVER

Since WebSocket services are inherently compatible with HTTP services, there is no necessity to construct a separate HTTP server when a WebSocket server is in place. However, as previously mentioned, a real-world backend environment typically comprises a vast cluster of specialized servers dedicated to specific functions. Therefore, the WebSocket server and HTTP server are implemented separately in the affiliated project of this thesis to align with the backend architecture that mirrors an actual real-world environment.

Building upon the foundational knowledge of constructing HTTP servers, a WebSocket server can be instantiated with a specified port number after incorporating the WS library. WebSocket connections exhibit multiple states, each requiring distinct callback functions to manage various scenarios effectively. The fundamental top-level design for this setup is illustrated in the following code snippet.

```
1. import "module-alias/register";

2. import { IncomingMessage } from 'http';

3. import * as WebSocket from 'ws';

4. import OnWSConnected from './onWSConnected'

5. import OnWSError from './onWSError'
```

```
6.    const wss = new WebSocket.Server({ port: 8383 });

7.    wss.on('connection', OnWSConnectedRoot);

8.    wss.on('error', OnWSError);

9.    function OnWSConnectedRoot(eachWebSocket: WebSocket, requestOfClient: Incoming
Message) { OnWSConnected(eachWebSocket, requestOfClient, wss); }
```

The methods prefixed with "On" represent callbacks executed in specific connection states, with their internal details not presented here. As illustrated in the example, the HTTP server operates on port 8080, while the WebSocket server uses port 8383. Although these ports are defined by the developer, it is typical for them to be the same. However, in this development setup, different port numbers are used to simulate two independent servers on a single machine. WebSocket communication involves intricate logic for interactions between the server and clients, as well as among clients themselves. The server must implement standardized response logic, and the client must correctly interpret and handle these responses. The most effective communication mechanism is one that is easily understood and processed by the client. The code below demonstrates the core functionality of the WebSocket server, detailing how the server responds to client requests. To maintain consistency with the response format of the HTTP protocol and facilitate unified processing by the client, the WebSocket server employs the JSON format for response information, adhering to the unified message structure described previously.

```
1.    import { IncomingMessage } from 'http';

2.    import * as WebSocket from 'ws';

3.    import { bytesToString, stringToBytes } from '../utils/Convertion'

4.    import SaveUserNetInfo from '../functions/saveUserNetInfo';

5.    import Dictionary from '@/utils/Dictionary';

6.    import { parseBool } from '@/utils/Convertion';

7.    import UpdateUserState from '../functions/updateUserState';

8.    import { UserState } from '@/entity/MessageStruct';

9.

10.   let gClientMap: Dictionary = new Dictionary();
```

```
11.  /**
12.   * On global WebSocket connected with a client.
13.   * @param aWebSocket Single WebSockets after connected with each clients;
14.   * @param WSServer Global WebSocket whic holds many clients after connection.
15.   */
16.  export default async function OnWSConnected(
17.      aWebSocket: WebSocket,
18.      clientRequest: IncomingMessage,
19.      WSServer: WebSocket.Server) {
20.      console.log('Connection established');
21.      let isUUIDStoredInDB: boolean = false;
22.      aWebSocket.on('message', (data) => {
23.          let reqDataStr: string = bytesToString(data);
24.          if (!isUUIDStoredInDB) {
25.              gClientMap.add(reqDataStr, aWebSocket); // Store current clients' uuid in a map
26.              writeClientIntoDB(clientRequest, reqDataStr);
27.              isUUIDStoredInDB = true;
28.          } else {
29.              let jsonResult = JSON.parse(reqDataStr);
30.              jsonResult["ResponseCode"] = 200;
31.              let userName = gClientMap.getKeyByValue(aWebSocket);
32.              (userName != null || userName != undefined) &&
33.                  UpdateUserState(userName, jsonResult["UserState"])
34.
35.              if (parseBool(jsonResult["HeartBeat"])) {
36.                  //console.log(jsonResult); // debugable here
37.                  let data2Send = JSON.stringify(jsonResult);
```

```
38.            aWebSocket.send(data2Send); // echo back to inform the client of network state
39.         }
40.         else {
41.            if (parseBool(jsonResult["Calling"])) {
42.               let calleeUUID = jsonResult["CalleeUUID"];
43.               jsonResult["Calling"] = false;
44.               jsonResult["Called"] = true;
45.               let data2Send = JSON.stringify(jsonResult);
46.               console.log("from Caller to Callee:" + data2Send); // debugable here
47.               sendMsgToClientByUUID(WSServer, calleeUUID, data2Send);
48.            } else {
49.               if (parseBool(jsonResult["Called"])) {
50.                  let callerUUID = jsonResult["CallerUUID"];
51.                  jsonResult["Calling"] = true;
52.                  jsonResult["Called"] = false;
53.                  let data2Send = JSON.stringify(jsonResult);
54.                  console.log("from Callee to Caller:" + data2Send); // debugable here
55.                  sendMsgToClientByUUID(WSServer, callerUUID, data2Send);
56.               } else {
57.                  console.log("Caller time out!");
58.                  // BOTH SIDE IS CURRENTLY NOT GOING TO CHATTING
59.               }
60.            }
61.         }
62.      }
63.   });
64.   aWebSocket.on('error', (err) => {
```

```
65.        console.log('Error: ' + err);
66.      });
67.      aWebSocket.on('close', (res, err) => {
68.        // callbacks after the connection with current WebSocket has been closed!
69.        console.log('Closed ' + res + (err ? (' in error: ' + err) : ' successfully!'));
70.      });}
71. function writeClientIntoDB(reqOfAClient: IncomingMessage, nickName: string) {
72.      const ip = reqOfAClient.socket.remoteAddress;
73.      const port = reqOfAClient.socket.remotePort;
74.      (ip != undefined && port != undefined)
75.         && SaveUserNetInfo(nickName, ip.toString(), port.toString(), UserState[UserState.Online]);
76. }
77. function sendMsgToClientByUUID(wsServer: WebSocket.Server, uuid: string, data: WebSocket.Data) {
78.      wsServer.clients.forEach((client) => {
79.         if (client.readyState === WebSocket.OPEN && client === gClientMap.find(uuid)) {
80.            client.send(stringToBytes(data.toString()));
81.         }  });  }
82. function setCurrenUserOffline(currentWebSocket: WebSocket,) {
83.      let userName = gClientMap.getKeyByValue(currentWebSocket);
84.      UpdateUserState(userName, "Offline");
85. }
```

Having completed the primary server setup for long-lived connections and the HTTP server for straightforward communication, the next step involves integrating the database and other backend modules to further enhance the backend architecture.

### 5.2.3 CONNECTING MySQL DATABASE

To use the MySQL database, you need to have MySQL service running, which does not impose strict requirements on the operating environment. You can download MySQL directly from the official website and install it. Prior to running MySQL, you must configure the database username and password and bind them to the server; otherwise, the server will not have the necessary permissions to manage the database. After installing and configuring MySQL, you can use the CREATE TABLE IF NOT EXISTS statement on the server to create tables without needing to manually set them up beforehand. Generally, the tables in the database are designed by database administrators for optimal server operation and usage.

For each entity in the MySQL database, you can define it in a separate script, which can then be compiled using TypeScript. For example, the script for the public live streaming room list information, when compiled, will automatically generate the specified fields in the MySQL database.

```typescript
1.  import { Entity, Column, PrimaryGeneratedColumn } from 'typeorm';
2.  @Entity()
3.  export class PublishedList {
4.    @PrimaryGeneratedColumn()
5.    id!: number;
6.    @Column()
7.    time!: Date;
8.    @Column()
9.    name!: string;
10.   @Column()
11.   title!: string;
12.   @Column()
13.   img_url!: string;
14.   @Column()
15.   ip!: string;
```

```
16.     @Column()
17.     port!: string;
18.     @Column()
19.     is_online!: boolean
20. }
```

CRUD operations can be performed at the server level, although the same effects can be achieved by directly executing operations on the MySQL database. Whether to use specific server interfaces for database CRUD operations or to interact with the database directly via SQL depends on the developer's preferences and the project's requirements. In traditional approaches, where the server does not offer a comprehensive database operation API, developers often use SQL directly to interact with the database. This approach eliminates the need for intermediate translation and can be more efficient at runtime. However, it may not be as developer-friendly due to the lack of code prompts and the verbosity of SQL, which can reduce development efficiency.

In contrast, the emergence of server-level database operation methods and tools has provided more convenient alternatives. This article demonstrates the use of an ORM (Object-Relational Mapping) library for database operations at the server level. ORM libraries abstract and simplify database interactions, making development more efficient and manageable. Below is an example of project code that utilizes an ORM library for database operations.

```
1.  async function CheckUserFromDB(username: string, password: string) {
2.      const connection = await createConnection();
3.      const repository = connection.getRepository(User);
4.      const result = await repository.createQueryBuilder("name")
5.          .where("name = :name", { name: username }).getOne();
6.      connection.close();
7.      if (result?.password === password) {
8.          return ResponseData.SuccessResponse;
```

```
9.         } else { return ResponseData.FailureResponse;} }
```

After employing a top-level encapsulation method for database operations, the code becomes more intuitive and easier to maintain. Additionally, developers can integrate Redis as an intermediary layer between the MySQL database and the server, serving as both a caching mechanism and an access accelerator. The setup and use of Redis are straightforward, similar to MySQL, with the added simplicity of its key-value storage model.

At this point, the fundamental backend structure is established. In many traditional approaches, backend development would typically conclude at this stage. However, in alignment with modern backend architecture, I have incorporated a content distribution network (CDN) to implement basic load balancing. Details on the specific implementation of a robust CDN will be provided in the subsequent section.

5.2.4 ESTABLISHING CDN

The fundamental principle of a Content Delivery Network (CDN) is reverse proxy. A reverse proxy involves using a proxy server to handle incoming connection requests from the Internet, forwarding these requests to an internal server, and then returning the server's response to the client. To the outside world, the proxy server appears as a node server. By deploying multiple reverse proxy servers, a reverse proxy server cluster is created, achieving a multi-node CDN setup.

The reverse proxy architecture illustrated below demonstrates that to enhance user experience, accelerate network speeds, and improve the security of user data and backend servers, a reverse proxy server is positioned in the middle layer of each set of Web servers (the affiliated project of this thesis includes both WebSocket and HTTP servers). The term "set" is used here to indicate that reverse proxy servers can be either singular or multiple. In large-scale commercial applications, multiple reverse proxy servers are typically deployed across various geographic locations—such as different countries, provinces, cities, and regions—to provide reliable proxy services to users worldwide.

The rationale for using one or more reverse proxy servers is to manage requests and user access information efficiently, especially in large-scale systems. To ensure service stability, additional reverse proxy servers are often added. This redundancy ensures that if one server fails, another server nearby can continue to forward requests to operational Web servers. The impact of a server failure in a commercial environment, where clients cannot access essential software or services, can be severe. The economic losses from user dissatisfaction and service interruptions can far exceed the costs of backend hardware. Therefore, while multiple servers can increase the financial burden, maintaining a stable and reliable service system is crucial for large-scale commercial projects. After all, consistent and robust service delivery is imperative for success.

The reverse proxy server in the middle layer primarily functions to forward user requests. This set of reverse proxy servers classifies users based on their IP addresses. In the affiliated project of this thesis, requests are further categorized by type. For instance, access requests for frequently accessed static files are routed to a specialized static file resource server, while requests related to client interface data are forwarded to a logic server dedicated to processing client data. This setup effectively manages user-side Web requests based on different IP segments and varying requirements.

This approach significantly enhances the stability and efficiency of the service running on the public network, while also improving the security of the public network server. By strategically handling different types of requests, it extends the service life of server hardware. Despite the increased cost associated with deploying multiple servers, this method alleviates server load and mitigates the impact of high concurrency on user requests.

In Figure 5.2.4-1, requests from clients with three distinct IP addresses are forwarded to different Web servers through the reverse proxy server. This method exemplifies a typical and effective request forwarding solution. For a practical analogy, when a user accesses the homepage of a well-known search engine, they do not connect directly to the homepage server. Instead, they are routed through the nearest reverse proxy server. Upon receiving the user's IP address and request content, the reverse proxy server evaluates the request and directs it to a relatively idle server for processing, based on this evaluation.

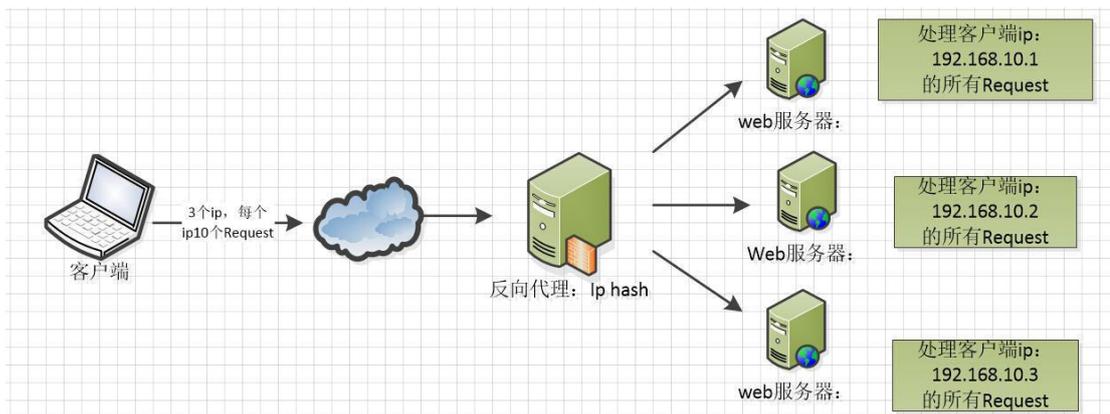

Figure 5.2.4-1 Backend architecture of the Nginx reverse proxy used in the project. [Figure Inserted from Original Thesis Source, 2021.7]

The reverse proxy mechanism fundamentally serves as a method of load balancing. Nginx is a widely adopted choice for reverse proxy servers due to its performance advantages, lightweight nature, stability, and cross-platform capabilities. In the affiliated project of this thesis, Nginx was employed to implement load balancing through reverse proxy.

In the experimental setup, the reverse proxy server was configured with a different IP address from the host server, achieving the same result as direct access to the host server's base address. Additionally, it is possible to deploy multiple Nginx servers as reverse proxies by adjusting port numbers. Given the limited resources (merely three hosts), this configuration provided a more realistic production environment, enhancing the final application's usability and maintainability. A straightforward content distribution network (CDN) can be established by deploying multiple Nginx servers. Nginx inherently supports reverse proxy functionality, making it convenient to use for this purpose. However, implementing a CDN requires multiple servers distributed across various locations. Using Docker Images allows for the creation of Nginx service containers, which can be quickly deployed across multiple servers. This approach streamlines the setup process and improves the efficiency of building a CDN by eliminating repetitive configuration tasks.To configure a single Nginx server for reverse proxy, follow these steps:

1. Install the Nginx Server: Nginx is versatile with respect to operating system requirements, and installation on Windows can be completed efficiently.

2. Configure Nginx: Access the configuration files in the Nginx installation directory. Set parameters such as the port number, forwarding routes, and request handling rules. Restart the server to apply the configurations.

Below is a basic example of an Nginx configuration file used in the affiliated project of this thesis:

```
1.    location / {
2.    expires 3d;
3.    proxy_set_header Accept-Encoding ";
4.    root /data/wwwroot/img.freehao123.com;
5.    proxy_store on;
6.    proxy_store_access user:rw group:rw all:rw;
7.    proxy_temp_path /data/wwwroot/img.freehao123.com/temp;
8.    if ( !-e $request_filename) {proxy_pass https://www.example.com;} }
```

Each property in the configuration can be customized according to specific requirements. For example, requests for retrieving image data can be directed to a dedicated image server, while requests for streaming video content can be forwarded to a streaming server. The configuration example provided above demonstrates how to forward all requests to a service hosted under the domain name "example.com", which processes the actual requirements. This example serves as a demonstration of configuration possibilities.

5.3 BACKEND R&D SUMMARY

This chapter begins with an overview of the backend design and elaborates on each module at the condition level of the project, progressing to the specific implementation methods of each module. It details the backend structure and implementation processes from high-level design to specific details, addressing both core components and auxiliary features. The chapter also uses the affiliated project of this thesis as a foundation to discuss more mature backend system architectures prevalent in the industry. Due to space constraints and the particular focus of the affiliated project of this thesis, many specific details and research findings related to the backend are not fully explored.

At this stage, all functionalities of the project have been implemented, culminating in a complete large-scale live video streaming system where each module has achieved a form suitable for commercial use. However, as the saying goes, "a journey of a thousand miles begins with a single step." Despite the completion of each part, ongoing optimization is necessary throughout the testing phase. Specifically, the performance optimization of live video streaming requires continuous efforts and long-term implementation. Maintenance and performance optimization will remain a crucial and ongoing endeavor.

## 5.4 LIMITATIONS

Since all components of the system have been developed and implemented, we will build upon the previous discussion by identifying the system's shortcomings and proposing corresponding improvement plans in this section. These improvements aim to enhance the system's performance and functionality in future implementations and maintenance. The identified shortcomings and their respective improvement plans are detailed in the table below:

| Functional Modules | Limitations | Potential Solutions |
| --- | --- | --- |
| Client | Using C++Encoding related provided by the underlying libraryAPIIt seems that the inter-frame coding efficiency is insufficient (or due to inappropriate usage) | C++The underlying library, or directly adopt a more efficient encoding algorithm in the industry, or use hardware encoding as appropriate. |
| | The client interface is not beautiful enough as a whole, and the pop-up window cannot be dragged freely | It is necessary to further improve the flexibility of pop-up windows according to user interaction habits, or even form a dedicated |

| | | underlying UISystem library for subsequent reuse. |
|---|---|---|
| Server CDNModules | Distributed storage is not implemented | The distributed storage methods in the industry need to be studied and referenced and incorporated intoCDNmodule, and add a mechanism to identify and prevent malicious requests before load balancing. |
| | Load balancing does not identify malicious requests | |
| Database | Relational databases do not fully reflect the relationship between tables | At present, due to the small amount of data to be stored and its low richness, it is necessary to further add relational models when subsequent projects expand and the data becomes complex. |

Table 5.4 Limitations of the proposed method and their potential solutions.

# 6. CONCLUSION

As the culmination of the my graduation thesis, this paper offers an in-depth exploration of various facets of the system, ranging from fundamental protocols to the advanced features of the client application and the comprehensive backend infrastructure. It delineates the research significance, objectives, implementation methodologies, encountered challenges, and devised solutions, culminating in a detailed account of the project's outcomes. This thesis functions both as a research document and as a developmental guide, providing insights into the system's design and operational aspects. Furthermore, it addresses practical issues encountered during development and presents strategies for their resolution. In the affiliated project of this thesis, I have conducted a thorough study, design, and implementation of a live video streaming client application along with its supporting backend system. The system leverages the RTSP protocol, H.264 encoding, Easy Movie Texture hardware decoding, and integrates multiple external video capture devices.

The primary aim of this thesis is to establish a theoretical foundation and practical methodologies for developing a low-latency, single-channel live video streaming system capable of supporting multiple external devices. Additionally, it summarizes the various challenges encountered and the corresponding solutions, offering a systematic approach to contemporary live streaming technologies. To address the limitations inherent in traditional multi-channel live streaming systems, especially with respect to multiple camera connections, this thesis proposes utilizing the U3D game engine's API to develop a low-latency, single-line live video streaming system. This approach effectively mitigates issues related to multi-camera integration while maintaining consistent delay across video streams. The system also introduces enhanced flexibility by allowing users to select and view video feeds from different cameras while ensuring uninterrupted data streaming. By employing U3D's Instantiated/Cloned GameObject concept, the system adapts dynamically to user-selected camera feeds, preserving the continuity and integrity of the live streaming. This implementation provides a robust theoretical framework for future advancements in video streaming technologies, promoting greater flexibility and adaptability in live streaming systems.

In the preceding chapters, I have extensively examined the live video streaming protocols and identified the RTSP protocol as the most suitable for the proposed system. Through rigorous testing, the H.264 encoding technology was selected as the optimal method for compressing video images due to its superior effectiveness compared to the custom packetization and unpacking strategies originally proposed. It is important to note that the choices made in this research represent the best solutions available at this stage and under the current conditions. While the quest for a universal solution is ongoing, the implementation of H.264 encoding proves to be a practical choice for the project's needs. The primary innovative contributions of this work lie in the integration of multiple external video capture devices and the real-time transmission of their video data. This area remains relatively underexplored in the market, providing a novel approach to live video streaming. The system can be analogously described as a "simple director" system, highlighting its ability to manage and display live streams from various camera sources. This analogy presents the system's functionality: similar to a director who orchestrates a live stream, the system enables users to view content controlled and mixed by the "director." The system supports a diverse range of external cameras, which can be seamlessly combined and configured, with the capability to display these combinations in one-to-one video calls.

By offering a more flexible and dynamic approach to video streaming, the proposed system enhances the scope and effectiveness of live streaming applications, allowing for greater customization and control over live video feeds.

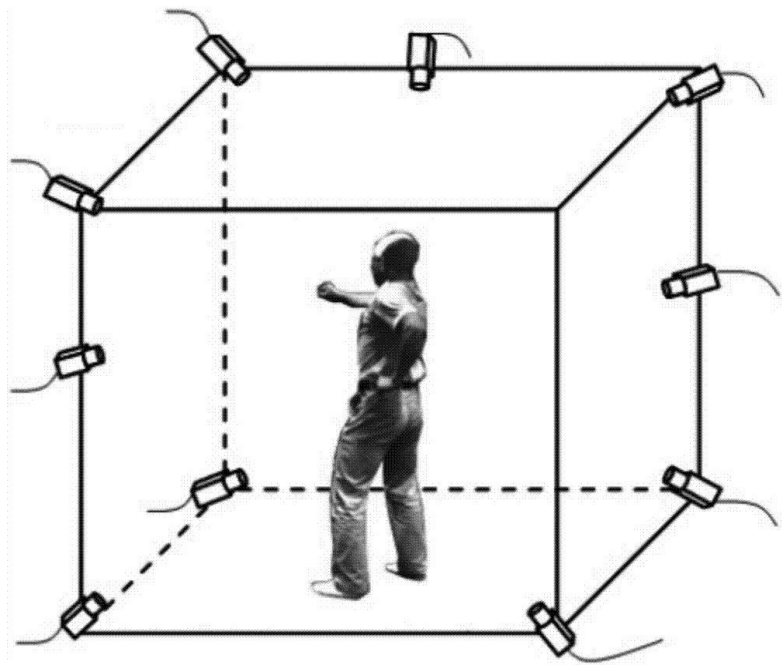

Figure 6.1 Schematic diagram of using multiple cameras to capture video in 3D reconstruction

To illustrate the practical applications of the proposed system, consider a scenario where User A and User B each connect four cameras in different directions to stream their content to one another. This setup allows each user to view different angles from the four cameras, enhancing the efficiency and directness of the video streaming. This example demonstrates a straightforward application of the system's capabilities. Moreover, the proposed system offers significant potential for advancing virtual reality (VR) technology, particularly in the realm of three-dimensional (3D) reconstruction. As depicted in Figure 6.1, current 3D reconstruction of real objects typically involves capturing detailed views from multiple camera angles. For real-time and efficient reconstruction, robust cloud services are essential. These services can integrate the multi-directional video streams transmitted by the clients and produce a real-time 3D model, which is then sent back to the clients.

The advantage of this approach lies in its elimination of redundant data transmission. Combined with the server's efficient real-time reconstruction capabilities, the system significantly enhances the efficiency of scene reconstruction on the client side. Clients require minimal equipment—just a few cameras and network cards to ensure stable transmission. The system allows clients to transmit multiple video feeds to the server, where machine learning techniques are employed to perform 3D reconstruction based on the incoming video data, even when resolution is insufficient. As illustrated in Figure 6.2, the server reconstructs real-world objects in real-time based on feature points and diverse video content, returning the reconstructed models to the clients.

This novel approach not only advances live video streaming but also holds promise for future developments in virtual reality technology, potentially shaping how 3D reconstruction and immersive experiences are achieved in the coming years.

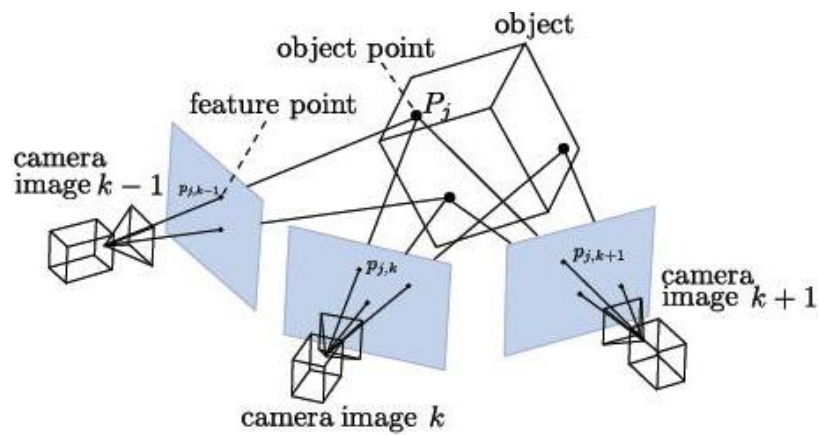

Figure 6.2 Illustration of cameras placement during a typical 3D reconstruction.

Additionally, the approach proposed in this paper offers more than just a solution for integrating live video streaming into the gaming industry; it also opens up extensive possibilities for virtual reality (VR) and augmented reality (AR) applications. As a live video streaming system designed and implemented using the U3D game engine, the proposed system leverages the theoretical foundations of real-time rendering from the gaming domain. The innovations and research findings presented in this paper are not only pertinent to the field of live video streaming but also extend their relevance to gaming and other VR-related fields. They provide a foundation for further development and expansion in these areas.

Future work could focus on enhancing the flexibility of live video streaming by allowing clients to control the content captured by multiple external devices in real-time during video streaming. This capability will make the live streaming experience more adaptable to various application scenarios. Expanding the project's scope to address diverse live video streaming needs will be a key direction for future development, building upon the current theoretical framework.

For this thesis, I have made all the my scripts open source [here](). The project page is [here]().

*Note that the references regarding commercial names are not included in this version.*